\documentclass[12pt,english]{article}
\usepackage[breaklinks,hypertexnames=false]{hyperref}
\hypersetup{colorlinks=true,linkcolor=blue,urlcolor=blue,citecolor=blue}
\usepackage{amsthm}
\usepackage{amsmath}
\usepackage{amssymb}
\usepackage{esint}
\usepackage[authoryear]{natbib}
\usepackage{har2nat}

\usepackage{subcaption} % For subfigures

\usepackage{appendix}
\usepackage[english]{babel}
\usepackage{enumitem}
\usepackage{booktabs}
\usepackage{multirow}
\usepackage{natbib}
\usepackage{textcomp,mathcomp}
\usepackage{adjustbox}
\usepackage{dcolumn}
\usepackage{booktabs}
\usepackage[normalem]{ulem}

\makeatletter
\usepackage{graphicx}
\usepackage{lscape}
\usepackage{rotating}
\usepackage[english]{babel}
\usepackage{latexsym}	
\usepackage{bbm}
\usepackage{tikz}
\usetikzlibrary{arrows,shapes,trees,positioning}
\usepackage{seqsplit}

\hypersetup{
	colorlinks=true,
	citecolor=blue
}

\makeatother

\newtheorem{assumption}{Assumption}

\newtheorem{remark}{Remark}

\newcommand{\defeq}{\vcentcolon=}

\input{tcilatex}
\usepackage{babel}

\begin{document}
	
	\title{Estimating Individual Responses when Tomorrow Matters\thanks{{We thank the Editor, the Associate Editor, four anonymous referees, Richard Blundell,} Adeline Delavande, Steven Durlauf, Pamela Giustinelli, Lars Hansen, Jim Heckman, Philip Heiler, Michael Keane, Arnaud Maurel, Pierre-Carl Michaud, Claudio Michelacci, Eduardo Morales, Chris Taber, Arne Uhlendorff, Wilbert Van der Klaauw, Daniel Wilhelm, and participants at various places for comments. Contact: \href{mailto:sbonhomme@uchicago.edu}{sbonhomme@uchicago.edu}; \href{mailto:angela.denis@bde.es}{angela.denis@bde.es}. The views expressed in this paper are those of the authors and do not necessarily reflect the position of the Banco de Espa\~na or the Eurosystem.}}
	\author{St\'{e}phane Bonhomme \\ University of Chicago \and Angela Denis \\Banco de Espa\~na}
	\date{Revised draft: \today
	}
	\maketitle
	\vskip 1cm
	
	\begin{abstract}
		
		We propose an approach to estimate how individuals' expectations influence their responses to a counterfactual change. The approach relies on average partial effects, which recover counterfactual impacts under conditions that we specify. We propose a three-step estimation method that relies on panel data on subjective expectations. We illustrate our approach in a model of consumption and saving, focusing on the impact of an income tax that not only changes current income but also affects beliefs about future income. Applying our approach to Italian survey data, we find that individuals' beliefs matter for evaluating the impact of tax policies on consumption decisions.

		%In an application to Italian survey data, we find that accounting for beliefs matters to assess how tax policies impact consumption decisions. 

	%	We propose a regression-based method to account for the role of expectations in how individuals respond to a policy that changes the economic environment. 
		
		\bigskip
		
		
	\end{abstract}
	
	\clearpage
	
	\global\long\def\ind{\mathbb{1}}
	\global\long\def\d{\mathrm{d}}
	\global\long\def\t{\intercal}
	\global\long\def\RR{\mathbb{R}}
	\global\long\def\defeq{:=}

	%% ==================================================================================
	%%                                     INTRO
	%% ==================================================================================

	\section{Introduction\label{secIntro}}
	
	Economists often seek to assess how changes in the economic environment affect individual decisions. A leading example is the \textit{ex ante} evaluation of policies that have not yet taken place. However, a key challenge is that, when the environment changes, individual decision rules are generally affected as well. In dynamic settings characterized by uncertainty, it is necessary to consider not only the immediate effect of the change but also its influence on expectations.
	
	%This question is at the core of the Lucas critique, and is of fundamental importance in macro and microeconomics.

	A common approach in applied work is to regress outcomes on covariates that one is interested in shifting in the counterfactual (e.g., under a new policy). Average partial effects based on regression estimates can be structurally interpreted as counterfactual policy effects under suitable conditions (\citealp{stock1989nonparametric}). However, underlying this interpretation is the assumption that the regression function remains invariant in the counterfactual. This invariance assumption can be restrictive in many settings where individuals' beliefs about the future matter. 
	
	Consider the introduction of a permanent income tax in a standard model of consumption and saving (see \citealp{deaton1992understanding}, for a textbook treatment). The effect of the tax can be estimated by regressing consumption on income (in logs), and by then computing an average partial effect associated with the tax change. However, such an effect is likely to be empirically misleading, since both current income and beliefs about future income will be affected by the tax. Not accounting for the change in beliefs will produce biased predictions of the effect of the tax, as emphasized by \citet{lucas1976econometric} in his influential critique. 
	
	As a second example, consider the effect of a change in the weather process in a model of agricultural production. Suppose that farmers choose dynamic inputs (such as irrigation or a fertilizer) based on their forecasts of future weather. In addition to affecting contemporaneous weather conditions, a change in the weather process will affect farmers' beliefs about future weather, which may lead them to modify their input choices. Not accounting for farmers' adaptation will bias calculations of the impact of a change in the weather process (\citealp{deschenes2007economic}, \citealp{burke2016adaptation}).

	%As an illustration, in the classic example of income and consumption in \citet{lucas1976econometric}, the effect of the introduction of a permanent income supplement affects consumption though two channels: current income, and income exectations. 
	
	In this paper, our aim is to study and estimate average partial effects in a dynamic framework that explicitly accounts for the role of individual expectations. In our intertemporal setup, individual beliefs are determinants of decisions, and they enter as additional state variables in the agent's decision problem. In this setting, we show how to assess the total effect of a counterfactual change by means of average partial effects calculations. In addition, we show how to decompose this total effect into a contemporaneous effect where beliefs are held fixed, and a purely dynamic effect that solely reflects the change in beliefs.
	
	To implement this approach we rely on data on subjective expectations. Belief data are increasingly available in a variety of settings (\citealp{manski2004measuring}). Given estimates of subjective probabilities based on survey responses, we account for beliefs in the definition and estimation of average partial effects. There are many examples of the use of expectations data on the right-hand side of a regression. Our contribution is to show how to interpret the estimates of such regressions, and to provide conditions under which those can be used for counterfactual prediction. 
	
	To interpret average partial effects, we propose a structural dynamic framework where agents choose actions based on their beliefs about the future. We use the framework to justify the use of average partial effects, yet we do not specify or estimate a structural model. As a result, the counterfactuals we focus on are restricted to changes in states of nature and beliefs about them, and our approach cannot answer other counterfactual questions related to changes in preferences or technology, for example. {Furthermore, we only study the impact of a counterfactual at time $t$ on outcomes at time $t$. Although the set of counterfactuals we focus on is limited, our approach to recover them relies on less restrictive assumptions than conventional approaches based on structural estimation.}

	In the structural framework that we outline, beliefs are time-varying state variables in the agent's decision problem. Variation in beliefs over time is crucial, since it allows us to control for preference heterogeneity, which we assume to be constant over time, {by allowing for individual unobserved heterogeneity}. {Variation in beliefs conditional on the other state variables is also key, in order to separately identify contemporaneous and dynamic effects.} We assume that current beliefs provide sufficient information to predict future beliefs, an assumption that we refer to as \emph{belief sufficiency}. We show this assumption is compatible with various popular models of belief formation, with and without rational expectations, including various forms of learning.

	The structural framework implies that the agent's decision rule is a function of exogenous state variables such as income or the weather, beliefs about them, and endogenous dynamic state variables {that depend on past actions}, such as assets or capital. We assess the effects of a counterfactual change by computing average partial effects which, unlike in the static case, account for changes in beliefs. Such effects correspond to well-defined structural counterfactuals under the assumption that the dynamic decision rule is invariant in the counterfactual. Hence, while we rely on a less restrictive invariance assumption than static average partial effects that do not allow for belief responses, a certain form of invariance is still needed to structurally interpret average partial effects in our setup.   

	To estimate average partial effects, we proceed in three steps that can be easily implemented given the availability of panel data on individual decisions and beliefs. In the first step, we estimate the subjective belief densities. This is straightforward in the case of beliefs about binary or discrete variables, in which case one can directly use the {empirical subjective} probabilities. For beliefs about continuous variables, to account for the fact that survey responses on subjective beliefs tend to be coarse, we assume that subjective densities depend on a low-dimensional parameter vector. In the second step, we estimate the regression function (i.e., the individual's decision rule). In the third step, we use these estimates to compute the impact on decisions of a counterfactual, given knowledge of how state variables and beliefs change under the counterfactual. Without additional assumptions, nonparametric identification is restricted to the empirical support of the conditioning variables. Moreover, the degree of individual heterogeneity that can be accounted for is limited by the length of the panel dimension.

	{To use our approach for counterfactual analysis, the researcher needs to specify the values that current exogenous state variables and beliefs about them would take in the counterfactual. We focus on changes involving a transformation of exogenous variables, such as an income tax, and assume full pass-through to the exogenous variable and the associated belief. For example, under a permanent proportional tax of 10\%, we assume that current income decreases by 10\% and that the income belief density is shifted downwards by the same amount. We discuss how this assumption could be relaxed with suitable data. }
	
	As an empirical illustration, we study how consumption decisions depend on current income and beliefs about future income. We rely on Italian data from the Survey on Household Income and Wealth (SHIW), which contains panel data on respondents' probabilistic income expectations for two consecutive waves. We then use our approach to predict the impact of various counterfactual income taxes, involving transitory or permanent increases in marginal tax rates, and a change in the degree of progressivity of the tax. Assuming that individuals fully incorporate the effects of the tax changes into their beliefs, we find that income beliefs shape consumption responses conditional on current income, and that they matter for predicting the effects of income taxes.

\paragraph{Related literature and outline.}

Subjective belief data are commonly included on the right-hand side of regressions. For example, \citet{guiso1999investment} study how a firm's investment depends on its beliefs about future demand; \citet{hurd2004effects} study the effects of subjective survival probabilities on decisions about retirement and social security claims; \citet{dominitz2007expected} analyze how beliefs about equity returns affect portfolio choice; \citet{bover2015measuring} studies how subjective expectations about future home prices affect car and secondary home purchases; and \citet{attanasio2019subjective} study how parental investment in children is influenced by beliefs about the production function. We provide assumptions under which such regressions can be interpreted structurally and used for counterfactuals within a dynamic framework.

\citet{manski2004measuring} (p. 1365) draws a distinction between expectations questions about unknown states of nature, which, combined with choice data, can be used to estimate econometric decision models, and questions about hypothetical choices under specified scenarios, which can be directly used to predict behavior. Our approach is designed for the first type of data (as in the examples mentioned in the previous paragraph), in the context of dynamic decision-making. This focus differs from a growing literature that relies on the second type of data, with the goal of providing methods for estimating heterogeneous treatment and policy effects using data on hypothetical choices (e.g., \citealp{arcidiacono2020ex}, \citealp{giustinelli2019seate}, \citealp{briggs2024identification}, \citealp{meango2023identification}, \citealp{bernheim2022causal}).

Our focus on the estimation of policy effects without a full structural model follows \citet{marschak1953economic},  \citet{ichimura2000direct,ichimura2002semiparametric}, and  \citet{keane2002estimating,keane2002estimating2}, among others; see also \citet{wolpin2013limits}. In our approach, we rely on subjective belief data and do not assume rational expectations. 

There is a growing literature on the combination of structural models and subjective belief data, see among others \citet{van2008social}, \citet{delavande2008pill},  \citet{van2012use}, \citet{stinebrickner2014major}, \citet{wiswall2015determinants}, \citet{an2021dynamic}, \citet{kocsar2022expectations}, \citet{o2023survival}, \citet{de2024evaluating}, and \citet{keiller2024production}; see also the recently released handbook on economic expectations (\citealp{bachmann2022handbook}). Our approach, which is tailored to specific counterfactuals, does not require to specify a full structural model.

Lastly, elicited beliefs about future income are increasingly available. Surveys with this information include the SHIW in Italy, the Survey of Economic Expectations and the Survey of Consumer Expectations in the US, the Survey of Household Finances in Spain, and the Copenhagen Life Panel in Denmark, among others. Previous contributions using income belief data include, among others, \citet{pistaferri2001superior}, \citet{guiso2002empirical}, and \citet{kaufmann2009disentangling}, who use data on income expectations in the SHIW in combination with models of consumption and saving; \citet{stoltenberg2022consumption}, who estimate a structural model with subjective income expectations using the same data; \citet{lee2023earnings}, who use data on subjective expectations and earnings realizations in Denmark to estimate a model where agents have partial information about earnings shocks; \citet{attanasio2020euler}, who combine data on subjective expectations with data on actual income and estimate an Euler equation for consumption; and \citet{arellano2024estimating}, who model and estimate the dynamic process of subjective income expectations using data from India and Colombia.
	
The outline is as follows. In Section \ref{sec_APE} we introduce average partial effects for dynamic settings. In Section \ref{sec_structural} we describe a structural framework and discuss the interpretation of average partial effects in this context. We present two examples in Section \ref{sec_examples}. We study identification and estimation in Section \ref{sec_ident1}, and we present our consumption application in Section \ref{sec_appli1}. Finally, in Section \ref{sec_conc} we describe some extensions of the approach. Replication files are available \href{https://drive.google.com/file/d/1ZGL6mDEhQ1JcV7cAkKt-hhxr3rkZofIj/view?usp=sharing}{online}.

\section{Average partial effects for dynamic settings\label{sec_APE}}

Suppose that a researcher has access to panel data on an individual outcome $y_{it}$ and some covariates $x_{it},z_{it}$, for {a cross-section of individuals $i$ and some time periods} $t=1,...,T$. To fix ideas, we will refer to the case where $y_{it}$ denotes consumption, $x_{it}$ is income, and $z_{it}$ includes other determinants such as assets. In addition, we assume the researcher has data about individual beliefs. We denote $i$'s subjective density of $x_{i,t+1}$ at time $t$ as $\pi_{it}$, and in this section we suppose that the researcher observes $\pi_{it}$. In practice, we have in mind situations where data about respondents' probabilistic expectations are available. Eliciting such responses is becoming increasingly common, see \citet{manski2004measuring} for a review. In Section \ref{sec_ident1} we will describe how we use elicited belief data to construct an empirical counterpart of the subjective density $\pi_{it}$. 

We postulate that, for some function $\phi_i$,
\begin{equation}y_{it}=\phi_i(x_{it},\pi_{it},z_{it})+\varepsilon_{it},\label{eq_novel}\end{equation}
where $\varepsilon_{it}$ has zero mean given $x_{it}$, $\pi_{it}$ and $z_{it}$. In the next section we will give conditions under which (\ref{eq_novel}) is obtained as the optimal decision rule for $y_{it}$ in a dynamic structural model. For example, in an intertemporal model of consumption and saving behavior, we will give conditions under which consumption $y_{it}$ depends, in addition to assets $z_{it}$ and current income $x_{it}$, on beliefs $\pi_{it}$ about next period's income $x_{i,t+1}$. 

Suppose the researcher is interested in documenting the impact, {in period $t$}, of an exogenous change in $x_{it}$ to some other value $x_{it}^{(\delta)}$, which in turn is associated with a change in beliefs from $\pi_{it}$ to $\pi_{it}^{(\delta)}$. An example is a proportional tax, corresponding (in logs) to $x_{it}^{(\delta)}=x_{it}+\delta$. {More generally, one may consider a transformation  $x_{it}^{(\delta)}=\delta(x_{it})$, with $\delta$ some function, in which case the whole distribution of $x_{it}$ changes in the counterfactual.} Then, $\pi_{it}^{(\delta)}$ is the belief about future log income $x_{i,t+1}$ under the tax. {However, we assume that the other factors $z_{it}$, which are predetermined, are not affected at time $t$ under the counterfactual, although they may change in subsequent periods. Hence, the tax has two distinct effects on period-$t$ outcomes: a contemporaneous effect associated with the change in $x_{it}$, and a dynamic effect associated with the change in beliefs $\pi_{it}$.}

To account for both impacts of the policy, we define the {period-$t$} total average partial effect, or TAPE, as
\begin{equation}{{\Delta_t^{\rm TAPE}(\delta)=\mathbb{E}\left[\phi_i\left(x_{it}^{(\delta)},\pi_{it}^{(\delta)},z_{it}\right)-\phi_i(x_{it},\pi_{it},z_{it})\right].}}\label{eq_delta_TAPE}
\end{equation}
We then further decompose this total effect as the sum of two terms: a contemporaneous APE (or CAPE), where beliefs are held constant, and a dynamic APE (or DAPE), which solely captures the change in beliefs. Formally, we decompose
\begin{align}{\Delta_t^{\rm TAPE}(\delta)}&{=\underset{=\Delta_t^{\rm CAPE}(\delta)}{\underbrace{\mathbb{E}\left[\phi_i\left(x_{it}^{(\delta)},\pi_{it},z_{it}\right)-\phi_i(x_{it},\pi_{it},z_{it})\right]}}+\underset{=\Delta_t^{\rm DAPE}(\delta)}{\underbrace{\mathbb{E}\left[\phi_i\left(x_{it}^{(\delta)},\pi_{it}^{(\delta)},z_{it}\right)-\phi_i\left(x_{it}^{(\delta)},\pi_{it},z_{it}\right)\right]}}.\label{eq_decompo}}\end{align}
{Note that these quantities measure the impacts of a policy introduced at time $t$ on outcomes at time $t$. In this paper we do not aim at recovering policy impacts on later outcomes, which would require additional assumptions.}

The structural framework in the next section  will allow us to transparently discuss the assumptions needed to structurally interpret these average partial effects (TAPE, CAPE and DAPE). The framework has two main features. First, $\pi_{it}$ is sufficient to predict future beliefs $\pi_{i,t+1}$, as formally defined in Assumption \ref{ass_beliefs} in the next section. This implies that $x_{it}$, $\pi_{it}$, and $z_{it}$ are the state variables in the economic model (in addition to some shocks subsumed in $\varepsilon_{it}$). This belief sufficiency assumption imposes restrictions on the belief formation process. However, we show it is satisfied in several popular models of beliefs. 

Second, structurally interpreting the average partial effects requires $\phi_i$ to be invariant to the policy change. In the structural model, $\phi_i$ depends on preferences, discounting, the law of motion of $z_{it}$, and the law of motion of the beliefs $\pi_{it}$. Consequently, one will need to assume that none of these quantities varies under the policy change. Assuming that the law of motion of the beliefs, which we denote as $\rho_i$, is invariant requires that, while agents account for the impact of the change on their beliefs about $x_{i,t+1}$, the way they update their beliefs after period $t+1$ is unaffected. Under this assumption, $\rho_i$ is an individual ``type'' that is invariant to the change. We will see that this assumption is automatically satisfied in a popular version of the consumption example.\footnote{Relaxing this assumption is conceptually straightforward in our framework, by defining $\pi_{it}$ in (\ref{eq_novel}) as beliefs about a sequence of future $x$'s, $x_{i,t+1},x_{i,t+2},...,x_{i,t+S}$. However, doing so imposes stronger demands on the data. We will return to this point in Section \ref{sec_conc}.}

It is informative to contrast our approach, which relies on the use of belief data and the dynamic decision rule (\ref{eq_novel}), to a static approach. Suppose instead that, for some function $g_i$, 
\begin{equation}y_{it}=g_i(x_{it},z_{it})+\varepsilon_{it},\label{eq_standard}\end{equation}
where $\varepsilon_{it}$ has zero mean given $x_{it}$ and $z_{it}$. A static average partial effect associated with the change in $x_{it}$ is 
\begin{equation}{\Delta_t^{\rm SAPE}(\delta)=\mathbb{E}\left[g_i\left(x_{it}^{(\delta)},z_{it}\right)-g_i(x_{it},z_{it})\right]}.\label{eq_APE}\end{equation}

To interpret $\Delta_t^{\rm SAPE}$ as the average impact on outcomes when $x_{it}$ changes to $x_{it}^{(\delta)}$, one needs to assume that the function $g_i$ in (\ref{eq_standard}) remains constant (\citealp{stock1989nonparametric}). This invariance assumption is often implausible in applications where dynamics matter. Indeed, in many settings where the current value of $x_{it}$ changes, beliefs about future $x_{it}$'s, which are implicit in the function $g_i$, are likely to change as well. For example, under a permanent income tax, both current income and beliefs about future income change. In contrast, in our approach based on (\ref{eq_novel}), we require $\phi_i$ to be invariant in the counterfactual. Although this assumption is not without loss of generality (and we will discuss it further in the context of a structural framework in the next section), it is weaker than the assumption that $g_i$ in (\ref{eq_standard}) is invariant to the change. The key difference is that, unlike (\ref{eq_standard}), (\ref{eq_novel}) explicitly accounts for variation in beliefs.

Finally, note that, when beliefs matter in (\ref{eq_novel}), an approach based on (\ref{eq_standard}) is incorrect for two reasons. The first one is that beliefs $\pi_{it}$, which are generally correlated with $x_{it}$ (though not collinear with $x_{it}$), are omitted variables in (\ref{eq_standard}). Hence, not accounting for $\pi_{it}$ gives incorrect contemporaneous APE estimands in general. The second reason is that relying on (\ref{eq_standard}) makes it impossible to recover the total APE, and to decompose it into contemporaneous and dynamic APEs. Hence, when (\ref{eq_novel}) holds, $\Delta_t^{\rm SAPE}$ defined in equation (\ref{eq_APE}) is not economically interpretable in general.  

 \section{Structural interpretation\label{sec_structural}}

In this section we describe a structural dynamic framework where individual decision rules take the form (\ref{eq_novel}), and we provide a structural interpretation for average partial effects.

\subsection{Economic environment\label{subsec_env}} 

Consider an individual $i$'s intertemporal decision making process in discrete time. In the presentation we first focus on a stationary infinite-horizon environment, and then show how to apply the framework to finite-horizon environments.

The timing is as follows. At the end of period $t-1$, the individual's information includes the history of exogenous state variables (i.e., states of nature) $x_{i,t-1},x_{i,t-2},...$, {which do not depend on past actions;} endogenous state variables $z_{i,t-1},z_{i,t-2},...$, {which depend on past actions;} actions $y_{i,t-1},y_{i,t-2},...$; and shocks (e.g., taste shocks) $\nu_{i,t-1},\nu_{i,t-2},...$. In addition, the individual may have observed other information, such as signals, that are relevant to her beliefs and future actions. Then, at the beginning of period $t$, $z_{it}$, $x_{it}$ and $\nu_{it}$ are realized and observed by the individual, and additional signals about future values $x_{i,t+1}$ may be observed as well. We denote the information set at that moment as $\Omega_{it}$. Given this information, the individual forms beliefs about $x_{i,t+1}$. Finally, she chooses the action $y_{it}$ based on the state variables in $\Omega_{it}$.

The individual's uncertainty about $x_{i,t+1}$ is represented by the subjective distribution of
$$	\left(x_{i,t+1}\,|\, y_{it},\Omega_{it}\right),$$conditional on her information set $\Omega_{it}$, and possibly contingent on her potential action $y_{it}$.\footnote{Here $y_{it}$ denotes a \emph{potential} action, contingent on which her beliefs are formed. In Assumption \ref{ass_beliefs0} we will rule out that beliefs may be contingent on actions. We will study the case of contingent beliefs in Section \ref{sec_conc}.} The belief distribution is subjective, and need not coincide with the realized distribution of $x_{i,t+1}$. In other words, we do not impose a rational expectations assumption. Our first assumption is that beliefs are not contingent on potential actions. Here and in the rest of this section, we use the shorthand $A\sim B$ to denote that $A$ and $B$ follow the same (subjective) distribution.\footnote{Throughout, densities are defined with respect to appropriate measures.}

\begin{assumption}{(beliefs)}\label{ass_beliefs0}
	\begin{equation*}
		\left(x_{i,t+1}\,|\, y_{it},\Omega_{it}\right)\sim \left(x_{i,t+1}\,|\, \Omega_{it}\right).
	\end{equation*} 
	We denote the corresponding conditional density as $\pi_{it}(x_{i,t+1})$.
\end{assumption}

We will refer to $\pi_{it}$, which is the individual subjective density of $x_{i,t+1}$, as the belief density, or simply as the ``beliefs''. $\pi_{it}$ is an element of $\Omega_{it}$, and it is a random function. Assumption \ref{ass_beliefs0} requires that beliefs about $x_{i,t+1}$, which are relevant to the action $y_{it}$, do not depend on $y_{it}$. In other words, beliefs are not contingent on potential actions. At the same time, Assumption \ref{ass_beliefs0} allows past choices $y_{i,t-1},y_{i,t-2},...$ to influence current beliefs $\pi_{it}$. In Section \ref{sec_conc}, we will outline a generalization of Assumption \ref{ass_beliefs0} where agents have so-called ``state-contingent'' beliefs{; for example, beliefs about wages contingent on working in a particular sector}. The framework is unchanged in that case, except for the fact that $\pi_{it}$ then consists of a set of conditional densities ({of, e.g., wages}) indexed by potential action values $y$ {(e.g., sector participation)}. 

 We make the following assumption regarding belief updating.

{
\begin{assumption}{(belief sufficiency)}\label{ass_beliefs}
	\begin{equation*}
		\left(\pi_{i,t+1}\,|\,  x_{i,t+1},y_{it},\Omega_{it}\right)\sim \left(\pi_{i,t+1}\,|\, x_{i,t+1},\pi_{it},x_{it}\right).
	\end{equation*}
We denote the corresponding conditional density as $\rho_i(\pi_{i,t+1}; x_{i,t+1},\pi_{it},x_{it})$.
\end{assumption}

We will refer to $\rho_i$ as the belief updating rule. Belief sufficiency, as stated by Assumption \ref{ass_beliefs}, is a key condition in our framework. It requires that current beliefs $\pi_{it}$, along with $x_{it}$ and $x_{i,t+1}$, be sufficient statistics for $\Omega_{it}$ when predicting future beliefs. Moreover, Assumption \ref{ass_beliefs} requires that beliefs are not affected by past actions, which may be plausible in some settings. For example, in a consumption model there may be no feedback from past consumption choices to future income beliefs. However, in other settings, it may be important to allow future beliefs $\pi_{i,t+1}$ to depend on past actions $y_{it}$. This is allowed for by the following generalization of Assumption \ref{ass_beliefs}.

\begingroup
	\renewcommand\theassumption{2$^{\prime}$}
\begin{assumption}{(belief sufficiency, extended)}\label{ass_beliefs_mod}
	\begin{equation*}
		\left(\pi_{i,t+1}\,|\,  x_{i,t+1},y_{it},\Omega_{it}\right)\sim \left(\pi_{i,t+1}\,|\, x_{i,t+1},y_{it},\pi_{it},x_{it},z_{it},\nu_{it}\right).
	\end{equation*}
	We then denote the corresponding conditional density as $\rho_i(\pi_{i,t+1}; x_{i,t+1},y_{it},\pi_{it},x_{it},z_{it},\nu_{it})$.
\end{assumption}
\addtocounter{assumption}{-1}
\endgroup

Assumptions \ref{ass_beliefs} and \ref{ass_beliefs_mod} have similar implications in terms of policy rules, and both can be used to justify decision rules of the form (\ref{eq_novel}). {We will discuss belief sufficiency further below and show that it is consistent with a variety of belief formation processes.} 
}

Next, we make the following assumption regarding the endogenous state variables $z_{it}$.

\begin{assumption}{(endogenous state variables)}\label{ass_zit}
	{	\begin{equation*}\left(z_{i,t+1}\,|\, x_{i,t+1},\pi_{i,t+1},y_{it},\Omega_{it} \right)\sim \left(z_{i,t+1}\,|\, z_{it},x_{it},y_{it} \right).\end{equation*}
	We denote the corresponding conditional density as $\gamma_i(z_{i,t+1}; z_{it},x_{it},y_{it})$.}
\end{assumption}

{Assumption \ref{ass_zit} nests cases where $z_{i,t+1}=\gamma_i(z_{it},x_{it},y_{it})$ is non-stochastic, such as a standard budget constraint. Moreover, $\gamma_i$ could additionally depend on $\pi_{it}$, $x_{i,t+1}$, or $\pi_{i,t+1}$, although we abstract from this dependence for conciseness.}

Lastly, we make the following assumption regarding the shocks $\nu_{it}$.

\begin{assumption}{(shocks)}\label{ass_shocks}
		\begin{equation*}
		\left(\nu_{i,t+1}\,|\, x_{i,t+1},\pi_{i,t+1},{z_{i,t+1}},y_{it},\Omega_{it} \right)\sim \nu_{i,t+1}.
	\end{equation*}
We denote the corresponding density as $\tau_i(\nu_{i,t+1})$. 
\end{assumption}

{The independence condition in Assumption \ref{ass_shocks} is commonly made in structural models where alternative-specific taste shocks are serially uncorrelated. The presence of serially correlated time-varying unobservables would invalidate this assumption.} 
	
{In this environment, we will focus on counterfactuals involving changes in variables $x_{it}$ and beliefs $\pi_{it}$, associated with counterfactual values $x_{it}^{(\delta)}$ and $\pi_{it}^{(\delta)}$, respectively. We assume that $x_{it}^{(\delta)}=\delta(x_{it})$ is a deterministic transformation of $x_{it}$. For example, $\delta(\cdot)$ is a tax schedule (e.g., proportional or progressive), or a transformation of temperature (e.g., a mean shift).\footnote{Note that the assumption that $x_{it}$ responds fully to $\delta(\cdot)$ is consistent with our framework where $x_{it}$ is an exogenous state variable.}

	Further, we assume that counterfactual beliefs are equal to the beliefs under the transformation $\delta$; that is, that $\pi_{it}^{(\delta)}$ is the subjective density of $$	\left(\delta(x_{i,t+1})\,|\,\Omega_{it}\right),$$
	which is simply the density of the transformed random variable $\delta(x_{i,t+1})$ for $x_{i,t+1}\sim \pi_{it}$. Consider as an example a permanent 10\% proportional tax change, where $\delta(x)=x-0.10$ (in logs). We assume that beliefs are equal to $\pi_{it}^{(\delta)}(x)=\pi_{it}\left(x+0.10\right)\equiv \pi_{it}^{(\delta,full)}(x)$. This amounts to assuming full pass-through of the tax onto the beliefs, which holds if individuals think of the change as being permanent. In Subsection \ref{subsubsec_beliefs} we will return to this point, and introduce a sensitivity analysis approach where we vary individuals' expectations about the counterfactual remaining in place in the future. In that case, $\pi_{it}^{(\delta)}$ will be a mixture between $\pi_{it}^{(\delta,full)}$ and the baseline $\pi_{it}$. Also, note that while here the tax only affects mean beliefs, other $\delta$ transformations may affect the entire belief density.

}

\subsection{Compatibility with belief formation models\label{subsec_belief_form}} 

We now illustrate that our belief sufficiency conditions, Assumptions \ref{ass_beliefs} and \ref{ass_beliefs_mod}, are consistent with several models of belief formation in economics, see \citet{pesaran2006survey} for references. 

\paragraph{Latent components.} As a first example, suppose that agents have rational expectations, and that $x_{it}=\eta_{it}+\varepsilon_{it}$ where $\eta_{it}$ follows a homogeneous first-order Markov process, and $\varepsilon_{it}$ is independent of $\eta_{it}$ with a stationary distribution. Suppose that agent $i$'s information set at time $t$ is 
$$\Omega_{it}=\{x_{it},x_{i,t-1},...,\eta_{it},\eta_{i,t-1},...\}.$$
An example is a permanent-transitory specification of the income process, as in our consumption example in Section \ref{sec_examples}. Note that $\pi_{it}$, which is the conditional density of $x_{i,t+1}$ given $\Omega_{it}$, coincides with the conditional density of $x_{i,t+1}$ given $\eta_{it}$. Given that $\eta_{it}$ follows an exogenous and homogeneous first-order Markov process, this implies that Assumption \ref{ass_beliefs} is satisfied. However, note that Assumption \ref{ass_beliefs} generally fails in this model if $\eta_{it}$ is not first-order Markov.   

\paragraph{Learning (exogenous beliefs).} As a second example, suppose that $x_{it}=\alpha_i+\varepsilon_{it}$. Suppose that agents do not know $\alpha_i$, and that they try to learn it given the observations $x_{it}$. Suppose in addition that $\varepsilon_{it}$ is i.i.d. Gaussian, and that agents are Bayesian decision-makers with Gaussian priors about $\alpha_i$ and rational expectations. We show in Appendix \ref{app_belform} that belief sufficiency, as stated by Assumption \ref{ass_beliefs}, holds. This follows from the form of the updating equations for the posterior mean and variance of $\alpha_i$, see (\ref{eq_mean})-(\ref{eq_var}) in Appendix \ref{app_belform}. Note that this example does not allow for learning from past choices, since beliefs are exogenous. 

%More generally, in our framowork we do not allow for agents to learn from other agents.

\paragraph{Learning (endogenous beliefs).} As a third example, consider a case where there are two possible choices $y_{it}=1$ and $y_{it}=0$. Suppose that the agent observes $x_{it}=\alpha_i+\varepsilon_{it}$ no matter what action she chooses, and that she observes an additional signal $s_{it}=\alpha_i+v_{it}$ only when choosing $y_{i,t-1}=1$. Suppose in addition that $(\varepsilon_{it},v_{it})$ is Gaussian {and i.i.d., that $\varepsilon_{it}$ and $v_{it}$ are independent}, and that agents have rational expectations and have a Gaussian prior about $\alpha_i$. We show in Appendix \ref{app_belform} that {Assumption \ref{ass_beliefs_mod} is satisfied.} This again follows from the form of the updating equations for the posterior mean and variance of $\alpha_i$, which here are conditional on the past action $y_{i,t-1}$; see (\ref{eq_var2})-(\ref{eq_mut}) in Appendix \ref{app_belform} for the case $y_{i,t-1}=1$. {Moreover,} in this example, beliefs are endogenous in the sense that they are affected by past choices.\footnote{{Note that} beliefs are not state-contingent in this example, and Assumption \ref{ass_beliefs0} holds. We will show in Section \ref{sec_conc} that our framework can be extended to allow for state-contingent beliefs, and we will provide a learning model as an illustration.} {Hence, while Assumption \ref{ass_beliefs_mod} holds, Assumption \ref{ass_beliefs} is not satisfied in this example.}  

{To see a case where Assumption \ref{ass_beliefs_mod} fails, consider the same setup but now with $v_{it}$ an AR(1) process, so signals $s_{it}$ are serially correlated. In this case, we show in Appendix \ref{app_belform} that $\pi_{i,t+1}$ is not independent of $s_{it}$ conditional on current beliefs $\pi_{it}$ and other state variables. In this example, beliefs $\pi_{it}$ are \emph{not sufficient} for future beliefs $\pi_{i,t+1}$, since signals have predictive power for future beliefs conditional on current beliefs and other state variables. Hence, Assumption \ref{ass_beliefs_mod} does not hold.}

\paragraph{Adaptive expectations.} Our setup is also compatible with some models of non-rational expectations. As an example, consider a simple model of adaptive expectations, where mean beliefs evolve as
\begin{equation}\mathbb{E}_{\pi_{it}}(x_{i,t+1})=\mathbb{E}_{\pi_{i,{t-1}}}(x_{it})+\lambda_i\left(x_{it}-\mathbb{E}_{\pi_{i,t-1}}(x_{it})\right).\label{eq_adaptive}\end{equation}
\citet{armona2019home} refer to individuals with $\lambda_i>0$ as ``extrapolators'', to those with $\lambda_i=0$ as ``non-updators'', and to those with $\lambda_i<0$ as ``mean reverters''. Assumption \ref{ass_beliefs} is satisfied if (\ref{eq_adaptive}) holds, and, say, beliefs are normally distributed with constant variance $\sigma_i^2$. More generally, Assumption \ref{ass_beliefs} is consistent with models of adaptive expectations where the entire belief density $\pi_{it}$ depends on $\pi_{i,t-1}$ and $x_{it}$.

This discussion provides several examples of belief formation models where belief sufficiency, as stated by Assumption \ref{ass_beliefs} {or Assumption \ref{ass_beliefs_mod}}, holds. Under either assumption, along with Assumptions \ref{ass_beliefs0}, \ref{ass_zit} and \ref{ass_shocks}, the vector $(x_{it},\pi_{it},z_{it},\nu_{it})$ contains all the relevant state variables when making the decision. An advantage of our approach is that, since beliefs $\pi_{it}$ are state variables, we can study counterfactuals that account for changes in beliefs without the need for a full-fledged structural model.

\subsection{Decisions and policy rule}

Let $u_i(y_{it},x_{it},z_{it},\nu_{it})$ denote period $t$'s contemporaneous payoffs.\footnote{Here $\pi_{it}$ are not payoff-relevant. However, the nonparametric decision rule in (\ref{eq_a}) will remain the same if payoffs $u_i(y_{it},x_{it},\pi_{it},z_{it},\nu_{it})$ depend on $\pi_{it}$.} Here the action may be continuous or discrete, so our framework covers structural dynamic discrete choice models as well as models with continuous choices. It also covers settings with vector-valued actions, including mixed discrete-continuous choices (e.g., \citealp{bruneel2022discrete}). {We consider a standard setup where individuals maximize the expected discounted sum of utilities, with a constant discount factor $\beta_i$.} The individual solves the infinite-horizon program
$$(y_{i,1},y_{i,2},...)=\limfunc{argmax}_{(y_1,y_2,...)}\, \mathbb{E}\left[\sum_{t=1}^{\infty}\beta_i^{t-1}u_i\left(y_{t},x_{it},z_{it},\nu_{it}\right)\right],$$
where the expectation is taken with respect to the process of $x_{it},\pi_{it},z_{it},\nu_{it}$ for given values $(y_1,y_2,...)$, as prescribed by Assumptions \ref{ass_beliefs0}, \ref{ass_beliefs}, \ref{ass_zit}, and \ref{ass_shocks}. 

Let $V_i(x,\pi,z,\nu)$ denote the value function associated with any given state $(x,\pi,z,\nu)$. Bellman's principle then implies\footnote{Here the integral in $(x_{t+1},\pi_{t+1},z_{t+1},\nu_{t+1})$ is taken relative to an appropriate measure.}
\begin{align}
&	V_i(x_t,\pi_t,z_t,\nu_t)=\limfunc{max}_{y_t}\, \bigg\{u_i(y_t,x_t,z_t,\nu_t)\notag\\
	&+{\beta_i \int V_i(x',\pi',z',\nu')\pi_{t}(x') {\rho}_i(\pi'; x',\pi_{t},x_t )\gamma_i(z';z_t,x_t,y_t)\tau_i(\nu')dx'd\pi'dz'd\nu'}\bigg\}.\notag\\
	\label{eq_Bellman}
\end{align}

{We assume that the policy rule for actions is a measurable function of state variables, that is,}\footnote{ {{See Chapter 9 in \citet{stokey1989recursive} for a formal analysis.}}} 
\begin{equation}y_{it}=\phi\left(x_{it},\pi_{it},z_{it},\nu_{it},\rho_i,u_i,\beta_i,\gamma_i,\tau_i\right),\label{eq_a}\end{equation}
for some function $\phi$. Then, let 
$$\phi_i\left(x_{it},\pi_{it},z_{it}\right)=\int \phi\left(x_{it},\pi_{it},z_{it},\nu_{it},\rho_i,u_i,\beta_i,\gamma_i,\tau_i\right)\tau_i(\nu_{it})d\nu_{it}$$
denote the average decision rule with respect to the shocks $\nu_{it}$. It follows from Assumption \ref{ass_shocks} that\footnote{We treat $\rho_i,u_i,\beta_i,\gamma_i,\tau_i$ as non-random quantities. {That is, in our setup agents are assumed to know their preferences and  discount factor, the law of motion of $z_{it}$, the belief updating rule, and the distribution of shocks.}} 
$$\phi_i\left(x_{it},\pi_{it},z_{it}\right)=\mathbb{E}\left[\phi\left(x_{it},\pi_{it},z_{it},\nu_{it},\rho_i,u_i,\beta_i,\gamma_i,\tau_i\right)\,|\, x_{it},\pi_{it},z_{it}\right].$$
Hence, (\ref{eq_novel}) holds for $\varepsilon_{it}=y_{it}-\phi_i\left(x_{it},\pi_{it},z_{it}\right)$, which has zero mean given $x_{it},\pi_{it},z_{it}$. In this framework, $\phi_i$ in (\ref{eq_novel}) can thus be interpreted as the individual's decision rule averaged over the shocks $\nu_{it}$.\footnote{It is straightforward to include additional state variables in (\ref{eq_a}), under the assumption that beliefs about them are constant and invariant in the counterfactual. Accounting for additional state variables can be empirically relevant, and we will include a number of such variables as controls in our application.} {Note that we have derived (\ref{eq_a}) under Assumption \ref{ass_beliefs}, but the same expression obtains under Assumption \ref{ass_beliefs_mod}.}

Lastly, the setup is readily adapted to a finite horizon environment. In this case, $t\in\{1,...,T_i\}$, and the Bellman equation (\ref{eq_Bellman}) becomes, for $t< T_i$ and some terminal value $V_{i,T_i}$, 
{	\begin{align*}&V_{it}(x_t,\pi_t,z_t,\nu_t){=}\limfunc{max}_{y_t}\, \bigg\{u_i(y_t,x_t,z_t,\nu_t)\notag\\
		&{+}\beta_i \int V_{i,t+1}(x',\pi',z',\nu')\pi_{t}(x') \rho_{it}(\pi'; x',\pi_{t},x_t )\gamma_{it}(z';z_{t},x_{t},y_{t})\tau_i(\nu')dx'd\pi'dz'd\nu'\bigg\},
	\end{align*}
where the transitions $\rho_{it}$ between $\pi_{it}$ and $\pi_{i,t+1}$ are time-specific, and $\gamma_{it}$ is the density of $z_{i,t+1}$ conditional on $z_{it},x_{it},y_{it}$.} Actions then take the form
	\begin{align}y_{it}&=\phi_{i}\left(x_{it},\pi_{it},z_{it},t\right)+\varepsilon_{it},\label{eq_phi_t}\end{align}
	where the dependence of $\phi$ on $i$ and $t$ stems from the presence of $u_i$, $\beta_i$, $\tau_i$, the terminal value $V_{i,T_i}$, and the $\rho_{is}$ and $\gamma_{is}$ in all periods $s\geq t$. Hence, by including $t$ (i.e., age) in $z_{it}$, (\ref{eq_phi_t}) takes the same form as (\ref{eq_novel}).

\subsection{Interpreting average partial effects}

Structurally interpreting an average partial effect as the impact of a counterfactual change requires $\phi_i$ to remain invariant in the counterfactual. We now discuss this invariance condition.

Keeping $u_i$ and $\beta_i$ constant requires assuming that $u_i$ (such as preferences) and $\beta_i$ (discounting) are invariant to changes in the environment. This is a common assumption in dynamic structural models. Invariance of the density of taste shocks $\tau_i$ is also commonly assumed. In turn, keeping $\gamma_i$ constant requires assuming that the process through which past actions and states feed back onto future $z_{it}$ values is invariant in the counterfactual. When $z_{it}$ is a stock that depreciates over time or an asset with some return, for example, this requires assuming away the presence of general equilibrium effects through which the return or the depreciation rate might change in the counterfactual. 

In addition, as our framework makes clear, structurally interpreting average partial effects generally requires assuming that the belief updating rule $\rho_i$ remains constant in the counterfactual. A change in $\rho_i$ corresponds to a steady-state or ``long-run'' counterfactual where the entire process of $x_{it}$, as perceived by the agent, changes. In our setup, we allow for policies or other counterfactuals to affect beliefs $\pi_{it}$, yet we assume that the belief updating rule $\rho_i$ is an individual characteristic that remains unaffected. In Section \ref{sec_conc} we will describe how to extend the approach to account for beliefs over multiple horizons, hence making the invariance assumption about $\rho_i$ less restrictive. Our focus on counterfactuals involving changes in $x_{it}$ and $\pi_{it}$, while $\rho_i$ is kept constant, can be viewed as an intermediate case between a static counterfactual where only $x_{it}$ varies, and a long-run, steady-state counterfactual where the entire long-run belief process, including the belief updating rule $\rho_i$, is allowed to vary.\footnote{To identify such long-run counterfactuals {based on average partial effects}, without taking a stand on all aspects of the structural model, one would need to recover the effect of the belief updating rule $\rho_i$ on decisions. This would require the availability of empirical counterparts for $\rho_i$, as well as suitable cross-sectional exogeneity assumptions (or a valid instrument for $\rho_i$). Both conditions would impose strong demands on the data. In particular, $\rho_i$ is a \emph{subjective} process perceived by the agent, which is not directly informed by responses to subjective expectations questions (since $\rho_i$ need not coincide with the process of realized beliefs $\pi_{it}$).}

{Lastly, in addition to $\rho_i$ being invariant, a separate requirement of our approach to compute average partial effects is knowledge of the values $\left(x_{it}^{(\delta)},\pi_{it}^{(\delta)}\right)$ in the counterfactual. Our baseline implementation is based on a full pass-through assumption.}

%In the dynamic counterfactuals that we study, tomorrow's uncertain realizations matter for today's decisions and outcomes. Since $\pi_{it}$ enters as an argument in $\phi$, a change in $\pi_{it}$ will have a direct effect on outcomes through an expectations channel. 

%In the remainder of this section we discuss two issues that are relevant for applications.

\section{Examples\label{sec_examples}}

In this section, we describe two examples of our framework. In the first one, we consider a model of consumption, savings, and income, with the aim to assess the effects on consumption of a change in the income process. In the second example, we outline a model of agricultural production that allows farmers to adapt to the weather, with the goal to document the effects of current and expected weather. Both examples fall into the class of structural models that we introduced in the previous section. However, the validity of {our approach does not depend on the details of these specific examples.}

\subsection{Consumption, saving, and income\label{subsec_cons}}

%\subsubsection{Model details}

In the first example, we consider a standard incomplete markets model of consumption and saving behavior. For simplicity, we focus on infinite-horizon environment, as in \citet{chamberlain2000optimal}, although the analysis can easily be adapted to a life-cycle environment. 

In the model, $y_{it}$ is household $i$'s log consumption in period $t$, and household utility over consumption is $u_i(y_{it},\nu_{it})$, where $u_i$ is an increasing utility function and $\nu_{it}$ are i.i.d. taste shocks with density $\tau_i$. Household $i$'s discount factor is $\beta_i$. Log income $x_{it}$ and beliefs $\pi_{it}$ about $x_{i,t+1}$ are exogenous, and Assumptions \ref{ass_beliefs0} and \ref{ass_beliefs} hold. Households can self-insure using a risk-free bond with constant interest rate $r_i$, and assets $z_{it}$ follow
\begin{equation}z_{i,t+1}=(1+r_i)(z_{it}+w_{it})-c_{it},\label{eq_budg2}\end{equation}
where $w_{it}=\exp(x_{it})$ and $c_{it}=\exp(y_{it})$ denote income and consumption, respectively. As in (\ref{eq_a}), the (log) consumption rule takes the form\footnote{In a finite-horizon environment, $\phi$ contains time $t$ (i.e., age) as an additional argument, as in (\ref{eq_phi_t}). }
\begin{align*}y_{it}&=\phi\left(x_{it},\pi_{it},z_{it},\nu_{it},\rho_i,u_i,\beta_i,r_i,\tau_i\right).\end{align*}

As a specific example for the income process perceived by the agent, consider a permanent-transitory model (e.g., \citealp{hall1982sensitivity}):
\begin{equation}x_{it}=\eta_{it}+u_{it},\quad \eta_{it}=\eta_{i,t-1}+v_{it},\label{eq_permtrans}\end{equation}
where $u_{it}\sim{\cal{N}}(0,\sigma_{iu}^2)$ and $v_{it}\sim{\cal{N}}(0,\sigma_{iv}^2)$ are independent over time and independent of each other at all leads and lags. At time $t$, the agent observes $x_{it}$ and $\eta_{it}$, but neither $x_{i,t+1}$ nor $\eta_{i,t+1}$. In this case, we have
\begin{equation}\pi_{it}(\widetilde{x})=\frac{1}{\sqrt{\sigma_{iu}^2+\sigma_{iv}^2}}\varphi\left(\frac{\widetilde{x}-\eta_{it}}{\sqrt{\sigma_{iu}^2+\sigma_{iv}^2}}\right),\label{pi_eq}\end{equation}
where $\varphi$ is the standard Gaussian density, and Assumption \ref{ass_beliefs} holds. In this specific example, only the mean of $\pi_{it}$ varies over time and its variance is constant. 

Suppose we wish to assess the impact on consumption at time $t$ of a proportional income tax $T(w)=(1-\exp(\delta))w$ introduced at time $t$, where recall that $w=\exp(x)$ denotes household income. Under the tax, log income is thus $x^{(\delta)}=x+ \delta$. Suppose households believe the tax will remain in place in the future, and they fully adjust their beliefs to the tax, {as described in Subsection \ref{subsec_env}}. When $\pi_{it}$ is given by (\ref{pi_eq}) in the absence of the tax, implementing the tax will lead to the new beliefs $$\pi_{it}^{(\delta)}(\widetilde{x})=\frac{1}{\sqrt{\sigma_{iu}^2+\sigma_{iv}^2}}\varphi\left(\frac{\widetilde{x}-\eta_{it}- \delta}{\sqrt{\sigma_{iu}^2+\sigma_{iv}^2}}\right).$$
Hence, the tax affects both current log income and the perceived conditional mean of future log income.

In this model, a proportional tax does not affect the belief updating rule $\rho_i$.\footnote{Indeed, the introduction of the tax is isomorphic to a change in the permanent component, from $\eta_{it}$ to $\eta_{it}^{(\delta)}=\eta_{it}+ \delta$. Moreover, the distribution of $(x_{i,t+1},\eta_{i,t+1})$ given $(x_{it},\eta_{it})$ does not change under the tax.} Hence, the total APE fully captures the effect of the tax on consumption. In this case, the contemporaneous APE corresponds to the effect of a purely transitory tax at $t$ that will disappear at $t+1$; equivalently, it is the effect of a $\delta$-shift in the transitory income shock $u_{it}$. In turn, the total APE corresponds to the effect of a $\delta$-shift in the permanent income shock $v_{it}$. The dynamic APE is the difference between the two.\footnote{The DAPE in (\ref{eq_decompo}) is evaluated at income $x^{(\delta)}$ after the tax, so that the CAPE and the DAPE add up to the TAPE. It is also possible to compute an alternative DAPE evaluated under income $x$ before the tax, ${\widetilde{\Delta}_t^{\rm DAPE}(\delta)=\mathbb{E}\left[\phi_i\left(x_{it},\pi_{it}^{(\delta)},z_{it}\right)-\phi_i(x_{it},\pi_{it},z_{it})\right]}$. In this case, the dynamic APE can be interpreted as the effect, {on period-$t$ outcomes,} of a tax that is announced at $t$ and will be implemented at $t+1$.}

The model in this subsection relies on specific assumptions about the income process, information, and beliefs. However, those assumptions could be incorrect; for example, agents might have different beliefs about future income. It is important to note that, in our approach, and in our empirical application in Section \ref{sec_appli1}, we do not assume that the consumption model with permanent-transitory income beliefs describes the data. {Irrespective of the details of the structural model, average partial effects can be interpreted as the structural effects of a counterfactual tax under the conditions we provide, including invariance of the belief updating rule $\rho_i$.}

\subsection{Weather and agricultural production\label{subsec_weather}}

In the second example, we consider a model of agricultural production with costly investment. Output $q_{i,t+1}=g_i(x_{i,t+1},k_{i,t+1})$ depends on the weather $x_{i,t+1}$ and on a dynamic input $k_{i,t+1}$ (such as capital). The weather $x_{it}$, and farmer $i$'s beliefs $\pi_{it}$ about $x_{i,t+1}$, satisfy Assumptions \ref{ass_beliefs0} and \ref{ass_beliefs}. The farmer can invest $y_{it}$ in the dynamic input $k_{it}$ at a cost $c_i(y_{it},\nu_{it})$, for some i.i.d. cost shifters $\nu_{it}$ with density $\tau_i$. The dynamic input follows the law of motion $k_{i,t+1}=(1-\delta_i)k_{it}+y_{it}$. The farmer decides on $y_{it}$ after observing today's weather $x_{it}$ and her beliefs $\pi_{it}$ about tomorrow's weather, but before observing $x_{i,t+1}$. Lastly, the instantaneous profit in period $t$ is $q_{it}-c_i(y_{it},\nu_{it})$, and the farmer's discount factor is $\beta_i$. 

The state variables of the decision problem are $x_{it}$, $\pi_{it}$, $k_{it}$, and $\nu_{it}$, and, under suitable regularity conditions, the optimal investment rule takes the form
\begin{align}y_{it}&=\phi\left(x_{it},\pi_{it},k_{it},\nu_{it},\rho_i,\beta_i,c_i,\delta_i,g_i,\tau_i\right),\label{eq_a_clim}
\end{align}
for some function $\phi$. Substituting (\ref{eq_a_clim}) into the output equation, output in period $t+1$ can thus be written as
\begin{align}q_{i,t+1}&=\widetilde{\phi}\left(x_{i,t+1},x_{it},\pi_{it},k_{it},\nu_{it},\rho_i,\beta_i,c_i,\delta_i,g_i,\tau_i\right),\label{eq_outcome_clim}
\end{align}
for some function $\widetilde{\phi}$. The presence of $\pi_{it}$ in (\ref{eq_a_clim}) and (\ref{eq_outcome_clim}) reflects that the farmer may {adapt} to the prospect of harmful weather in the future by investing  today.\footnote{Farmers' adaptation has been studied in the literature using various approaches. \citet{burke2016adaptation} rely on a long-difference approach to account for farmers' responses to a changing climate. \citet{shrader2020improving} proposes a framework to account for adaptation in a model where, in contrast with our dynamic framework, the firm's current choice does not affect outcomes (i.e., profit) in later periods. See also \citet{dell2014we} and \citet{keane2020climate}. Other approaches in the literature rely on specific aspects of the production model, such as envelope condition arguments (\citealp{hsiang2016climate}, \citealp{lemoine2018estimating}, \citealp{gammans2020reckoning}).}

{In this application, one may be interested in studying investment, through the policy rule (\ref{eq_a_clim}), or in studying an outcome that depends on investment, such as output in (\ref{eq_outcome_clim}).} {In particular, Equation} (\ref{eq_outcome_clim}) motivates regressing output on current and past weather and on {past} weather beliefs. Exploiting changes over time in $x_{it}$ and $\pi_{it}$, within farmer, is robust to the presence of individual heterogeneity. As an application, one can estimate our belief-augmented average partial effects to assess the impact of a change in the weather process that affects both weather realizations and weather beliefs. In this case as well, structurally interpreting the total APE as reflecting the total effect of such a change relies on the assumption that the belief updating process $\rho_i$ is invariant. While this assumption may be tenable in the short or medium run, the total APE will not capture the full impact of long-run changes in the climate under which $\rho_i$ could be affected.

\subsection{Simulated tax counterfactuals {in a structural model}}

{In the last part of the section, we illustrate how the structural approach and our approach relate to each other in the context of a consumption model.} For this purpose, we simulate a large sample from a life-cycle model of consumption and savings based on \citet{Kaplan_Violante_2010}, where identical, risk-averse households save to smooth consumption while facing borrowing constraints. We entertain two different processes of belief formation. We use this exercise to compare and contrast {our approach with the structural approach} to counterfactual prediction.

{Relative to the model we presented in Subsection \ref{subsec_cons}, we make several changes.} First, we impose no borrowing. Second, we specify two different processes for households' expectations. In the first case, we assume that expectations are rational, and coincide with (\ref{eq_permtrans}). In the second case, we still assume that (\ref{eq_permtrans}) describes the realized income process, but we specify households' expectations as adaptive, similarly to (\ref{eq_adaptive}). In both cases, income beliefs, which are key state variables in the model, can be summarized by their time-varying means, which follow a first-order Markov process jointly with log income. Except for having different expectations processes, the two models have exactly the same structure and primitive parameters. See Appendix \ref{app_struct} for details. The structural model has no time-invariant household heterogeneity.

%%%%%%%%%%%%%%%%%%%%%%%%%%%%%%%%%%%%%%%%%%%%%%%%%%%%%%%%%%%%%%%%%%%%%%%%%%%%%%%%%%%%%%%%%%%%%%%%%%%%%%%%%%%%%%%%%%%%%%%%%%%%%
\begin{table}[tbp]
	\begin{center}
		\caption{Simulated tax counterfactuals under rational and adaptive expectations}
		\label{Table_sim_decom}
		\adjustbox{max width=\linewidth}{
			\resizebox{\linewidth}{!}{	
				\begin{tabular}{lccccccccc}\hline\hline
					& \multicolumn{4}{c}{Rational expectations} & \multicolumn{4}{c}{Adaptive expectations} \\ \cmidrule(lr){2-5} \cmidrule(lr){6-9}
					& \multirow{2}{*}{Structural} & \multicolumn{3}{c}{APE} & \multirow{2}{*}{Structural} & \multicolumn{3}{c}{APE}  \\\cmidrule(lr){3-5}\cmidrule(lr){7-9}
					&  & Linear & Quadratic & Spline &  & Linear & Quadratic & Spline \\ \hline
					% latex table generated in R 4.1.1 by xtable 1.8-4 package
% Wed Feb 15 17:20:39 2023
%  \hline
CAPE & -0.0163 & -0.0151 & -0.0150 & -0.0150 & -0.0122 & -0.0344 & -0.0191 & -0.0133 \\ 
  DAPE & -0.0802 & -0.0917 & -0.0863 & -0.0860 & -0.0496 & -0.0518 & -0.0512 & -0.0513 \\ 
  TAPE & -0.0965 & -0.1068 & -0.1013 & -0.1010 & -0.0618 & -0.0863 & -0.0704 & -0.0646  %\cmidrule(lr){2-5} \cmidrule(lr){6-9}
%  Observations & \multicolumn{4}{c}{11,171,428} & \multicolumn{4}{c}{11,492,522} %\\ 
%  Observations & \11,171,428 & 11,171,428 & 11,171,428 & 11,171,428 & 11,492,522 & 11,492,522 & 11,492,522 & 11,492,522 %\\ 
%   \hline
  \\ \hline\hline
		\end{tabular}}}
	\end{center}
	\footnotesize{\textit{Notes: Effects of a 10\% permanent income tax on log consumption in two model economies, where households have rational (in the left panel) or adaptive expectations (in the right panel), respectively. In both economies, log income follows a permanent-transitory process. For the structural counterfactuals we compute the effect of the tax under the model. For {the calculation of APE,} we regress log consumption on log income, income belief and its interaction with log income, age, age squared, and a function of log assets (linear, quadratic, or 20-knot spline depending on the specification). Households with positive assets, age 26--49.}}
\end{table}

Under both versions of the model, we compute the true effect of a 10\% permanent proportional income tax, and we decompose it under the model into a contemporaneous effect due to current income and a dynamic effect due to beliefs. Then, we compare these counterfactual predictions with our average partial effects (TAPE, CAPE, and DAPE), which we obtain by estimating consumption regressions in the simulated sample. Since the model has a finite horizon, the consumption function $\phi$ is age-dependent, and we proxy for this dependence by controlling for age and its square in the regressions. Note that, as we discussed in Subsection \ref{subsec_cons}, the belief updating rule $\rho_i$ is invariant under the counterfactual in the rational expectations version of the model. In the adaptive expectations version we assume that invariance is satisfied as well. We provide details about the model, parameter values, and calculation of counterfactuals in Appendix \ref{app_struct}.

We report the counterfactual calculations in Table \ref{Table_sim_decom}. We use a large number of simulated draws, so that variability due to the simulation is negligible. Focusing first on the version with rational expectations (in the left panel), the model predicts a decrease in log consumption of $-0.097$, which is almost one-for-one with the tax increase, as is expected in this model, and a large part can be attributed to a change in beliefs. The {average partial effects}, which do not rely on the knowledge of the structure and the parameter values of the structural model but are computed using regressions, come close to these numbers. We report the results of three specifications, where we control for linear, quadratic, or spline functions of log assets, and all of them give comparable results in this case. 

Turning next to the version with adaptive expectations (in the right panel), the model predicts a smaller effect of the tax ($-0.062$), given the expectations process that we assume. {As a result, a researcher incorrectly assuming rational expectations in this setting, even if she had recovered the other primitive parameters of the model, would overestimate the effect of the tax. This illustrates that, when using a structural approach to predict counterfactuals, correctly specifying belief formation is key. In contrast, our approach does not require knowledge of the belief formation process (e.g., rational or adaptive expectations). Indeed, the right panel in Table \ref{Table_sim_decom} shows that the {average partial effects}, which do not rely on correct specification of the model (including the belief formation part of the model), again come close to the tax effects, albeit in this case only when the regression specification is flexible enough (i.e., quadratic or spline).}\footnote{This reflects the fact that the linear approximation to the consumption policy rule is less accurate in the structural model with adaptive expectations than in the model with rational expectations. In Appendix Figure \ref{fig_sim_policy_rules} we report the policy rules at several ages, for both rational and adaptive expectations. In Appendix Table \ref{Table_sim_decom_age} we present the tax counterfactual results for different ages.}

\section{Estimating average partial effects\label{sec_ident1}}

In this section we study identification  and estimation of average partial effects based on (\ref{eq_novel}).

\subsection{Identification}

\subsubsection{Beliefs}

%Type of belief data
%
%
%Point forecast - not us
%
%Probabilistic forecast $m_{it}=m(\pi_{it})$. 
%
%Binary variable. $m_{it}$ is $\pi_{it}(1)$. Sufficient
%
%Continuos variable, Our application. We use a parametetric model 

Our approach to the measurement of beliefs $\pi_{it}$ relies on data about respondents' expectations. It is increasingly common to elicit responses in a probabilistic manner, by asking respondents to report their subjective probabilities about future events (see \citealp{manski2004measuring}). Responses to questions about subjective probabilistic expectations provide information about some features of $\pi_{it}$. Typically, the responses can be interpreted as some functionals $m_{it}=m(\pi_{it})$, such as the mean, variance, or some other moments of $\pi_{it}$. {We assume that such data are available for a sample of individuals $i=1,...,n$ and time periods $t=1,...,T$.} In this section, we abstract from measurement error in responses, but we will account for measurement error in our empirical application.  

When beliefs concern a binary variable $x_{i,t+1}\in\{0,1\}$ (e.g., job loss), the subjective probability $\pi_{it}(1)=\Pr(x_{i,t+1}=1\,|\, \Omega_{it})$ provides all the required information in the sense that, under Assumption \ref{ass_beliefs} or \ref{ass_beliefs_mod}, it is a sufficient statistic for decisions. {One can thus directly use the elicited subjective probability in our approach.} However, when beliefs are about a continuous variable, such as income in our application, the subjective density $\pi_{it}$ is a function. At the same time, expectations data are often coarse. A common strategy in such a case is to assume that $\pi_{it}$ belongs to a parametric family. For example, in the 1995 and 1998 waves of the SHIW in Italy, respondents are asked about the minimum and maximum earnings that they expect to receive if employed in the following year, together with the probability that their earnings will be below the mid-point between those two values. \citet{kaufmann2009disentangling} assume that income beliefs follow a triangular distribution conditional on employment.

We will assume that $\pi_{it}$ is parametrically specified; that is, that there exists a finite-dimensional vector $\theta_{it}$ such that
\begin{equation}\label{eq_pi_theta}
	\pi_{it}=\pi(\cdot;\theta_{it}),
\end{equation}
where $\pi(\cdot;
\theta)$ is known given $\theta$. When $x_{it}$ is binary or discrete, this assumption is without loss of generality.\footnote{Note that a special case of our parametric assumption is $\theta_{it}=m_{it}$. In this case, the key assumption is that the mapping $\pi\mapsto m(\pi)$ is injective, so that $m_{it}$ uniquely determines $\pi_{it}$.} However, when $x_{it}$ is continuous the assumed parametric family may be misspecified. {In Appendix \ref{app_extens}, we discuss how one could relax the parametric specification on $\pi_{it}$ with rich enough data on beliefs.}

\subsubsection{Decision rule}

We impose the following mean independence condition,
\begin{align}\mathbb{E}[\varepsilon_{it}\,|\, x_{it},\pi_{it},z_{it}]=0.\label{eq_epsilon}
\end{align}
Note that (\ref{eq_epsilon}) is satisfied in the structural framework of Section \ref{sec_structural}. To enhance the plausibility of this condition in applications, one can control for additional time-varying regressors (which can be interpreted as additional state variables), as well as for time-invariant fixed-effects. We will account for both factors in our empirical application.\footnote{In certain applications, (\ref{eq_epsilon}) may not be plausible, but one may have access to instruments $w_{it}$ (e.g., instruments that exploit some policy variation in sample) such that $\mathbb{E}[\varepsilon_{it}\,|\, w_{it}]=0$. Identification of $\phi_i$ then requires suitable relevance conditions (see \citealp{newey2003instrumental}).}

Given (\ref{eq_novel}), (\ref{eq_pi_theta}). and (\ref{eq_epsilon}), we have
\begin{equation}\phi_i(x_{it},\pi_{it},z_{it})=\mathbb{E}\left[y_{it}\,|\, x_{it},\theta_{it},z_{it}\right].\label{eq_phi_identification}\end{equation}
It follows that, in an environment with a growing number of time periods (i.e., $T$ tends to infinity), the individual-specific decision rule $\phi_i(x,\pi,z)$ is identified for all $x$, $\pi$, $z$ in the empirical support of $x_{it}$, $\pi_{it}=\pi(\cdot;\theta_{it})$, and $z_{it}$. {It is worth emphasizing that, in order to separately identify the contemporaneous effect of $x_{it}$ and the dynamic effect of $\pi_{it}$, it is crucial that beliefs $\pi_{it}$ vary over time conditional on $x_{it}$ and $z_{it}$. Such empirical variation reflects changes in the agent's information set $\Omega_{it}$ over and beyond the changes in the covariates $(x_{it},z_{it})$ that the econometrician observes.}

In many empirical settings, however, belief data are only available on a short panel. In that case, the individual-specific function $\phi_i$ is no longer identified. We follow the literature on nonlinear panel data models and impose structure on heterogeneity via a latent variable, or ``type'', $\alpha_i$. Specifically, we assume that, for a function $\phi$ and a latent variable $\alpha_i$, we have
\begin{align}{\phi}_i\left(x_{it},\pi_{it},z_{it}\right)&={\phi}\left(x_{it},\pi_{it},z_{it},\alpha_i\right).\label{eq_action_empirics_short_gen}
\end{align} 
In the structural model of Section \ref{sec_structural}, the type $\alpha_i$ could index primitive parameters such as preferences, for example.

	Identifying and estimating a non-separable model of the form (\ref{eq_action_empirics_short_gen}) raises three challenges: the presence of the latent variable $\alpha_i$, the nonlinearity of the function $\phi$, and the fact that the $z_{it}$'s depend on past actions, hence are not strictly exogenous in a panel data sense. The literature has only recently begun to analyze these three issues simultaneously (see \citealp{bonhomme2025moment}).

	One avenue to tackle these challenges is to suppose, in addition to {$x_{it}$ and} beliefs $\pi_{it}$ being strictly exogenous, that the law of motion of $z_{it}$, as represented by $\gamma_i$, is the same for all individuals. This assumption of a homogeneous feedback process simplifies the model structure, as shown by \citet{kasahara2009nonparametric} and \citet{bonhomme2023identification}, and as we illustrate in Subsection \ref{subsec_estimation}. In effect, the researcher can proceed as if $z_{it}$ were strictly exogenous, despite their dependence on past actions. Existing techniques to show identification of mixture models with strictly exogenous covariates can then be used, for example based on a finite-type assumption that is popular in structural models (e.g., \citealp{hall2003nonparametric}, \citealp{kasahara2009nonparametric}). Moreover, while the plausibility of the homogeneous feedback assumption is context-specific, it appears natural in our consumption application provided agents face a common budget constraint, e.g., a common interest rate.

\subsubsection{Average partial effects\label{subsubsec_beliefs}}

{Consider a counterfactual change $\delta$, leading to $\left(x_{it}^{(\delta)},\pi_{it}^{(\delta)}\right)$. Average partial effects require knowledge of those counterfactual values. In the absence of data on those, a possibility is to assume that individuals fully incorporate the effect of the change in $x_{it}$ and $\pi_{it}$, as we outlined in Subsection \ref{subsec_env}.} 
	
	{To implement this assumption in practice, we suppose that beliefs remain in the same parametric family in the counterfactual.} Hence, for some parameter $\theta_{it}^{(\delta)}$, $$\pi_{it}^{(\delta)}=\pi\left(\cdot;\theta_{it}^{(\delta)}\right).$$ Then, we propose to set \begin{equation}\label{eq_delta_change}
		\theta_{it}^{(\delta)}=\underset{{\theta}}{\limfunc{argmax}}\,\, \mathbb{E}_{\theta_{it}}\left[  \log\left(\pi\left(x_{i,t+1}^{(\delta)};{\theta}\right)\right)\right],
	\end{equation}
	where the expectation is with respect to the baseline belief density, $x_{i,t+1}\sim \pi(\cdot;\theta_{it})$.\footnote{That is, $	\theta_{it}^{(\delta)}={\limfunc{argmax}}_{{\theta}}\,\,\int  \log\left(\pi\left(\delta(x);{\theta}\right)\right)\pi(x;\theta_{it})dx$.}

	{As an example, consider the introduction of a permanent proportional income tax. Let $x_{it}$ denote log income without the counterfactual tax, and let $x_{it}^{(\delta)}=x_{it}+\delta$ denote log income net of the tax. Suppose $\pi_{it}$ is normal with mean $\mu_{it}$ and variance $\sigma_{it}^2$, so $\theta_{it}=(\mu_{it},\sigma_{it}^2)$. Under (\ref{eq_delta_change}), $\pi_{it}^{(\delta)}$ remains normal under the tax, with mean and variance $\theta_{it}^{(\delta)}=(\mu_{it}+\delta,\sigma_{it}^2)$.}

		Lastly, given actual and counterfactual values of $x_{it}$ and $\pi_{it}$, when $\phi_i$ is identified on the empirical support, average partial effects (TAPE, CAPE and DAPE) are all identified, provided the support of covariates in the counterfactual lies within their empirical support. {In short panels, identification of $\phi$ on the right-hand side of (\ref{eq_action_empirics_short_gen}), along with identification of the type distribution, guarantee identification of average partial effects.}
		
		\begin{remark}\label{rem1}		
	{To assess the impact of individuals not fully incorporating $\delta(\cdot)$ into their beliefs, one can assume that individuals have a common subjective probability $(1+\xi)^{-1}$ that the counterfactual will remain in place next period, 
		and replace (\ref{eq_delta_change}) by
		\begin{equation}
			\theta_{it}^{(\delta)}=\underset{{\theta}}{\limfunc{argmax}}\,\,  \mathbb{E}_{\theta_{it}}\left[  \log\left(\pi\left(x_{i,t+1}^{(\delta)};{\theta}\right)\right)+\xi\log\left(\pi\left(x_{i,t+1};{\theta}\right)\right)\right].\label{eq_delta_change3}
		\end{equation}
	In Appendix \ref{app_sensiv} we show that $	\theta_{it}^{(\delta)}$ in (\ref{eq_delta_change3}) minimizes the Kullback-Leibler (KL) divergence between the parametric family $\pi(\cdot;\theta)$ and the mixture density $\frac{1}{1+\xi}\pi_{it}^{(\delta,full)}+\frac{\xi}{1+\xi}\pi_{it}$, where $\pi_{it}^{(\delta,full)}$ is the subjective density of the transformed $\delta(x_{i,t+1})$ for $x_{i,t+1}\sim \pi_{it}$.\footnote{For example, consider a change $x_{it}^{(\delta)}=x_{it}+\delta$. If $\pi_{it}$ is normal with mean and variance $\theta_{it}=(\mu_{it},\sigma_{it}^2)$, then $\pi_{it}^{(\delta)}$ has mean and variance $\theta_{it}^{(\delta)}=\left(\mu_{it}+\frac{\delta}{1+\xi},\sigma_{it}^2+\xi \left(\frac{\delta}{1+\xi}\right)^2\right)$.} Likewise, $	\theta_{it}^{(\delta)}$ in (\ref{eq_delta_change}) minimizes the KL divergence between $\pi(\cdot;\theta)$ and $\pi_{it}^{(\delta,full)}$. If individuals view the counterfactual as only applying this period ($\xi=\infty$), then $\theta_{it}^{(\delta)}=\theta_{it}$ is unchanged, while if they believe the change will be permanent ($\xi=0$) then $\theta_{it}^{(\delta)}$ is given by (\ref{eq_delta_change}).}
		\end{remark}
	
	\begin{remark}{To learn about $\pi_{it}^{(\delta)}$, an alternative approach is to elicit individual expectations under various policy counterfactual scenarios. Such data could also be used to learn about common or heterogeneous $\xi$ parameters in (\ref{eq_delta_change3}), for example. This is a promising avenue, although data on beliefs under counterfactual policies are not commonly available yet (see \citealp{roth2023effects} for a recent exception).}
		\end{remark}

%		For example, assuming that individuals face a cost of adjusting their beliefs that is proportional to the Kullback-Leibler divergence between the baseline and counterfactual beliefs, one can replace (\ref{eq_delta_change}) by
%	\begin{equation}
%		\theta^{(\delta)}=\underset{\widetilde{\theta}}{\limfunc{argmax}}\,\, \left\{ \mathbb{E}\left[  \log\left(\pi\left(x_{t+1}^{(\delta)};\widetilde{\theta}\right)\right)\right]-\xi {\rm KL}\left(\widetilde{\theta},\theta\right)\right\},\label{eq_delta_change3}
%	\end{equation}
%	where ${\rm KL}\left(\widetilde{\theta},\theta\right)=\mathbb{E}\left[ \log\left(\frac{\pi(x_{t+1};\theta)}{\pi\left(x_{t+1};\widetilde{\theta}\right)}\right)\right]$. According to (\ref{eq_delta_change3}), $\theta^{(\delta)}$ is given by (\ref{eq_delta_change}) when the adjustment cost $\xi$ is zero, $\theta^{(\delta)}=\theta$ is unchanged when the cost is infinite, and the individual partially adjusts her beliefs for intermediate values of $\xi$.
%	

\subsection{Estimation\label{subsec_estimation}}

%\subsubsection{Three-step estimation}

For estimation we proceed in three steps. 

\paragraph{First step.} First, we estimate the parameters $\theta_{it}$ that govern the belief density. Assuming that subjective expectations responses $m_{it}=m(\pi_{it})$ are available, a minimum-distance estimator solves $$\widehat{\theta}_{it}=\underset{\theta}{\limfunc{argmin}} \,\,\, d\left(m_{it},m(\pi(\cdot;\theta))\right),$$
where $d$ is some distance function (e.g., Euclidean). Under the assumption that beliefs are elicited without error, i.e., $m_{it}=m(\pi_{it})$, this step involves no sampling uncertainty.

\paragraph{Second step.} 

In the second step, we estimate $\phi$ in (\ref{eq_action_empirics_short_gen}). Various approaches are available. For example, \citet{stock1989nonparametric} proposes a partially linear semiparametric approach. There are also various ways of incorporating unobserved heterogeneity. In the application, which is based on a two-period panel, we will {rely on the following specification:}
	\begin{align}\phi_{\beta}(x,\theta,z,\alpha)=\sum_{r=1}^R\beta_{r}(\alpha)P_r(x,\theta,z),\label{eq_varphi_est_short}
\end{align}
where $P_r$ is a family of functions, such as polynomials, and $R$ is the number of terms. {In our main specification, latent types $\alpha_i\in\{1,...,,K\}$ have finite support, so the parameters $\beta_{r}(\alpha)$, for $\alpha\in\{1,...,K\}$, take a finite number $K$ of values.} 

{In our implementation, we will postulate a parametric model for the zero-mean error $\varepsilon_{it}$ given $x_{it},\theta_{it},z_{it}$, indexed by a parameter $\mu$}, as well as a parametric specification for $\alpha_i$ given $x_i=(x_{i1},...,x_{iT})$, $\theta_i=(\theta_{i1},...,\theta_{iT})$, and $z_{i1}$, indexed by {another parameter} $\eta$. Under the assumption that households face a common budget constraint (i.e., that feedback is homogeneous), estimation can be based on the partial log-likelihood function
\begin{equation}L(\beta,\mu,\eta)=\sum_{i=1}^n\log\left(\sum_{\alpha=1}^K\prod_{t=1}^Tf_{\beta,\mu}(y_{it}\,|\, x_{it},\theta_{it},z_{it},\alpha_i=\alpha)f_{\eta}(\alpha\,|\,x_{i},\theta_{i},z_{i1} )\right).\label{eq_Lbetaeta}\end{equation}
Notice that the law of motion of $z_{it}$, which does not depend on $\alpha_i$ under homogeneous feedback, does not appear in this expression.\footnote{Denoting the (homogeneous) feedback process as $f(z_{it}\,|\, z_{i,t-1},x_{i,t-1},y_{i,t-1})$, the log-likelihood function is
	\begin{align*}&\sum_{i=1}^n\log\left(\sum_{\alpha=1}^K\prod_{t=1}^Tf_{\beta,\mu}(y_{it}\,|\, x_{it},\theta_{it},z_{it},\alpha_i=\alpha)\prod_{t=2}^Tf(z_{it}\,|\, z_{i,t-1},x_{i,t-1},y_{i,t-1})f_{\eta}(\alpha\,|\, x_{i},\theta_{i},z_{i1} )\right)\\
		&	=L(\beta,\mu,\eta)+\sum_{i=1}^n\sum_{t=2}^T\log\left(f(z_{it}\,|\, z_{i,t-1},x_{i,t-1},y_{i,t-1})\right),
	\end{align*}
	where the second term on the right-hand side does not depend on $(\beta,\mu,\eta)$, and the first term is a partial log-likelihood in the sense of \citet{cox1975partial}.} {Given observations $y_{it},x_{it},z_{it}$ and estimates $\widehat{\theta}_{it}$, for $i=1,...,n$ and $t=1,...,T$, we estimate $\beta$, $\mu$, and $\eta$ by maximizing the partial log-likelihood function in (\ref{eq_Lbetaeta}) where we replace $\theta_{it}$ by their estimates $\widehat{\theta}_{it}$.}

{In our application, the two-period nature of the data limits the flexibility of the model that we can reliably estimate. For example, nonparametric estimation methods of latent type models requires at least three periods (e.g., \citealp{hall2003nonparametric}, \citealp{bonhomme2016estimating}, \citealp{hu2015microeconomic}). This underscores the usefulness of collecting belief data over time. In our context, we will explore the robustness of our empirical findings based on a parametric finite mixture model by estimating several variations of it, including specifications where only the intercept is heterogeneous in (\ref{eq_varphi_est_short}) and all other $\beta_r$ parameters are common across individuals yet rich nonlinearities in $x_{it},\theta_{it},z_{it}$ can be entertained. }

\paragraph{Third step.}

{Lastly, in the third step we estimate counterfactuals.  We plug in the estimates $\widehat{\theta}_{it}$, $\widehat{\beta}_{r}(\alpha)$, and $\widehat{\eta}$, and the counterfactual values $x^{(\delta)}_{it}$ and $\widehat{\theta}^{(\delta)}_{it}$, in the APE formulas.\footnote{Note that $\widehat{\mu}$ is not needed.} For example, we estimate the total APE, averaged across periods $t=1,...,T$, as 
\begin{equation}
	\widehat{\Delta}^{\rm TAPE}(\delta)=\frac{1}{nT}\sum_{i=1}^n\sum_{t=1}^T\sum_{\alpha=1}^K\left(\phi_{\widehat{\beta}}\left(x_{it}^{(\delta)},\widehat{\theta}_{it}^{(\delta)},z_{it},\alpha\right)-\phi_{\widehat{\beta}}(x_{it},\widehat{\theta}_{it},z_{it},\alpha)\right)f_{\widehat{\eta}}(\alpha\,|\,x_i,\widehat{\theta}_{i},z_{i1} ).\label{eq_est_APE_mixt}
\end{equation} 
$\widehat{\Delta}^{\rm TAPE}(\delta)$ is a standard multi-step estimator, for which inference methods are available (e.g., \citealp{newey1994large}).}

\section{Income, consumption, and income expectations\label{sec_appli1}}

In this section we apply our approach to empirically study how consumption depends on current and expected income, and to conduct various tax counterfactuals.

\subsection{Data}

The Italian Survey on Household Income and Wealth (SHIW) is a cross-sectional survey that collects information on annual consumption, disposable income, and wealth of Italian families. Since 1989, it includes a panel component. We use the 1989--1991 waves and the 1995--1998 waves, which include questions about income expectations asked to a subsample of households.

The expectations questions differ in both sets of waves. However, as we show in Appendix \ref{App_consumption}, the results are qualitatively similar when analyzing the waves separately, so we pool them together to increase power. In 1989 and 1991, individuals are asked about the probability their income growth will fall within a set of predetermined intervals. In 1995 and 1998, individuals are asked the maximum and minimum amounts they expect to earn if employed, and the probability of earning less than the mid-point between the maximum and minimum. We assume beliefs about log income in the following year follow a normal distribution. In Appendix \ref{app_shiw} we describe our approach to estimate the mean $\mu_{it}$ and standard deviation $\sigma_{it}$ of the beliefs for each individual and time period, which follows \citet{arellano2021income}. We will also comment on robustness checks obtained under different assumptions and estimation strategies.

We focus on employed household heads, while excluding the self-employed. Our cross-sectional sample with information on beliefs has 7,796 household-year observations, and our panel sample with data on beliefs in two consecutive waves for the same head has 1,646 observations. In Appendix Tables \ref{Table_desc_expq_89} and \ref{Table_desc_expq_95} we report descriptive statistics about income expectations questions. In Appendix Table \ref{Table_desc_bel} we provide descriptive statistics about income, consumption, assets, and the estimated means and variances of log income beliefs. Belief questions are about individual income, while consumption, assets, and current income are reported at the household level. We will account for this discrepancy in our construction of average partial effects, and we will also report estimates that control for spousal beliefs when available. Another issue with the belief data in the SHIW is that expectations questions about income in the next 12 months are asked a few months after the end of the calendar year. We will return to this issue in the next subsection. As a preliminary validation check for the expectations questions, in Appendix Table \ref{Table_income} we document that beliefs have explanatory power for future log income, even conditional on current log income and other controls, in line with what \citet{kaufmann2009disentangling} found for the 1995-1998 waves.

\begin{table}[h!]
	\begin{center}
		\caption{Regression of log consumption on log income and income beliefs}
		\label{Table_cons}
		\adjustbox{max width=\linewidth}{
			\resizebox{0.9\linewidth}{!}{	\begin{tabular}{lrrrrrr}\hline\hline
					                &\multicolumn{1}{c}{(1)}&\multicolumn{1}{c}{(2)}&\multicolumn{1}{c}{(3)}&\multicolumn{1}{c}{(4)}&\multicolumn{1}{c}{(5)}\\
\midrule
Mean expected log income &         &    0.235&    0.238&    0.229&    0.231\\
                &         &  (0.094)&  (0.095)&  (0.093)&  (0.093)\\
\addlinespace
(Mean expect. log income)$\cdot$(Log family income)&         &         &         &    0.104&    0.104\\
                &         &         &         &  (0.061)&  (0.061)\\
\addlinespace
Var expected log income &         &         &   -2.590&         &   -2.613\\
                &         &         &  (1.876)&         &  (1.941)\\
\addlinespace
(Var expect. log income)$\cdot$(Log family income)&         &         &         &         &   -1.144\\
                &         &         &         &         &  (3.499)\\
\addlinespace
Log family income &    0.584&    0.439&    0.439&    0.439&    0.440\\
                &  (0.070)&  (0.089)&  (0.089)&  (0.089)&  (0.089)\\
\addlinespace
Log family assets&    0.010&    0.018&    0.018&    0.019&    0.018\\
                &  (0.023)&  (0.023)&  (0.023)&  (0.023)&  (0.023)\\
\midrule
Household fixed effect&      Yes&      Yes&      Yes&      Yes&      Yes\\
Controls        &      Yes&      Yes&      Yes&      Yes&      Yes\\
N observations  &    1,536&    1,536&    1,536&    1,536&    1,536\\
N households    &      768&      768&      768&      768&      768\\
R-squared       &     0.24&     0.26&     0.26&     0.26&     0.26\\
Pvalue F beliefs&         &     0.01&     0.03&     0.02&     0.05\\\hline

		\end{tabular}}}
	\end{center}
	
	%\vspace{0.1cm}
	\footnotesize{\textit{Notes: SHIW, 1989--1991 and 1995--1998. Regression for household heads. The belief variables (mean and variance) and log family income are centered around the weighted average in the sample. Controls include age and age squared, existence of a spouse, marital status, family size, number of children 0-5, 6-13, 14-17 years old in the household, number of children outside the household, number of income earners in the household, and a wave indicator. Regression results are weighted using survey weights. Standard errors (shown in parentheses)  are clustered at the household level.}}
\end{table}

As a first look at the data, in Table \ref{Table_cons} we report estimates of a regression of log consumption on log income and income beliefs. In addition to log income and the mean and variance of log income beliefs, covariates include log assets as well as a variety of controls (including age, household composition, and a wave indicator). Parameters are estimated in first differences in both sets of waves. In the table we show standard errors clustered at the household level.\footnote{Standard errors in Table \ref{Table_cons} do not account for the estimation of the means and variances of beliefs, in line with our baseline assumption that beliefs are elicited without error. We will study the impact of measurement error in beliefs at the end of this section.} The estimates in columns (2) and (3) show that the mean of log income beliefs influences consumption decisions significantly over and beyond current income, while the variance of the beliefs has an insignificant effect. 

%BETTER TO REPORT BOOTSTRAP STD

It is also interesting to compare the estimates in column (2) with those in column (1) that do not account for beliefs. When including beliefs, the coefficient of family income decreases from $0.58$ to $0.44$. This finding is consistent with the presence of an upward omitted variable bias in column (1). In column (4) of Table \ref{Table_cons}, we interact the mean income beliefs with current income. While the estimates suggest the effect of the mean belief tends to be larger for higher-income households, the interaction effect is only marginally significant. Lastly, in column (5) we add the variance of beliefs and its interaction with income. We find small differences compared to column (4), with insignificant coefficients associated with the variance of beliefs. 

We have verified that the empirical correlations reported in Table \ref{Table_cons} are not driven by specific choices we made. Regression coefficients are similar when (i) using various approaches to construct the mean and variance of beliefs; (ii) controlling for spouse' beliefs about their own income; (iii) running regressions for the two sets of waves 1989--1991 and 1995--1998 separately -- with some quantitative differences; and (iv) using different approaches to control for assets. Details can be found in Appendix \ref{App_consumption}.

%-some robustness (shorter)

%CONDITIONAL ON EMPLOYMENT IN 95-98

%\footnote{In 1989 and 1991, the expectations questions were asked to employed individuals and pensioners. In 1995 and 1998, the expectations questions were asked to employed and unemployed individuals (including first-job seekers) but excluding pensioners.}

%Finally, it is important to mention there is a miss-match in the timing of the beliefs questions; while consumption, income and assets refer to past calendar year\footnote{Assets refer to assets held at the end of the calendar year.}, the expectations questions refer to expectations held at the moment of the survey, which is a few months after the calendar year has ended. Though it would be ideal to have the timing of the questions math for the interpretation of the results, we expect the expectations questions to convey a large part of the expectations held just a few months before. 

\subsection{Main estimates}

{In this subsection we report our main estimates based on a parametric non-separable model with finite-type heterogeneity, as in (\ref{eq_Lbetaeta}). The covariates specification is as in Table \ref{Table_cons}. In addition, we let type probabilities depend on an intercept, average income, and average mean beliefs across the two periods. The mean belief coefficient, the coefficient of log family income, and the intercept, are all allowed to vary with the latent type in the main equation. We model error terms in the consumption equation to be i.i.d. normal (with a variance that does not depend on the type), and the type probabilities as following a multinomial logit specification. We report results for $K=2$ (which is the optimal number of types according to the Bayesian Information Criterion) and $K=3$ types. We also estimated specifications with $K=4$ but found estimates to be more unstable. We provide details about the empirical specification and estimation, including how we deal with multiple local optima of the likelihood function, in Appendix \ref{App_types}.}

In Table \ref{Table_cons_types} we report the coefficient estimates in the log consumption function $\phi$. We find that current income and mean income beliefs both affect consumption positively. However, we also find heterogeneity across households. In the specification with $K=2$ types, the first group has a large coefficient (0.817) on current income and a small (0.017) and insignificant coefficient of mean income beliefs. In contrast, the second group has a lower coefficient (0.536) on current income and a larger (0.172) coefficient on mean income beliefs. Likewise, the specification with $K=3$ types shows a negative association between impacts of current and expected income across types. In both specifications, mean beliefs and current income interact positively, and the variance of beliefs has a positive though insignificant impact.

%%%%%%%%%%%%%%%%%%%%%%%%%%%%%%%%%%%%%%%%%%%%%%%%%%%%%%%%%%%%%%%%%%%%%
\begin{table}[h!]
	\begin{center}
		\caption{Estimates of the log consumption function}
		\label{Table_cons_types}
		\adjustbox{max width=\linewidth}{
			\resizebox{\linewidth}{!}{	
				\begin{tabular}{lccccccccccccc}\hline\hline
					& \multicolumn{2}{c}{$K=2$ types} & \multicolumn{3}{c}{$K=3$ types} &\\\cmidrule(lr){2-3}\cmidrule(lr){4-6}
					& $k=1$ & $k=2$& $k=1$ & $k=2$ & $k=3$ \\\hline
					% latex table generated in R 4.4.3 by xtable 1.8-4 package
% Sun Oct 12 08:18:48 2025
Intercept & 10.474 & 10.324 & 10.063 & 10.267 & 10.394 \\ 
   & (0.133) & (0.131) & (0.143) & (0.162) & (0.170) \\ 
  Mean expected log income &  0.017 &  0.172 &  0.589 &  0.231 &  0.024 \\ 
   & (0.055) & (0.038) & (0.246) & (0.050) & (0.107) \\ 
  (Mean expected log income)$\cdot$(Log family income) &  0.055 &  0.055 &  0.114 &  0.114 &  0.114 \\ 
   & (0.030) & (0.030) & (0.095) & (0.095) & (0.095) \\ 
  Var expected log income &  2.012 &  2.012 &  1.629 &  1.629 &  1.629 \\ 
   & (1.805) & (1.805) & (1.875) & (1.875) & (1.875) \\ 
  (Var expected log income)$\cdot$(Log family income) & -2.030 & -2.030 & -2.152 & -2.152 & -2.152 \\ 
   & (3.321) & (3.321) & (4.132) & (4.132) & (4.132) \\ 
  Log family income &  0.817 &  0.536 &  0.132 &  0.549 &  0.855 \\ 
   & (0.054) & (0.033) & (0.145) & (0.042) & (0.123) \\ 
  Log family assets &  0.013 &  0.013 &  0.014 &  0.014 &  0.014 \\ 
   & (0.005) & (0.005) & (0.005) & (0.005) & (0.005) 
 \\\hline\hline
		\end{tabular}}}
	\end{center}
	\vspace{0.1cm}
	\footnotesize{\textit{Notes: SHIW, 1989--1991 and 1995--1998. Estimates of the conditional mean of log consumption in the finite mixture model with types $k\in\{1,...,K\}$. See Table \ref{Table_cons} for the full list of covariates, and see Appendix \ref{App_types} for the exact specification. Estimates are weighted using survey weights. Analytical standard errors are shown in parentheses.}}
\end{table}

We next use our framework, and our estimates of the consumption function reported in Table \ref{Table_cons_types}, to assess the effects of a counterfactual income tax on consumption. We assume that the tax schedule takes the parametric form $T(w_g)=w_g-\lambda w_g^{1-\tau}$, where $w_g$ denotes gross income (e.g., \citealp{benabou2002tax}). To define a baseline level of the tax, we rely on the estimates obtained by \citet{holter2019tax} for Italy, averaged over family characteristics in our sample. 

We consider three counterfactuals, corresponding to changes in the $\lambda$ and $\tau$ parameters that index the tax schedule. In the \emph{transitory tax} and \emph{permanent tax} counterfactuals, we increase the average tax by 10 percentage points by decreasing $\lambda$, only for one period in the former case and in all subsequent periods in the latter. In the \emph{regressivity} counterfactual, we set the parameter $\tau$ to its value in the French tax system (which is somewhat less progressive than the Italian one) while at the same time decreasing $\lambda$ such that the tax change is neutral in terms of total tax revenue. We provide details about counterfactual calculations in Appendix \ref{app_counter}.

In Figure \ref{fig_decompo_cons_types} we report the average effects TAPE, CAPE, and DAPE associated with the three tax counterfactuals, aggregated across types. In the calculations for the permanent tax and regressivity counterfactuals, we assume that individuals fully adjust their beliefs to the new tax; i.e., we implement the formula in (\ref{eq_delta_change}). In Appendix Table \ref{tab_counter_types} we report APE estimates and standard errors. The top panel in Figure \ref{fig_decompo_cons_types} shows average partial effects based on the estimates for $K=2$, while the bottom panel corresponds to estimates for $K=3$. On the left graphs we show the effects on log consumption of a 10\% transitory tax. 
The overall effect is $-0.072$ for $K=2$, and $-0.067$ for $K=3$. In addition, in both specifications there is only moderate variation along income quantiles (indicated on the x axis).

On the middle graphs we show the effect of a 10\% permanent tax. Note that the contemporaneous average partial effect (CAPE) coincides with the effect of a transitory tax (compare with the left graphs). Added to this contemporaneous effect, we find sizable dynamic effects. The dynamic APE (DAPE), which reflects the impact of a changes in beliefs, contributes an additional $-0.012$ for $K=2$, and $-0.021$ for $K=3$. The total change in consumption, which is $-0.084$ for $K=2$ and $-0.089$ for $K=3$, is less than the 10\% decrease in income, as is expected if households are only partially insured against income changes (\citealp{blundell2008consumption}). Moreover, the estimates from both specifications indicate that dynamic effects are larger for higher-income households. Lastly, on the right graphs we show the effect of a revenue-neutral decrease in the progressivity of the tax. While the total effects averaged over all households are relatively small ($-0.013$ for both specifications), they show substantial heterogeneity along the income distribution: reducing progressivity tends to favor the rich, and it hurts the log consumption of the poor proportionally more.

{However, the aggregate numbers shown in Figure \ref{fig_decompo_cons_types} mask important heterogeneity. To see this, we show average effects by latent types in Figure \ref{fig_decompo_cons_by_types} (and report the corresponding point estimates and standard errors in Appendix Tables \ref{tab_counter_types_K2} and \ref{tab_counter_types_K3}). Under the permanent tax, for the two-types model, one of the types has a larger contemporaneous effect ($-0.09$ versus $-0.06$), yet virtually no dynamic effect. This type accounts for slightly more than a third of households. The three-types model shows even more heterogeneity: one of the types still has a large contemporaneous effect and non dynamic effect, but the other two show different patterns. In particular, type 1, which accounts for 19\% of households, has a low contemporaneous effect ($-0.01$) and a large dynamic effect ($-0.06$). These differences could reflect behavioral heterogeneity, e.g., between ``hand-to-mouth'' consumers and ``permanent income'' consumers, although sharpening this interpretation would require a structural model.}

\begin{figure}[h!]
	\begin{center}
		\caption{Average partial effects for various tax counterfactuals}
		\label{fig_decompo_cons_types}
		\begin{tabular}{ccc}
			\multicolumn{3}{c}{A. $K=2$ types } \\ 
			(a) Transitory tax & (b) Permanent tax & (c) Regressivity\\
			\includegraphics[width=5.35cm]{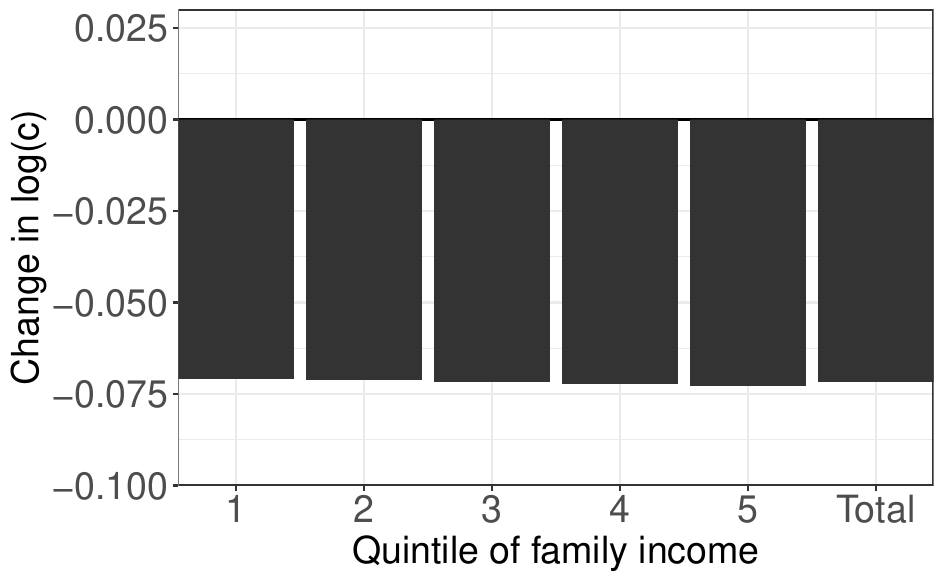}& 
			\includegraphics[width=5.35cm]{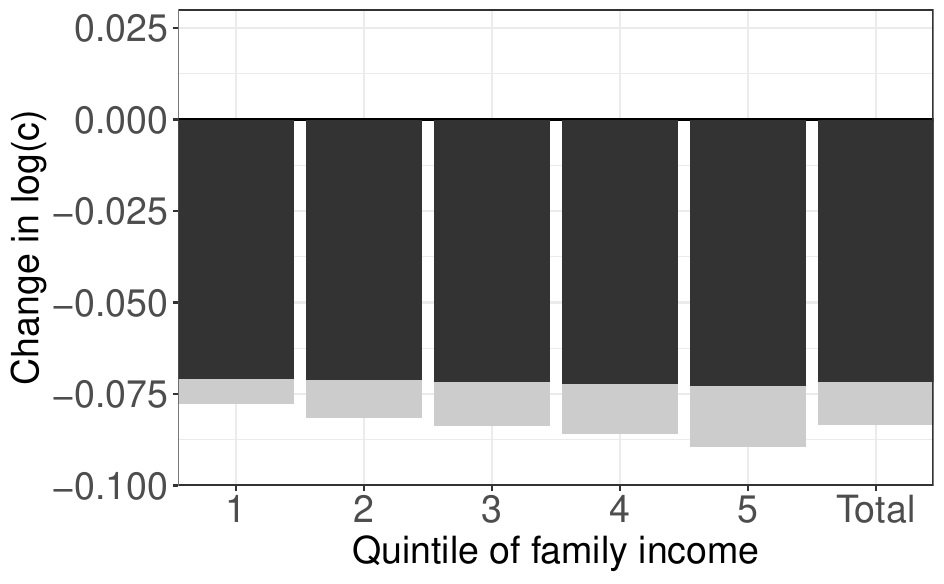}& 
			\includegraphics[width=5.35cm]{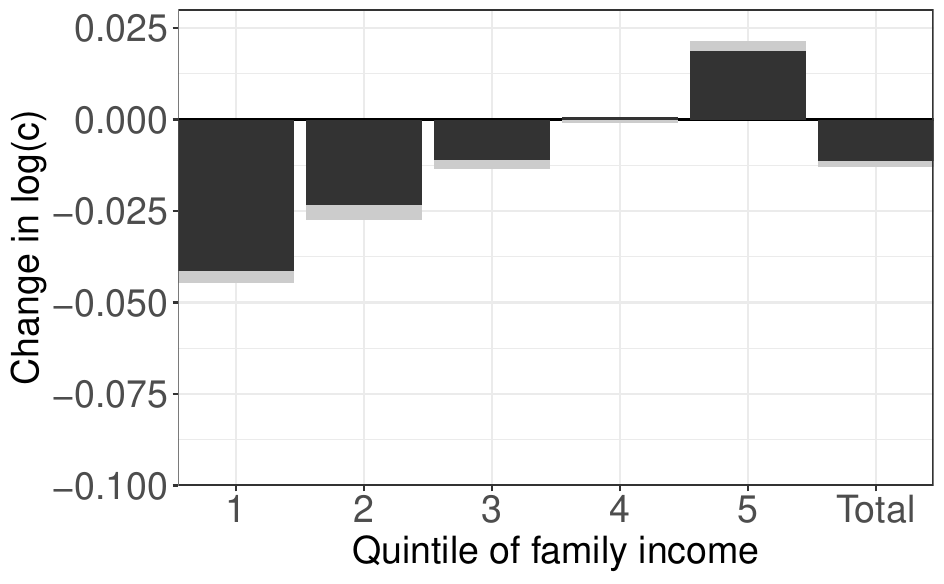}\\
			\multicolumn{3}{c}{B. $K=3$ types} \\ 
			(d) Transitory tax & (e) Permanent tax & (f) Regressivity\\
			\includegraphics[width=5.35cm]{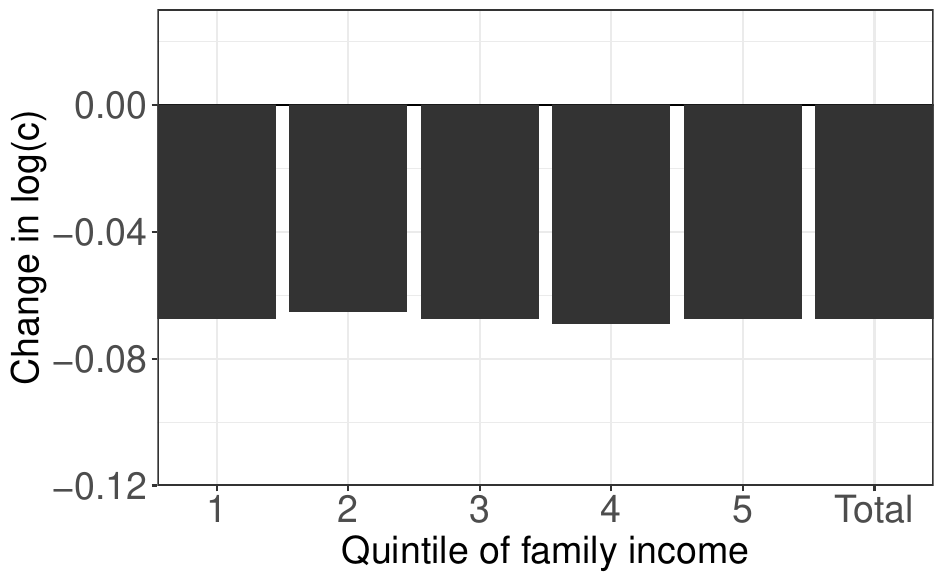}& 
			\includegraphics[width=5.35cm]{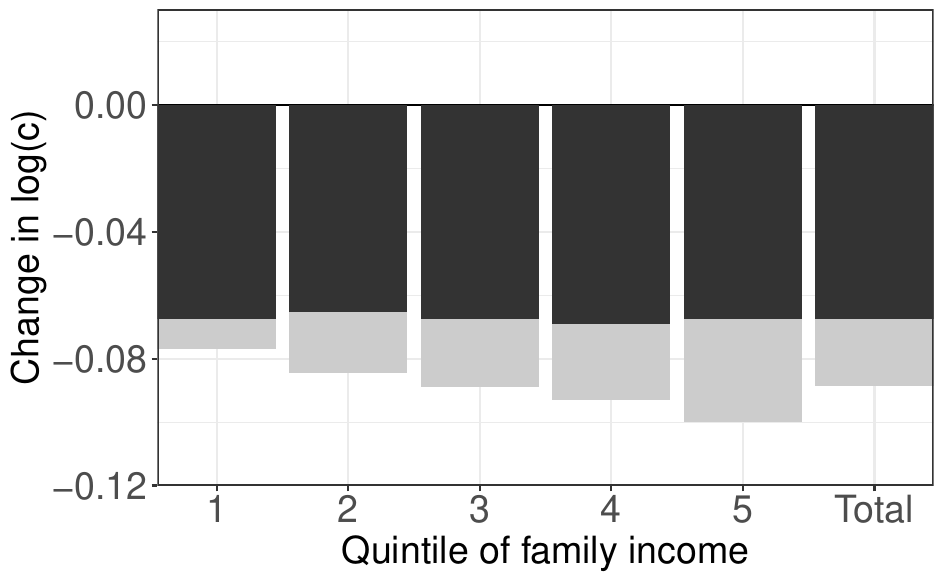}& 
			\includegraphics[width=5.35cm]{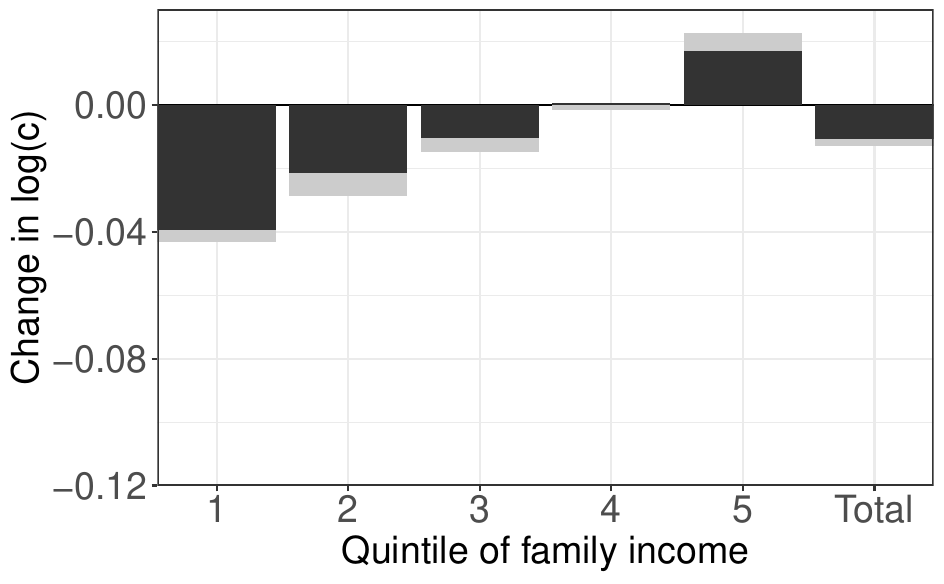}\\
			
		\end{tabular}
	\end{center}
	
	{\footnotesize \textit{Notes: SHIW, 1989--1991 and 1995--1998, cross-sectional sample. Black bars correspond to contemporaneous APE and grey bars correspond to dynamic APE. Total APE are the sums of CAPE and DAPE. Overall average effects, by quintile of family income. Estimates based on a parametric model with finite types: two types in the top panel, and three types in the bottom panel.}}
\end{figure}

\begin{figure}[h!]
	\begin{center}
		\caption{Average partial effects by type}
		\label{fig_decompo_cons_by_types}
		\begin{tabular}{ccc}
			\multicolumn{3}{c}{A. $K=2$ types } \\ 
			(a) Transitory tax & (b) Permanent tax & (c) Regressivity\\
			\includegraphics[width=5.35cm]{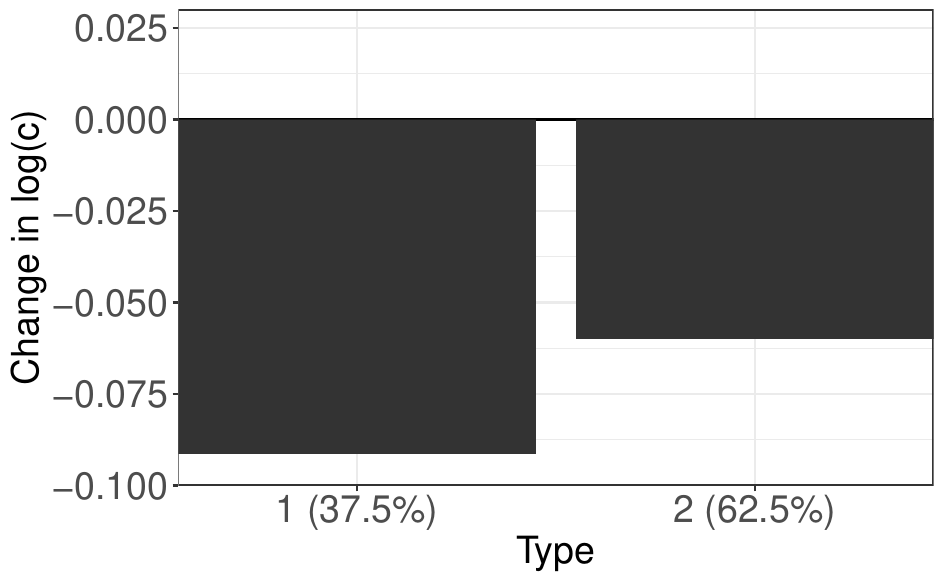}& 
			\includegraphics[width=5.35cm]{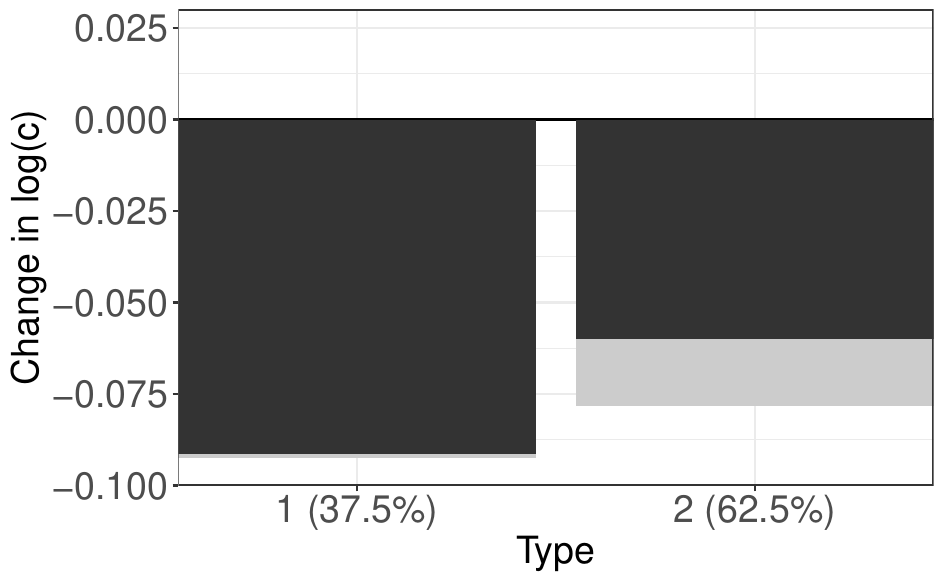}& 
			\includegraphics[width=5.35cm]{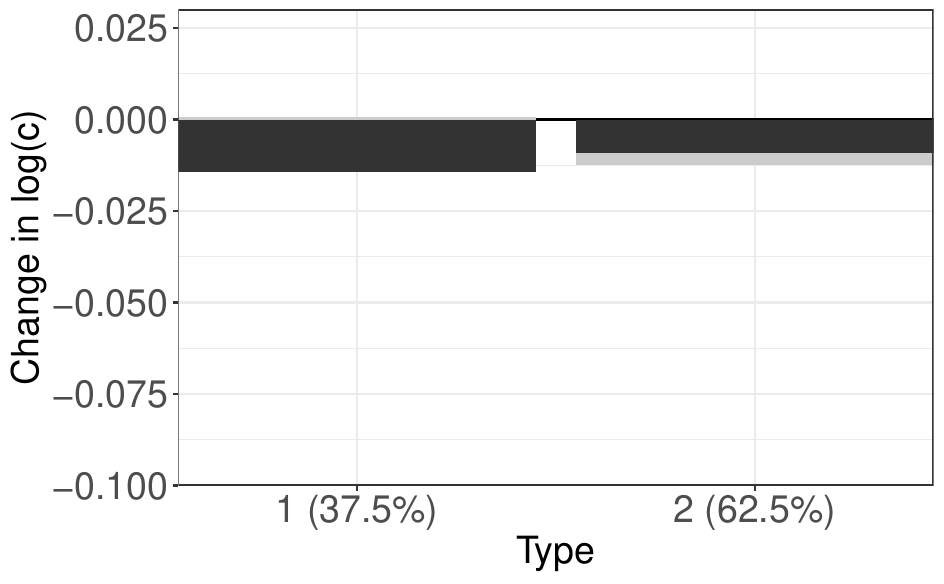}\\
			\multicolumn{3}{c}{B. $K=3$ types} \\ 
			(d) Transitory tax & (e) Permanent tax & (f) Regressivity\\
			\includegraphics[width=5.35cm]{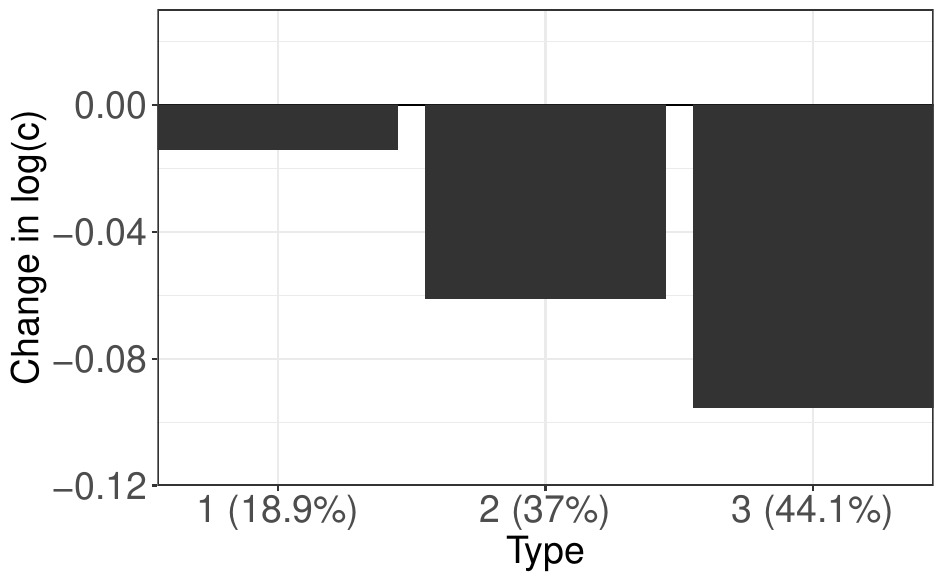}& 
			\includegraphics[width=5.35cm]{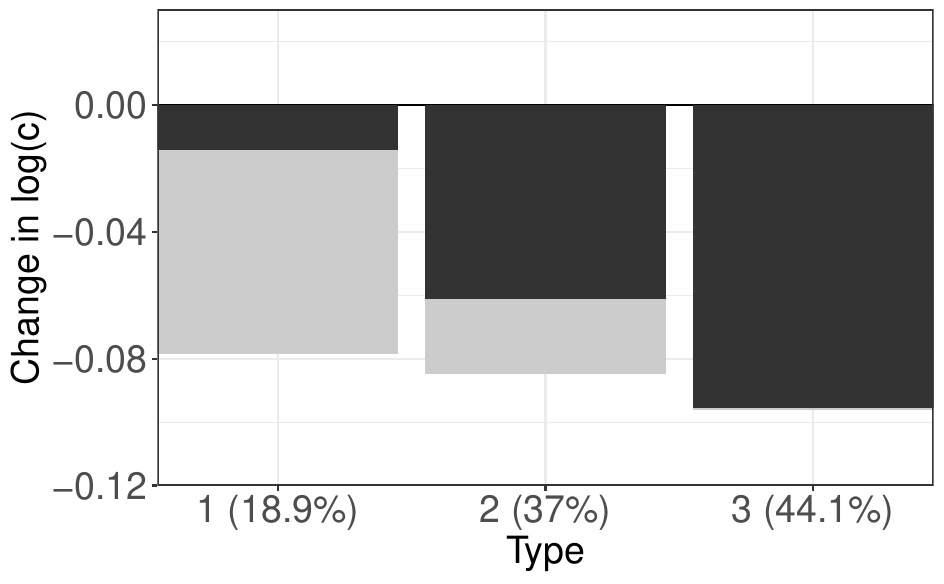}& 
			\includegraphics[width=5.35cm]{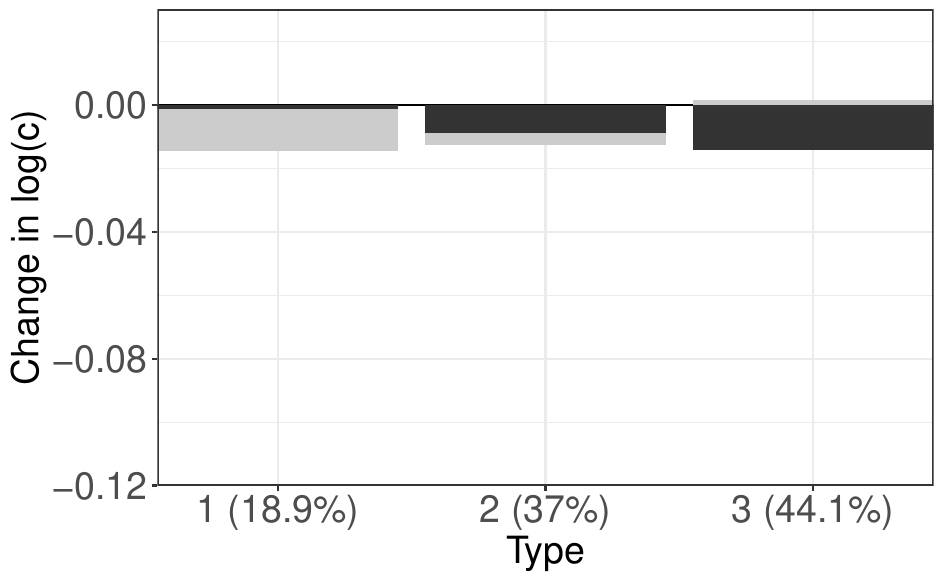}\\
			
		\end{tabular}
	\end{center}
	
	{\footnotesize \textit{Notes: See the notes to Figure \ref{fig_decompo_cons_types}. Average effects by types. Type proportions in the cross-sectional sample are indicated on the x-axis.}}
\end{figure}

{It is interesting to compare these estimates to average partial effects calculations that do not account for the role of beliefs, that is, which rely on a finite-type model yet exclude the belief-related covariates.} In that case, focusing on $K=3$ the average consumption effect over all households of a 10\% permanent income tax is $-0.077$. This is larger than the contemporaneous effect ($-0.067$) in Figure \ref{fig_decompo_cons_types}, consistently with beliefs being an omitted yet relevant regressor in the specification without beliefs. However, this is lower than the total effect in the same figure that accounts for both contemporaneous and dynamic margins ($-0.089$). See Appendix \ref{app_no_beliefs} for additional results based on this specification. These differences underscore the need to account for beliefs when computing average partial effects. In addition, note that an estimation method that does not include beliefs cannot account for the difference in impact between a permanent tax and a transitory one. 

Lastly, it is worth emphasizing that two conditions are needed in order to interpret the average partial effects in Figures \ref{fig_decompo_cons_types} and \ref{fig_decompo_cons_by_types} as structural tax counterfactuals. The first one is that individual beliefs respond one-to-one to the tax. The extension outlined in Remark \ref{rem1} can be useful to relax this assumption, given plausible values for the $\xi$ parameters. The second condition is that the belief updating rule $\rho_i$ is invariant under the tax. When tax changes have a long-lasting effect, changes in $\rho_i$ may occur and induce a third margin of response, beyond contemporaneous and dynamic effects (i.e., beyond CAPE and DAPE). While this third margin may be small or zero in certain cases (as in the permanent-transitory model with a proportional tax, see Subsection \ref{subsec_cons}), accounting for it may be important in other cases. The extension to beliefs over longer horizons that we outline in Section \ref{sec_conc} provides a possible way forward.

\subsection{Other specifications}

The finite-type specifications that we rely on to produce counterfactual estimates depends on parametric assumptions, both for the shape of the consumption function and the functional form and serial dependence of the error term. In theory, the mixture specification could be made richer in both dimensions by estimating semi- or nonparametric specifications. However, the fact that the panel component in the SHIW is restricted to two periods and has a relatively low sample size makes such extensions impractical. For example, nonparametric identification results based on finite mixture models in the literature require at least three consecutive periods (e.g., \citealp{hall2003nonparametric}, \citealp{hu2015microeconomic}). 

In this subsection we report results based on three alternative specifications. In the first two, we restrict coefficient heterogeneity while relying on less restrictive assumptions about the error terms and allowing for nonlinear effects. The specifications are based on the assumption that only the intercept in $\phi_i$ is heterogeneous. The first specification is based on the same covariates that appeared in Table \ref{Table_cons_types}, while the second specification relies on a dictionary including interactions and powers of the covariates up to the third order. We report estimates of TAPE, CAPE, and DAPE obtained using OLS for the former specification, and the Lasso for the latter, where we rely on the double/debiased Lasso method introduced by \citet{belloni2014inference}. In a third specification, we model the conditional mean function $\phi_i$ as linear in heterogeneous coefficients $\beta_i$  (which are not restricted to be discrete), and specify the mean of $\beta_i$ as a linear function of household-specific averages of covariates in the spirit of \citet{mundlak1978pooling}.\footnote{We provide details about the double/debiased Lasso approach in Appendix \ref{App_dlasso}, and about the correlated random-effects approach in Appendix \ref{app_CRE}.	Unlike in the finite-type model, in all three specifications that we study in this subsection we treat all covariates -- including assets -- as strictly exogenous. In Appendix Table \ref{Table_cons_assets} we report various robustness checks.} These three specifications are non-nested with our baseline parametric finite mixture model, since in particular they do not restrict the form of error terms except for mean independence conditions.

%\begin{align}{\phi}_i\left(x_{it},\pi_{it},z_{it}\right)&={\phi}\left(x_{it},\pi_{it},z_{it}\right)+\alpha_i,\label{eq_action_empirics_short}
%\end{align}  

\begin{figure}[h!]
	\begin{center}
		\caption{Average partial effects for various tax counterfactuals, other specifications }
		\label{fig_decompo_cons}
		\begin{tabular}{ccc}
			\multicolumn{3}{c}{A. Average partial effects based on OLS estimates} \\ 
			(a) Transitory tax & (b) Permanent tax & (c) Regressivity\\
			\includegraphics[width=5.35cm]{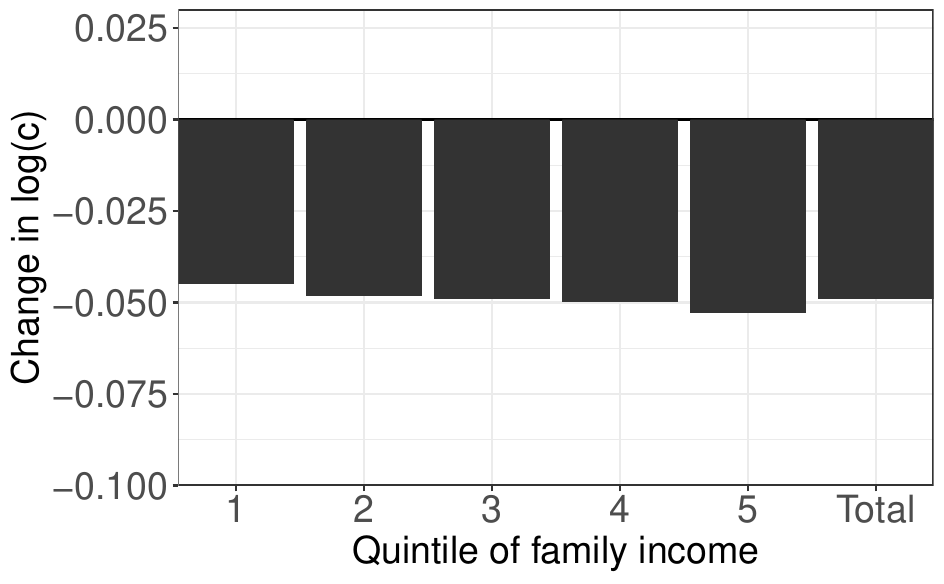}& 
			\includegraphics[width=5.35cm]{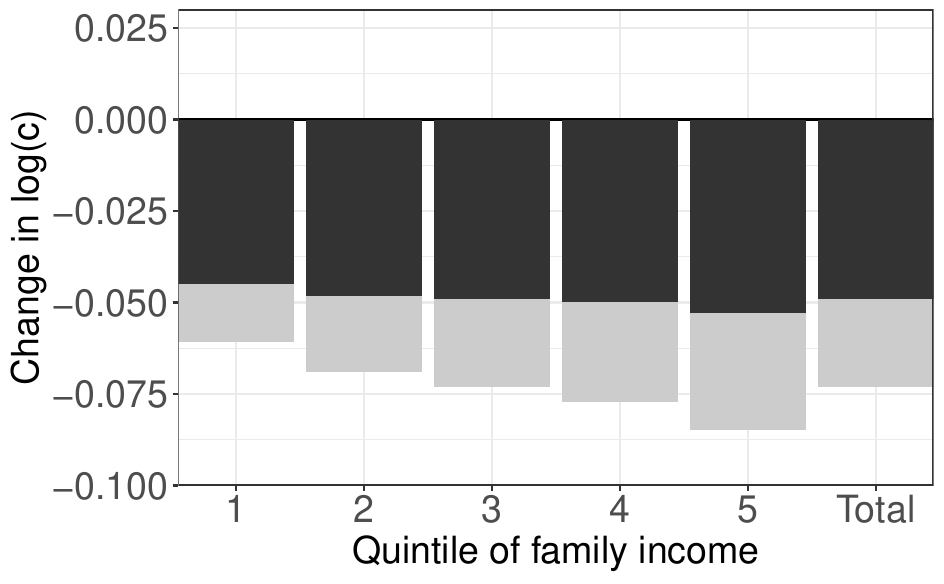}& 
			\includegraphics[width=5.35cm]{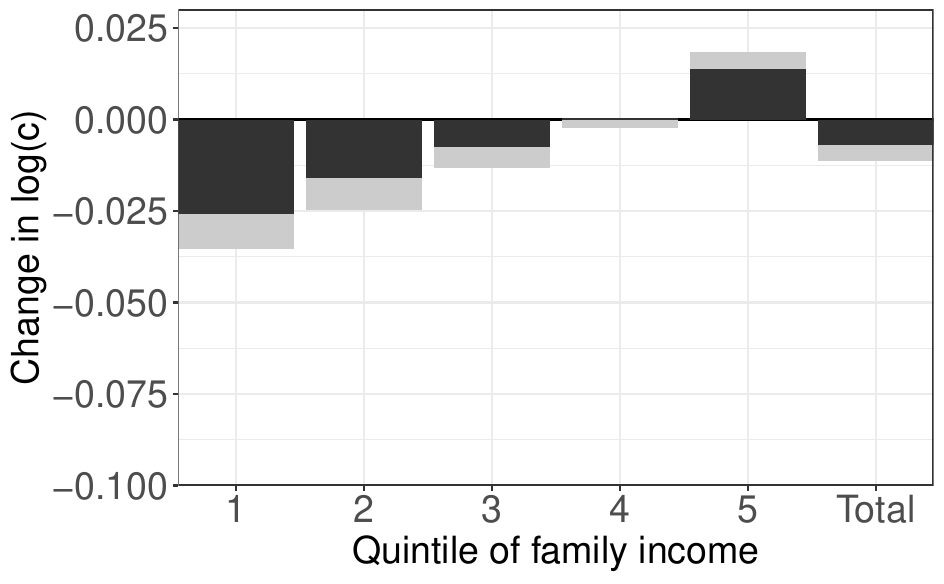}\\
			\multicolumn{3}{c}{B. Average partial effects based on the Lasso} \\ 
			(d) Transitory tax & (e) Permanent tax & (f) Regressivity\\
			\includegraphics[width=5.35cm]{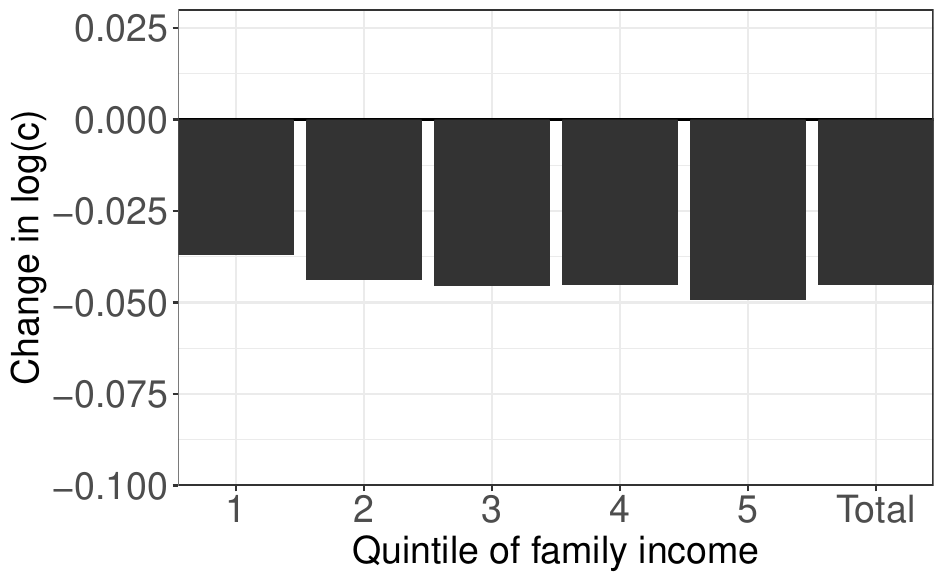}& 
			\includegraphics[width=5.35cm]{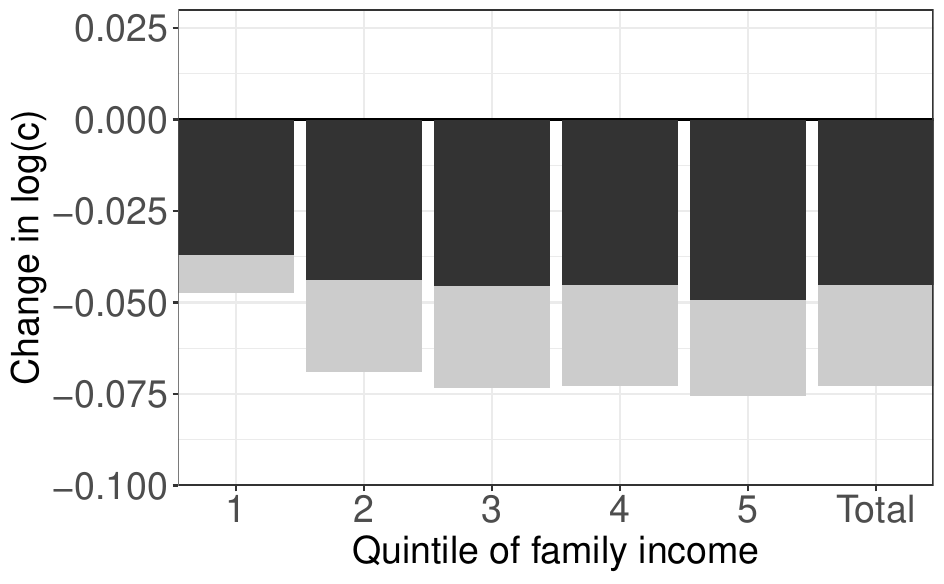}& 
			\includegraphics[width=5.35cm]{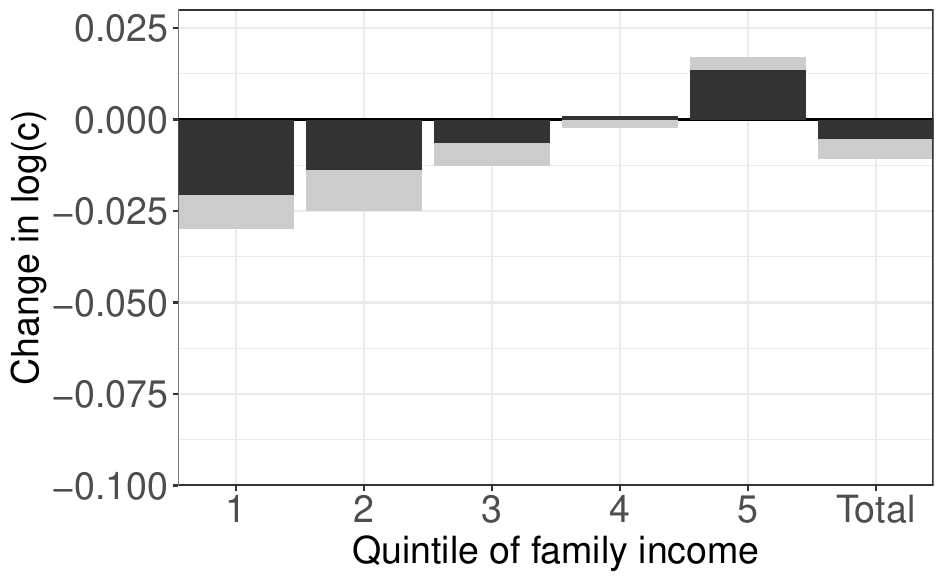}\\
			\multicolumn{3}{c}{C. Average partial effects based on correlated random-effects estimates} \\ 
				(a) Transitory tax & (b) Permanent tax & (c) Regressivity\\
			\includegraphics[width=5.35cm]{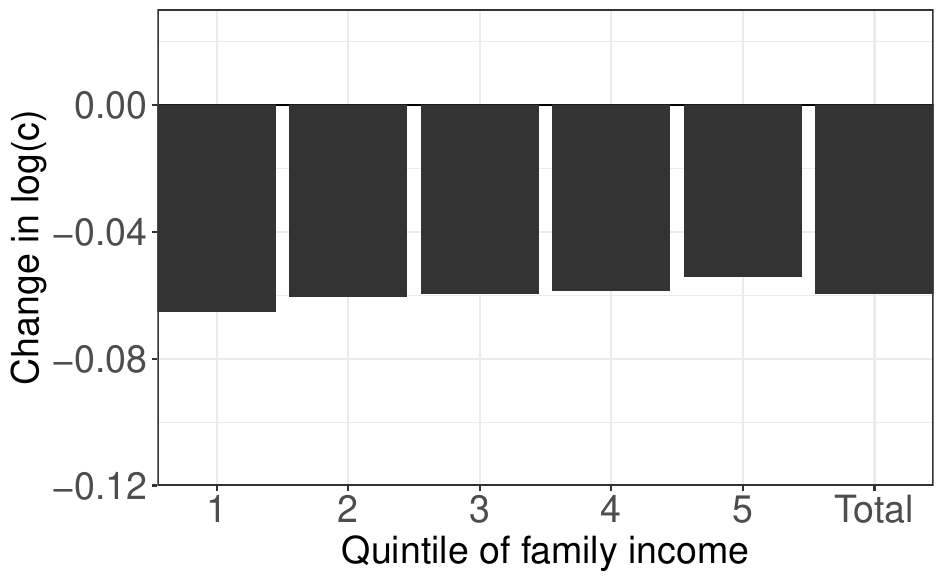}& 
			\includegraphics[width=5.35cm]{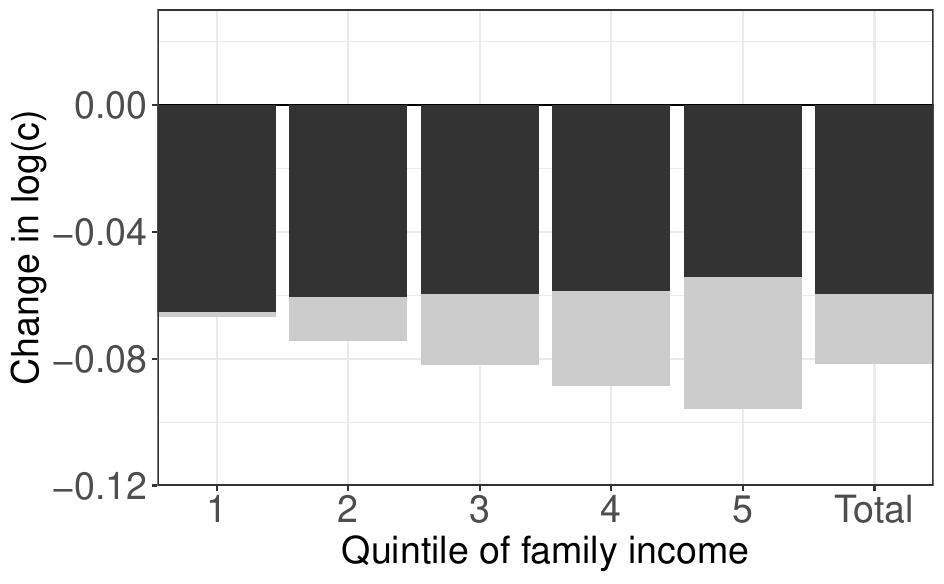}& 
			\includegraphics[width=5.35cm]{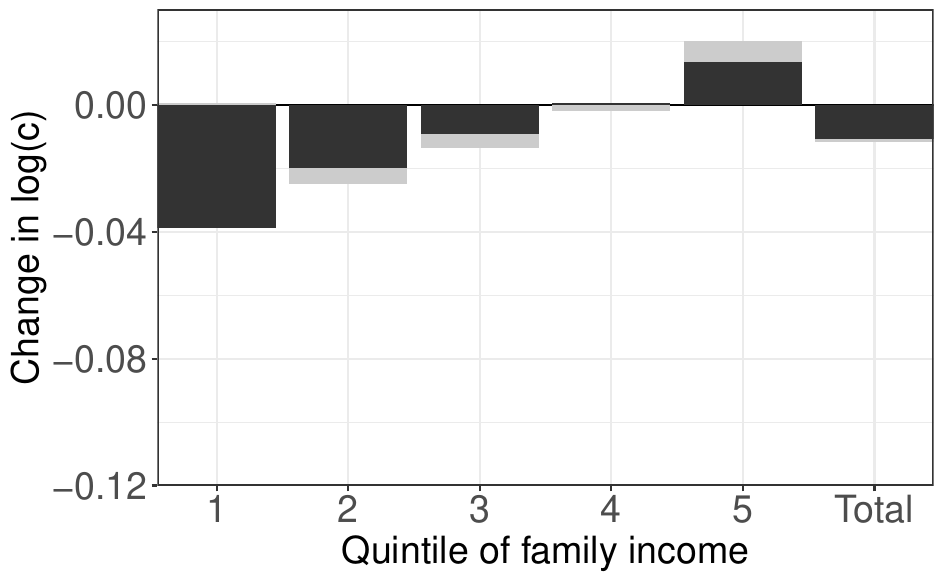}\\

		\end{tabular}
	\end{center}
	
	{\footnotesize \textit{Notes: See the notes to Figure \ref{fig_decompo_cons_types}. In the top panel we report results based on OLS estimates, see column (5) in Table \ref{Table_cons}. In the middle panel we report estimates based on the double/debiased Lasso, for a dictionary including interactions and powers of the covariates up to the third order. See Appendix Figure \ref{fig_counter_lasso} for results corresponding to second and fourth order interactions and powers. In the bottom panel we report estimates based on a correlated random-effects specification described in Appendix \ref{app_CRE}.}}
\end{figure}

The top two panels in Figure \ref{fig_decompo_cons} show average partial effects based on OLS and the Lasso, respectively.\footnote{We report point estimates and standard errors based on the bootstrap in Appendix Table \ref{tab_counter}.} The effect on log consumption of a 10\% transitory tax based on OLS is $-0.049$, and it is similar according to the Lasso. Moreover, there is only moderate variation along income quantiles. These effects are lower than the one obtained in our finite-type model with $K=3$ ($-0.067$). For both specifications we find sizable dynamic effects in the permanent tax counterfactual, which contribute an additional $-0.024$ according to OLS, and $-0.028$ according to the Lasso. This is slightly larger than the estimate based on the finite-type specification ($-0.021$). Lastly, the right graphs show the effect of a revenue-neutral decrease in the progressivity of the tax. The total effects averaged over all households are around $-0.011$, similar to the estimates based on the finite-type models. 

We find that the finite-type specifications that feature response heterogeneity, and the OLS and Lasso specifications that restrict heterogeneity to only affect the intercept, tend to give quite similar, albeit not identical, counterfactual estimates on the data. While this is not a general result and there might be larger discrepancies across these methods in other settings, we conducted a small Monte Carlo experiment to assess the sign and magnitude of APE estimates based on OLS in first differences in a data generating process based on our non-separable finite-mixture model with three types. The results, shown in Appendix \ref{app_mcarlo}, indicate overall agreement between the OLS estimates based on a misspecified model and the true counterfactual effects in this case. At the same time, an important advantage of the finite-type model is that it can capture heterogeneous responses across households, as shown by Figure \ref{fig_decompo_cons_by_types}.

As an additional robustness check for our estimates based on a finite-type model, in the bottom panel in Figure \ref{fig_decompo_cons} we report estimates based on a random coefficients model under a correlated random-effects specification that allows for continuously distributed household unobserved heterogeneity.\footnote{We report point estimates and standard errors of the coefficients in Appendix Table \ref{Table_CRE}, and of the counterfactual effects in panel C of Appendix Table \ref{tab_counter}.} The estimates show contemporaneous effects similar to the ones based on the finite-type model (compare with Figure \ref{fig_decompo_cons_types}), and somewhat larger than the effects based on specifications without coefficient heterogeneity. In turn, dynamic effects in the permanent tax counterfactual are slightly lower than for the OLS and Lasso estimates, and again close to the estimates based on the finite-type model -- although the correlated random-effects specification implies somewhat higher variation of DAPE along the income distribution. Lastly, both specifications with heterogeneous coefficients (i.e., discrete or continuous) predict similar effects under the regressivity counterfactual. The close agreement between finite-type and correlated random-effects specifications suggests that the estimates are not driven primarily by the parametric choices we made when implementing the finite-type estimator.

Lastly, a possible concern with the estimates in this section is measurement error in belief data. To explore this issue, in Appendix \ref{app_merror} we report alternative sets of estimates based on a specific model of measurement error, where we assume that respondents do not report their subjective probabilities exactly but estimates of these probabilities based on a given number $M$ of draws (or income growth ``scenarios'') from their subjective distributions. Using a ``small-$\sigma$'' approximation in the spirit of \citet{evdokimov2022simple}, we find that estimates of log consumption regression coefficients are relatively insensitive to this form of (non-classical) measurement error. However, it is important to note that this exercise is based on a specific model, and addressing the possibility of other forms of measurement error is an important area for future work.\footnote{{{For example, one reason for measurement error}} could be experimenter demand effects, as studied by \citet{mummolo2019demand} and \citet{de2025subjective}.}

\section{Extensions\label{sec_conc}}

We discuss possible extensions of our framework, and conclude with a discussion of implications for belief data collection.

\subsection{Multiple-horizons}

A key assumption in our framework is that, while beliefs about next period's state variables change in the data and counterfactual, the belief updating rule $\rho_i$ is time-invariant in sample and invariant to the counterfactual change. This assumption can be relaxed by introducing beliefs over multiple horizons.

If one had access to data on the sequence of beliefs about $x_{i,t+1},x_{i,t+2},...$ into the far future, accounting for those as determinants of the decision, and shifting them in the counterfactual, would provide valid predictions without the need for an invariance assumption about some $\rho_i$ process. To go one step in this direction, one can elicit beliefs over multiple horizons $x_{i,t+1},x_{i,t+2},...,x_{i,t+S}$ (as in \citealp{kocsar2025workers}), and account for variation in those beliefs in estimation and counterfactuals.

To describe such an approach, let us replace Assumption \ref{ass_beliefs0} by the following, for some $S\geq 1$:\begin{equation}
	\left(x_{i,t+S},...,x_{i,t+1}\,|\, y_{it},\Omega_{it}\right) \sim \left(x_{i,t+S},...,x_{i,t+1}\,|\, \Omega_{it}\right),\label{ass_2_prime}
\end{equation}
and denote the corresponding conditional density as $\pi_{it}(x_{i,t+S},...,x_{i,t+1})$. In this case, (\ref{eq_Bellman}) becomes
{\begin{align*}&V_i(x_t,\pi_t,z_t,\nu_t)=\limfunc{max}_{y_t}\, \bigg\{u_i(y_t,x_t,z_t,\nu_t)\notag\\
	&+\beta_i \int V_i(x',\pi',z',\nu')\pi_{t}^{(1)}(x') {\rho}_i(\pi'; x',\pi_{t},x_t)\gamma_i(z';z_t,x_t,y_t)\tau_{i}(\nu')dx'd\pi'dz'd\nu'\bigg\},
\end{align*}}where $\pi_{t}^{(1)}$ denotes the marginal of $\pi_{t}$ corresponding to period-$t+1$ outcomes. This implies that equation (\ref{eq_a}) is satisfied for the $\pi_{it}$ corresponding to (\ref{ass_2_prime}). Hence our approach is unchanged, except for the use of a multivariate subjective belief density.

\subsection{State-contingent beliefs}

It is interesting to allow for ``state-contingent'' beliefs, where beliefs are contingent on {potential choices} $y_{it}$, and Assumption \ref{ass_beliefs0} does not hold. For example, in a model of occupational choice, individual income beliefs contingent on occupational choice may be available (e.g., \citealp{patnaik2020role}, \citealp{arcidiacono2020ex}). In that case, our framework is unchanged except for the fact that the state-contingent beliefs enter as arguments in the decision rule. 

To see this, suppose for simplicity that actions $y_{it}$ belong to a finite set ${\cal{Y}}$ with $n$ elements. {In this case, one can define $\pi_{it}=\{\pi_{it}(\cdot;y)\, :\, y\in {\cal{Y}}\}$ to be a set of $n$ densities where, for all $y\in{\cal{Y}}$, $\pi_{it}(\cdot;y)$ is the subjective density of $\left(x_{i,t+1}\,|\, y_{it}=y,\Omega_{it}\right)$.} With this new definition of $\pi_{it}$, and the associated change in the definition of $\rho_i$ in {Assumption \ref{ass_beliefs_mod}}, the framework is unchanged relative to Section \ref{sec_structural}. In particular, the decision rule is still given by (\ref{eq_a}), so actions depend on the $n$ belief densities $\pi_{it}(\cdot;y)$. 

As an example of a model with state-contingent beliefs, suppose $x_{it}=\alpha_i+\varepsilon_{it}(k)$ when $y_{i,t-1}=k$, for $k\in\{0,1\}$.\footnote{This is equivalent to assuming the individual only observes $x_{it}(k)=\alpha_i+\varepsilon_{it}(k)$ when $y_{i,t-1}=k$. As an extension, $\alpha_i$ may also depend on $k$ (for example, $\alpha_i$ may represent a vector of occupation-specific abilities), and $x_{it}(k)=\alpha_i(k)+\varepsilon_{it}(k)$. In that case, the updating formulas (\ref{eq_mean3})-(\ref{eq_var3}) need to be adjusted to vector-valued $\mu_{it}$ and matrix-valued $\sigma^2_{it}$. See \citet{arcidiacono2025college} for an example.} Suppose in addition that $\varepsilon_{it}(k)\sim {\cal{N}}(0,\sigma_{\varepsilon_i(k)}^2)$, independent across $i$ and $t$, and that agents are Bayesian with a normal prior on $\alpha_i$. {At the beginning of period $t$, the posterior distribution of $\alpha_{i}$ when $y_{i,t-1}=k$ is then} ${\cal{N}}(\mu_{it},\sigma_{it}^2)$, where $\mu_{it}$ and $\sigma^2_{it}$ are functions of $k$ satisfying
\begin{eqnarray}
	\mu_{it} &=& \mu_{i,t-1} + \frac{\sigma^2_{it}}{\sigma^2_{\varepsilon_i}(k)}\bigg(x_{it}-\mu_{i,t-1}\bigg), \label{eq_mean3}\\
	(\sigma^2_{it})^{-1} &=& (\sigma^2_{i,t-1})^{-1} + (\sigma^2_{\varepsilon_i(k)})^{-1}. \label{eq_var3}
\end{eqnarray}
{When deciding to choose $y_{it} \in \{0,1\}$, $i$ also considers the belief distribution about the yet unobserved $x_{i,t+1}$, which is the sum of two independent normal variables. The first, $\alpha_i$, has an expected distribution with mean and variance given by (\ref{eq_mean3}) and (\ref{eq_var3}), respectively. The second variable, $\varepsilon_{i,t+1}(j)$, has zero mean and a variance that depends on her current (not yet taken) choice, $y_{it}=j$.} We then define beliefs as $\pi_{it}=(\pi_{it}(0),\pi_{it}(1))$, where $\pi_{it}(j)$ is the normal density with mean $\mu_{it}$ and variance $\sigma_{it}^2+\sigma_{\varepsilon_i(j)}^2$ for $j\in\{0,1\}$. It follows from (\ref{eq_mean3})-(\ref{eq_var3}) that {Assumption \ref{ass_beliefs_mod}}, for these beliefs $\pi_{it}$, is satisfied.

\subsection{Implications for belief data collection}

In this paper we provide a method to account for the role of individual expectations in assessing the impact of policies and other counterfactuals. {Our approach is justified} under dynamic structural assumptions, yet implementing the method does not require full specification and estimation of a structural model. A key input to our approach is the use of data on subjective beliefs. Belief elicitation is an active research area. Our approach motivates more work on this front, in several directions. 

First, in this section we have shown the usefulness of eliciting belief responses over multiple horizons, and how to incorporate such beliefs to our approach. Research along this line (see, e.g., \citealp{kocsar2025workers}) should be particularly useful to understand dynamic responses under less restrictive invariance conditions, and to study counterfactual questions at various horizons in the future. 

Second, we have discussed the usefulness of collecting data on state-contingent beliefs (e.g., \citealp{patnaik2020role}, \citealp{arcidiacono2020ex}), and shown that such data can easily be incorporated into our approach. We have also discussed the usefulness of eliciting beliefs under counterfactual policy scenarios (e.g., \citealp{roth2023effects}), to directly measure how beliefs may or may not change in a counterfactual situation.

Lastly, we have highlighted the usefulness of having longitudinal information on individual beliefs. While many data sets with elicited beliefs such as the SHIW have a panel component, the panel dimension often tends to be short, which {puts constraints on} the researcher's ability to allow for individual heterogeneity. Collecting longer longitudinal information {exhibiting more variation in beliefs over time} is important for harnessing the power of belief data. 

\clearpage

{\footnotesize
\bibliographystyle{econometrica}
\bibliography{biblio}
}
\clearpage

\appendix
\renewcommand{\thesection}{\Alph{section}}

\setcounter{figure}{0}\renewcommand{\thefigure}{\arabic{figure}}

\setcounter{table}{0}\renewcommand{\thetable}{\arabic{table}}

\setcounter{footnote}{0}\renewcommand{\thefootnote}{\arabic{footnote}}

\setcounter{assumption}{0}\renewcommand{\theassumption}{A\arabic{assumption}}

\setcounter{equation}{0}\renewcommand{\theequation}{A\arabic{equation}}

\setcounter{lemma}{0}\renewcommand{\thelemma}{A\arabic{lemma}}

\setcounter{proposition}{0}\renewcommand{\theproposition}{A\arabic{proposition}}

\setcounter{corollary}{0}\renewcommand{\thecorollary}{A\arabic{corollary}}

\setcounter{theorem}{0}\renewcommand{\thetheorem}{A\arabic{theorem}}

\hypersetup{colorlinks=true,linkcolor=blue,urlcolor=blue,citecolor=blue}

\setcounter{figure}{0}\renewcommand{\thefigure}{\thesection\arabic{figure}}

\setcounter{table}{0}\renewcommand{\thetable}{\thesection\arabic{table}}

\vskip 3cm

%%%%%%%%%%%%%%%%%%%%%%%%%%%%%%%%%%%%%%%%%%%%%%%%%%%%%%%%%%%%%%%%%%%%%%%%%%%%%%%%%%%%%%%%%%%%%%%%%%%%%%
%%%%%%%%%%%%%%%%%%%%%%%%%%%%%%%%%%%%%%%%%%%%%%%%%%%%%%%%%%%%%%%%%%%%%%%%%%%%%%%%%%%%%%%%%%%%%%%%%%%%%%
%%%%%%%%%%%%%%%%%%%%%%%%%%%%%%%%%%%%%%%%%%%%%%%%%%%%%%%%%%%%%%%%%%%%%%%%%%%%%%%%%%%%%%%%%%%%%%%%%%%%%%

\clearpage

\setcounter{figure}{0}\renewcommand{\thefigure}{\thesection\arabic{figure}}

\setcounter{table}{0}\renewcommand{\thetable}{\thesection\arabic{table}}

\begin{center}
	{ {\LARGE ONLINE APPENDIX } }
\end{center}

\section{Belief formation models with learning\label{app_belform}}

In this section of the appendix we describe two models of belief formation with learning that we mentioned in Subsection \ref{subsec_belief_form}. 

%%%%%%%%%%%%%%%%%%%%%%%%%%%%%%%%%%%%%%%%%%%%%%%%%%%%%%%%%%%%%%%%%%%%%%%%%%%%%%%%%%%%%%%%%%%%%%%%%%%%%%
\subsection{Exogenous beliefs}

We start with the model where beliefs are not affected by past actions. Suppose that
$$x_{it}=\alpha_i+\varepsilon_{it},$$
where $\varepsilon_{it}$ are i.i.d. ${\cal{N}}(0,\sigma_{\varepsilon_i}^2)$. Suppose agents have rational expectations, with information set $\Omega_{it}=\{x_{it},x_{i,t-1},...\}$, which does not include $\alpha_i$. Furthermore, assume agents are Bayesian learners with prior beliefs about $\alpha_i$ that are normally distributed. Then, by Bayes rule, posterior beliefs about $\alpha_i$ over time are also normally distributed with mean $\mu_{it}$ and variance $\sigma^2_{it}$ satisfying
\begin{eqnarray}
	\mu_{it} &=& \mu_{i,t-1} + \frac{\sigma^2_{it}}{\sigma^2_{\varepsilon_i}}\bigg(x_{it}-\mu_{i,t-1}\bigg), \label{eq_mean}\\
	(\sigma^2_{it})^{-1} &=& (\sigma^2_{i,t-1})^{-1} + (\sigma^2_{\varepsilon_i})^{-1}. \label{eq_var}
\end{eqnarray}
Then, $\pi_{it}$ is a normal density with mean $\mathbb{E}_{\pi_{it}}(x_{i,t+1})=\mu_{it}$ and variance $\limfunc{Var}_{\pi_{it}}(x_{i,t+1})=\sigma^2_{it}+\sigma^2_{\varepsilon_i}$. Hence, by (\ref{eq_mean})-(\ref{eq_var}) the belief process satisfies
Assumption \ref{ass_beliefs}. Note that the mean beliefs in (\ref{eq_mean}) are as in the adaptive expectations case, see (\ref{eq_adaptive}), but with a parameter $\lambda_{it}=\frac{\sigma^2_{it}}{\sigma^2_{\varepsilon_i}}$ that is time-varying and converges to zero over time.

%%%%%%%%%%%%%%%%%%%%%%%%%%%%%%%%%%%%%%%%%%%%%%%%%%%%%%%%%%%%%%%%%%%%%%%%%%%%%%%%%%%%%%%%%%%%%%%%%%%%%%
\subsection{Endogenous beliefs}

We now describe a variation of the previous model, where actions $y_{it}\in\{0,1\}$ are binary, and the agent observes an additional signal about $\alpha_i$,
$$s_{it}=\alpha_i+v_{it},$$
only when $y_{i,t-1}=1$. We assume that $v_{it}$ are i.i.d. ${\cal{N}}(0,\sigma_{v_i}^2)$, independent of $\varepsilon_{it}$ at all leads and lags. The posterior distribution of $\alpha_i$ is ${\cal{N}}(\mu_{it},\sigma_{it}^2)$, where now $\mu_{it}$ and $\sigma_{it}^2$ depend on $y_{i,t-1}$. When $y_{i,t-1}=0$, $\mu_{it}$ and $\sigma_{it}^2$ are given by (\ref{eq_mean})-(\ref{eq_var}), while when $y_{i,t-1}=1$ they are given by
\begin{eqnarray}
	\mu_{it} &=& \mu_{i,t-1} + \frac{\sigma^2_{it}}{\sigma^2_{\varepsilon_i}}\bigg(x_{it}-\mu_{i,t-1}\bigg)+ \frac{\sigma^2_{it}}{\sigma^2_{v_i}}\bigg(s_{it}-\mu_{i,t-1}\bigg), \label{eq_mean2}\\
	(\sigma^2_{it})^{-1} &=& (\sigma^2_{i,t-1})^{-1} + (\sigma^2_{\varepsilon_i})^{-1}+ (\sigma^2_{v_i})^{-1}. \label{eq_var2}
\end{eqnarray}
Now, denoting $\widetilde{\sigma}_{it}^2=\left[(\sigma^2_{i,t-1})^{-1} + (\sigma^2_{\varepsilon_i})^{-1}\right]^{-1}$, we have
\begin{equation}(s_{it}\,|\, x_{it},y_{i,t-1}=1,\Omega_{i,t-1})\sim {\cal{N}}\left(\mu_{i,t-1} + \frac{\widetilde{\sigma}^2_{it}}{\sigma^2_{\varepsilon_i}}\bigg(x_{it}-\mu_{i,t-1}\bigg),\widetilde{\sigma}_{it}^2+\sigma^2_{v_i}\right).\label{eq_sig_update}\end{equation}
Hence, by (\ref{eq_mean2}), 
\begin{equation}(\mu_{it}\,|\, x_{it},y_{i,t-1}=1,\Omega_{i,t-1})\sim {\cal{N}}\left(\mu_{i,t-1} + \left(\frac{\sigma^2_{it}}{\sigma^2_{\varepsilon_i}}+ \frac{\sigma^2_{it}}{\sigma^2_{v_i}}\frac{\widetilde{\sigma}^2_{it}}{\sigma^2_{\varepsilon_i}}\right)\bigg(x_{it}-\mu_{i,t-1}\bigg),\frac{\sigma_{it}^4}{\sigma_{v_i}^4}\left(\widetilde{\sigma}_{it}^2+\sigma^2_{v_i}\right)\right).\label{eq_mut}\end{equation}
It thus follows from (\ref{eq_var2})-(\ref{eq_mut}) in the case $y_{i,t-1}=1$, and from (\ref{eq_mean})-(\ref{eq_var}) in the case $y_{i,t-1}=0$, that $\pi_{it}$, which is the normal density with mean  $\mu_{it}$ and variance $\sigma_{it}^2+\sigma^2_{\varepsilon_i}$, {satisfies Assumption \ref{ass_beliefs_mod}. Note that, in this case, beliefs $\pi_{it}$ depend on past actions $y_{i,t-1}$, so Assumption \ref{ass_beliefs} does not hold.}

{Suppose now that $v_{it}=\psi v_{i,t-1}+\zeta_{it}$, where $\zeta_{it}$ are i.i.d. ${\cal{N}}(0,\sigma_{\zeta_i}^2)$, independent of $\varepsilon_{it}$ at all leads and lags. If $\psi\neq 0$, then, when $y_{i,t-1}=y_{i,t-2}=1$, $s_{it}$ is no longer independent of $s_{i,t-1}$ given $\pi_{i,t-1},x_{it},x_{i,t-1}$ (in contrast with (\ref{eq_sig_update})). Indeed, if $y_{i,t-2}=1$ then $\Omega_{i,t-1}$ contains $s_{i,t-1}$, and $v_{it}$ and $v_{i,t-1}$ are not independent conditional on $\pi_{i,t-1},x_{it},x_{i,t-1}$.}

%%%%%%%%%%%%%%%%%%%%%%%%%%%%%%%%%%%%%%%%%%%%%%%%%%%%%%%%%%%%%%%%%%%%%%%%%%%%%%%%%%%%%%%%%%%%%%%%
%%%%%%%%%%%%%%%%%%%%%%%%%%%%%%%%%%%%%%%%%%%%%%%%%%%%%%%%%%%%%%%%%%%%%%%%%%%%%%%%%%%%%%%%%%%%%%%%
%%%%%%%%%%%%%%%%%%%%%%%%%%%%%%%%%%%%%%%%%%%%%%%%%%%%%%%%%%%%%%%%%%%%%%%%%%%%%%%%%%%%%%%%%%%%%%%%
\setcounter{figure}{0}\renewcommand{\thefigure}{\thesection\arabic{figure}}
\setcounter{table}{0}\renewcommand{\thetable}{\thesection\arabic{table}}
\section{Relaxing parametric assumptions on beliefs}
\label{app_extens}

The parametric approach we adopt in our application is motivated by the coarse belief information available in the SHIW. In other applications with richer information, a nonparametric treatment of the belief density $\pi_{it}$ may be feasible. \citet{poczos2013distribution} propose a nonparametric regression estimator that, given a nonparametric estimate $\widehat{\pi}_{it}$, can be used to consistently estimate $\phi_i$ and average partial effects. However, their estimator suffers from a slow convergence rate in general. An alternative is to assume that $\phi_i$ in (\ref{eq_novel}) is linear, or more generally polynomial, in beliefs, as in the literature on functional regression (see, e.g., \citealp{ramsay1991some}, and \citealp{yao2010functional}). Under linearity in beliefs, there exists a function $\varphi_i$ such that
\begin{align}\phi_i(x,\pi,z)&=\int \varphi_i(x,\widetilde{x},z)\pi(\widetilde{x})d\widetilde{x},\label{eq_linear_in_pi}\end{align}
and one can estimate $\varphi_i$ using functional regression estimators based on principal components analysis or Tikhonov regularization (\citealp{hall2007methodology}). However, all these methods require large samples and the availability of rich information about $\pi_{it}$. 

When subjective data are too coarse, the information in the expectations responses $m_{it}$ may not be sufficient to point-identify $\pi_{it}$ nonparametrically. One possibility is to impose parametric assumptions, as we do in our application. An alternative approach is to follow a partial identification strategy. To illustrate this approach, let us omit the reference to $x$ and $z$ for conciseness. The conditional mean ${\phi}_i(\pi_{it})=\mathbb{E}[y_{it}\,|\, \pi_{it}]$ is bounded as follows:
$$\underset{=B^L_i(m_{it};{\phi}_i)}{\underbrace{\underset{\pi\in \Pi(m_{it})}{\limfunc{inf}}\, {\phi}_i(\pi)}}\leq \mathbb{E}[y_{it}\,|\, \pi_{it}]\leq  \underset{=B^U_i(m_{it};{\phi}_i)}{\underbrace{\underset{\pi\in \Pi(m_{it})}{\limfunc{sup}}\, {\phi}_i(\pi)}},$$
where $\Pi(m_{it})=\left\{\pi\,:\, m(\pi)=m_{it}\right\}$. These bounds imply the following moment inequalities on ${\phi}_i$:
$$\mathbb{E}\left[y_{it}-B^L_i(m_{it};{\phi}_i)\,|\, m_{it}\right]\geq 0, \quad \mathbb{E}\left[y_{it}-B^U_i(m_{it};{\phi}_i)\,|\, m_{it}\right]\leq 0.$$ We do not pursue such a strategy here, and leave it as an avenue for future work.

%%%%%%%%%%%%%%%%%%%%%%%%%%%%%%%%%%%%%%%%%%%%%%%%%%%%%%%%%%%%%%%%%%%%%%%%%%%%%%%%%%%%%%%%%%%%%%%%
%%%%%%%%%%%%%%%%%%%%%%%%%%%%%%%%%%%%%%%%%%%%%%%%%%%%%%%%%%%%%%%%%%%%%%%%%%%%%%%%%%%%%%%%%%%%%%%%
%%%%%%%%%%%%%%%%%%%%%%%%%%%%%%%%%%%%%%%%%%%%%%%%%%%%%%%%%%%%%%%%%%%%%%%%%%%%%%%%%%%%%%%%%%%%%%%%
\setcounter{figure}{0}\renewcommand{\thefigure}{\thesection\arabic{figure}}
\setcounter{table}{0}\renewcommand{\thetable}{\thesection\arabic{table}}
\section{Sensitivity analysis }
\label{app_sensiv}

{We start by noting that, using a change in variables,
	\begin{align*}
	\mathbb{E}_{\theta_{it}}\left[  \log\left(\pi\left(x_{i,t+1}^{(\delta)};{\theta}\right)\right)\right]&=\int  \log\left(\pi\left(\delta(x);{\theta}\right)\right)\pi(x;\theta_{it})dx\\
	&=\int  \log\left(\pi\left(x;{\theta}\right)\right)\pi^{(\delta,full)}(x;\theta_{it})dx,		
\end{align*}
where $\pi^{(\delta,full)}(\cdot;\theta_{it})$ is the density of the transformed random variable $\delta(x_{i,t+1})$ for $x_{i,t+1}\sim \pi(\cdot;\theta_{it})$. Maximizing this quantity with respect to $\theta$ is equivalent to minimizing the KL divergence between $\pi(\cdot;\theta)$ and $\pi^{(\delta,full)}(\cdot;\theta_{it})$.

We next note using a similar argument that
\begin{align*}
	&\mathbb{E}_{\theta_{it}}\left[ \frac{1}{1+\xi} \log\left(\pi\left(x_{i,t+1}^{(\delta)};{\theta}\right)\right)+\frac{\xi}{1+\xi} \log\left(\pi\left(x_{i,t+1};{\theta}\right)\right)\right]\\
	&=\int 	\left( \frac{1}{1+\xi} \log\left(\pi\left(\delta(x);{\theta}\right)\right)+\frac{\xi}{1+\xi} \log\left(\pi\left(x;{\theta}\right)\right)\right)\pi(x;\theta_{it})dx\\
	&=\int 	\left( \frac{1}{1+\xi} \log\left(\pi\left(\delta(x);{\theta}\right)\right)\right)\pi(x;\theta_{it})dx+\int 	\left( \frac{\xi}{1+\xi} \log\left(\pi\left(x;{\theta}\right)\right)\right)\pi(x;\theta_{it})dx\\
	&=\int 	\left( \frac{1}{1+\xi} \log\left(\pi\left(x;{\theta}\right)\right)\right)\pi^{(\delta,full)}(x;\theta_{it})dx+\int 	\left( \frac{\xi}{1+\xi} \log\left(\pi\left(x;{\theta}\right)\right)\right)\pi(x;\theta_{it})dx\\
	&=\int  \log\left(\pi\left(x;{\theta}\right)\right)	\left( \frac{1}{1+\xi} \pi^{(\delta,full)}(x;\theta_{it})+ \frac{\xi}{1+\xi} \pi(x;\theta_{it})\right)dx.
\end{align*}
Maximizing this quantity with respect to $\theta$ is equivalent to minimizing the KL divergence between $\pi(\cdot;\theta)$ and $\frac{1}{1+\xi}\pi^{(\delta,full)}(\cdot;\theta_{it})+\frac{\xi}{1+\xi}\pi(\cdot;\theta_{it})$.

%%%%%%%%%%%%%%%%%%%%%%%%%%%%%%%%%%%%%%%%%%%%%%%%%%%%%%%%%%%%%%%%%%%%%%%%%%%%%%%%%%%%%%%%%%%%%%%%
%%%%%%%%%%%%%%%%%%%%%%%%%%%%%%%%%%%%%%%%%%%%%%%%%%%%%%%%%%%%%%%%%%%%%%%%%%%%%%%%%%%%%%%%%%%%%%%%
%%%%%%%%%%%%%%%%%%%%%%%%%%%%%%%%%%%%%%%%%%%%%%%%%%%%%%%%%%%%%%%%%%%%%%%%%%%%%%%%%%%%%%%%%%%%%%%%
\setcounter{figure}{0}\renewcommand{\thefigure}{\thesection\arabic{figure}}
\setcounter{table}{0}\renewcommand{\thetable}{\thesection\arabic{table}}
\section{Belief data }
\label{app_shiw}

In this section of the appendix we describe the income belief questions in the SHIW, and explain how we estimate the parameters of the belief densities.

%%%%%%%%%%%%%%%%%%%%%%%%%%%%%%%%%%%%%%%%%%%%%%%%%%%%%%%%%%%%%%%%%%%%%%%%%%%%%%%%%%%%%%%%%%%%%%%%
\subsection{Expectations questions in the SHIW}
The SHIW includes questions about income expectations in waves 1989--1991 and 1995--1998; however the expectations questions differ in the two sets of waves. 

The 1989--1991 waves include a question about expected income growth:

\noindent \textit{Thinking now of your total income from work or retirement and its evolution [for the next 12 months]\ldots
	Which categories would you exclude? Suppose you have 100 points to distribute among the remaining categories, how many would you give to each?} 

The possible categories are more than 25\%, between 20\% and 25\%, between 15\% and 20\%, between 13\% and 15\%, between 10\% and 13\%, between 8\% and 10\%, between 7\% and 8\%, between 6\% and 7\%, between 5\% and 6\%, between 3\% and 5\%, between 0\% and 3\%, or less than 0\%, and in that case, by how much. In Table \ref{Table_desc_expq_89} we report descriptive statistics corresponding to this question. 

%In these two waves, the survey also includes a similar question about inflation expectations.

The 1995--1998 waves include three questions about expected income level:

\noindent Minimum amount expected to earn: \textit{Assuming that you remain in or find employment in the next 12 months, can you say what is the minimum overall annual amount you expect to earn, net of taxes, including overtime, bonuses, fringe benefits, etc?}

\noindent Maximum amount expected to earn: \textit{Assuming again that you remain in or find employment in the next 12 months, can you say what is the maximum overall annual amount you expect to earn, net of taxes, including overtime, bonuses, fringe benefits, etc?}

\noindent Probability of earning less than half: \textit{What is the probability that you will earn less than X (the amount
	obtained for (maximum + minimum)/2)? If you had to give a score of between 0 and 100 to the chances of earning less than X, what would it be? (``0'' if certain of earning more than X, ``100'' if certain of earning less than X)}. 

In Table \ref{Table_desc_expq_95} we report descriptive statistics corresponding to these questions. In these two waves, the survey also includes a question about the probability of being employed next year that we use in a robustness check specific to those waves.

%%%%%%%%%%%%%%%%%%%%%%%%%%%%%%%%%%%%%%%%%%%%%%%%%%%%%%%%%%%%%%%%%%%%%%%%%%%%%%%%%%%%%%%%%%%%%%%%%%%%
\subsection{Estimation of income beliefs\label{app_measbel}}

We assume log income beliefs are normally distributed, with mean $\mu_{it}$ and variance $\sigma^2_{it}$, and use the expectations questions to estimate these two parameters for each individual and wave. In this subsection, we omit the reference to $i$ and $t$ for ease of notation.

\paragraph{First two waves.}

For the 1989--1991 waves, we use the survey expectations questions to estimate the mean and variance of the beliefs of log income growth, which are normally distributed under our assumptions, with mean $\mu_g=\mu-x$ (where $x$ is the current log income), and variance $\sigma^2_g=\sigma^2$. Given estimates of $\mu_g$ and $\sigma_g^2$, we then recover estimates of $\mu$ and $\sigma^2$. 

Let $\widehat{p}_j$ denote the fraction of points the respondent assigns to bin $j$ (out of 100 points), for $j=1,..., J$, where $J=12$. For each bin, one could interpret $\widehat{p}_j$ as the probability that a ${\cal{N}}(\mu_g,\sigma^2_g)$ draw takes values within the interval corresponding to that bin. Under this interpretation, one could estimate $\mu_{g}$ and $\sigma_{g}$ using maximum likelihood or minimum distance given the fractions $\widehat{p}_j$. However, this approach does not work well in practice since many of the $\widehat{p}_j$'s are exactly 0 or 1.

Instead of assuming that respondents report exact, normal-based probabilities, we follow \citet{arellano2021income} and assume that, when answering the survey expectations questions, individuals sample $M$ draws from their underlying ${\cal{N}}(\mu_{g},\sigma^2_{g})$ distribution, and use those draws to provide their answers $\widehat{p}_j$. Given that in the survey, individuals are asked to distribute 100 points among the 12 bins, we take $M=100$ as our baseline. Hence, the answers $\widehat{p}_j$ are obtained from $M=100$ trials from a multinomial distribution with true probabilities $p_j$. 

To estimate the $p_j$, we assume an uninformative (Jeffreys) prior on $(p_1,...,p_J)$. It then follows that the posterior means of the $p_j$ are
\begin{eqnarray}
\widetilde{p}_j &=& \frac{\widehat{p}_j+ \frac{1}{2M}}{1+\frac{J}{2M}}, \quad j=1,..., J.\label{eq_ptilde}
\end{eqnarray}
The estimates $\widetilde{p}_j$ are regularized counterparts to the $\widehat{p}_j$. An advantage is that they take values in the open interval $(0,1)$, which allows one to implement minimum distance or maximum likelihood estimation strategies based on them. We have performed robustness checks using other regularization devices, including different $M$ values, and found only minor impacts on the results (see Section \ref{App_consumption} of this appendix).  

%In Appendix \ref{app_merror}, we also use this model to assess measurement error. \\

Given the regularized responses $\widetilde{p}_j$ in (\ref{eq_ptilde}), we then construct the cumulative probabilities, $\widetilde{c}_j= \sum_{k=1}^{j}{\widetilde{p}_k}$, and estimate $\mu_g$ and $\sigma_g$ based on the following system of linear equations:
\begin{eqnarray}
\Phi^{-1}(\widetilde{c}_j)\cdot\sigma_{g}+\mu_{g} &=& v_{j}, \quad  j=1,..., J-1, \label{eq_meas}
\end{eqnarray}
where $v_{j}$ correspond to the right endpoint of the $j$-th bin, and $\Phi$ denotes the standard normal cdf. Since the first and last bins in the survey question are unbounded, we add bounds to those bins (-10\% for the bin below 0\%, and 35\% for the bin above 25\%).\footnote{We verified that our estimates of log consumption regression coefficients remain similar when using different bounds, and  when excluding observations that assign all points to the first or last bin.} This amounts to working with 14 bins in total. We then estimate $\mu_g$ and $\sigma_g$ using OLS based on a subset of the equalities in (\ref{eq_meas}). Specifically, we use all the bins $j$ for which $\widehat{p}_j>0$, and use in addition one unbounded bin to the left and one unbounded bin to the right.  The reason for only using a subset of the restrictions in (\ref{eq_meas}) is to reduce the influence of the regularization for bins with $\widehat{p}_j=0$.\footnote{We found that using all bins with $\widehat{p}_j=0$ tended to artificially increase the variance of estimated beliefs.}

As an example, consider an individual who assigns 60 points to the 5--6\% bin, and 40 points to the 6--7\% bin. In this case we use the intervals (0.05,0.06) and (0.06,0.07), both of which have positive $\widehat{p}_j$, and we add the intervals ($-\infty$,0.05) and (0.07,$+\infty$), to the left and to the right, respectively. We then compute the sums of the $\widetilde{p}_j$ in (\ref{eq_ptilde}), in each of these four intervals. Lastly, we use these cumulative probabilities to estimate $\mu_g$ and $\sigma_g$ by OLS. Since, in the fourth interval, the cumulative probability is equal to 1, in this example we only rely on three independent linear restrictions to estimate $\mu_g$ and $\sigma_g$.

\paragraph{Last two waves.}

For the 1995--1998 waves, we use the survey expectations questions to estimate the mean $\mu$ and variance $\sigma^2$ of log income beliefs directly (since in these waves the questions are about income levels, not income growth). We interpret the answers as probabilities assigned to two bins (between the minimum and the mid-point, and between the mid-point and the maximum). As in the 1989--1991 waves, we add two additional bins, one below the reported minimum and another one above the reported maximum, which amounts to be working with $4$ bins in total. These additional bins have a positive but low probability $\widetilde{p}_j=\frac{1}{2M+4}$, which might reflect that respondents interpret the minimum and maximum questions as asking them to report quantiles of their distributions (see \citealp{Delavande2011measuring}). In the 1995--1998 waves, the locations and widths of the bins come from individuals' responses, providing more information to capture beliefs, in particular beliefs with very small variance. For example, when the reported minimum and maximum coincide, the implied estimate of $\sigma$ is equal to zero. 

\paragraph{Descriptives and predictive power.}

In Table \ref{Table_desc_bel} we provide descriptive statistics about the beliefs that we estimate and the main variables in the consumption equation. 

In Table \ref{Table_income} we assess the predictive power of these beliefs: we regress $\log(w_{i,t+1})$ in columns (1) to (4), and $ \log(w_{i,t+1})- \log(w_{it})$ in columns (5) to (8), as functions of the estimated mean beliefs $\mu_{it}$ and other controls. In this table, we use log individual income as our dependent variable. The estimates suggest that individual beliefs predict future income, even conditional on current income.

%%%%%%%%%%%%%%%%%%%%%%%%%%%%%%%%%%%%%%%%%%%%%%%%%%%%%%%%%%%%%%%%%%%%%%%%%%%%%%%%%%%%%%%%%%%%%%%%%%%%%%%%%%%%%%%%%%%%%%%%%%%%
%%%%%%%%%%%%%%%%%%%%%%%%%%%%%%%%%%%%%%%%%%%%%%%%%%%%%%%%%%%%%%%%%%%%%%%%%%%%%%%%%%%%%%%%%%%%%%%%%%%%%%%%%%%%%%%%%%%%%%%%%%%%
%%%%%%%%%%%%%%%%%%%%%%%%%%%%%%%%%%%%%%%%%%%%%%%%%%%%%%%%%%%%%%%%%%%%%%%%%%%%%%%%%%%%%%%%%%%%%%%%%%%%%%%%%%%%%%%%%%%%%%%%%%%%
\setcounter{figure}{0}\renewcommand{\thefigure}{\thesection\arabic{figure}}
\setcounter{table}{0}\renewcommand{\thetable}{\thesection\arabic{table}}
\section{Robustness checks for Table \ref{Table_cons}}
\label{App_consumption}

In this section of the appendix we provide several robustness checks for the estimation of the consumption regression (see Table \ref{Table_cons}), focusing on the specification with mean beliefs interacted with log current income. 

Our main estimates are obtained using a particular approach to construct the mean and variance of log income beliefs. We first probe the robustness of our estimates to different assumptions about the distribution of beliefs, and to different construction methods for the mean and variance of beliefs. In columns (1) and (2) in Table \ref{Table_cons_rob} we show the estimates are robust to relying on different distributional assumptions for beliefs: a discrete distribution for waves 1989--1991 (as in \citealp{pistaferri2001superior}), and a triangular distribution for waves 1995-1998 (as in \citealp{kaufmann2009disentangling}). In columns (3) to (6) we show that estimates are robust to the value of $M$ used for estimation (see (\ref{eq_ptilde}), where the baseline corresponds to $M=100$). 

While consumption and income correspond to households, the income beliefs questions correspond to individual income. In the baseline results we only use the beliefs of household heads (and adjust our counterfactual calculations). In columns (7) and (8) in Table \ref{Table_cons_rob} we also control for the spouse's beliefs about their own income, when available.\footnote{When spousal beliefs are not available, we set the variable to zero and add binary indicators for missingness, distinguishing between spouses that are homemakers, employed, or other labor status. Note that only 32\% (resp., 17\%) of the 768 households are households where data on spousal beliefs are available in at least one wave (resp., both waves).} Estimates remain virtually unchanged, and spousal beliefs don't appear to play a major role in household consumption for this sample.

The estimates in Table \ref{Table_cons} are obtained by pooling two sets of waves, 1989--1991 and 1995--1998. Economic conditions, as well as the belief elicitation strategies, differ between these two periods. In Table \ref{Table_cons_by_wave} we estimate the consumption regression, separately for waves 1989--1991 and 1995--1998.\footnote{In each pair of waves, we also control for other expectations questions available: inflation expectations in 1989--1991, and expectations about future employment in 1995--1998.} The point estimates are different in the two samples, with a larger effect of beliefs in the 1995-1998 waves. However, in both cases beliefs play a significant role in household consumption.\footnote{Using the 1995--1998 waves, we also estimated the consumption regression including unemployed household heads in the sample and controlling for beliefs about future employment, and found similar results. In the 1989--1991 waves expectations questions were not asked to the unemployed.} 

Lastly, although assets are important determinants of consumption, their measurement in the SHIW is imperfect. Indeed, respondents are asked about end-of-year assets, while the state variable in the consumption function is beginning-of-period assets. We assess the robustness of our results in this dimension in two ways. First, following \citet{stoltenberg2022consumption} we construct an alternative measure of assets by subtracting yearly savings from end-of-year assets. A concern with this specification in our context is that savings in the SHIW are constructed by netting out consumption expenditures from total income, so measurement error in consumption might bias our regression coefficients. Given this, we also report the results of a second specification where we do not include any control for assets. In addition to these checks, we also report results based on an IV strategy that relies on first-period assets and income as instruments for current assets. In Table \ref{Table_cons_assets} we present estimates obtained under different approaches for dealing with assets. Estimates of the coefficients of current income and income beliefs are quite similar across specifications, although we see some quantitative differences, especially in the case of the IV specification in columns (3) and (4).  

%%%%%%%%%%%%%%%%%%%%%%%%%%%%%%%%%%%%%%%%%%%%%%%%%%%%%%%%%%%%%%%%%%%%%%%%%%%%%%%%%%%%%%%%%%%%%%%%%%%%%%%%%%%%%%%%%
%%%%%%%%%%%%%%%%%%%%%%%%%%%%%%%%%%%%%%%%%%%%%%%%%%%%%%%%%%%%%%%%%%%%%%%%%%%%%%%%%%%%%%%%%%%%%%%%%%%%%%%%%%%%%%%%%
%%%%%%%%%%%%%%%%%%%%%%%%%%%%%%%%%%%%%%%%%%%%%%%%%%%%%%%%%%%%%%%%%%%%%%%%%%%%%%%%%%%%%%%%%%%%%%%%%%%%%%%%%%%%%%%%%
\section{Estimation in the finite-type model\label{App_types}}

In this section of the appendix we describe the main specification considered in our application, and our strategy to estimate the parameters and compute standard errors. 
	
	%%%%%%%%%%%%%%%%%%%%%%%%%%%%%%%%%%%%%%%%%%%%%%
	\paragraph{Specification.} We assume that log family income $x_{it}$ and beliefs $\pi_{it}$ are strictly exogenous, there is a homogeneous feedback process for assets $z_{it}$, and a finite number of unobserved types $\alpha_i\in\{1,...,K\}$. We further assume that, given a type, log consumption $y_{it}$ is normally distributed with a mean that depends linearly on $x_{it}$, $\pi_{it}$ (mean and variance), and $z_{it}$ (which includes the controls in column (5) of Table \ref{Table_cons}), and a constant variance, i.i.d. over time and across households. In turn, types are determined according to a multinomial logit specification where type probabilities depend on a linear index in an intercept, average income, and average mean beliefs across the two periods. 
	
	%%%%%%%%%%%%%%%%%%%%%%%%%%%%%%%%%%%%%%%%%%%%%%
	\paragraph{Likelihood.} 
	Let $\tau_{ki}$ denote the probability that individual $i$ is of type $k=1,\ldots K$. Then, the log-likelihood (weighted using survey weights $w_i$) is given by
	\begin{eqnarray}
		&&\sum_{i=1}^n w_i \log \left(\sum_{k=1}^K\prod_{t=1}^T \tau_{ki}  f_k(y_{it}| x_{it},\pi_{it},z_{it})\prod_{t=2}^Tf(z_{it}| z_{i,t-1},x_{i,t-1},y_{i,t-1})\right)\nonumber\\
		&=&\sum_{i=1}^n w_i \log\left(\sum_{k=1}^K\prod_{t=1}^T \tau_{ki}  f_k(y_{it}| x_{it},\pi_{it},z_{it})\right) +\sum_{i=1}^n\sum_{t=2}^T w_i \log \left(f(z_{it}| z_{i,t-1},x_{i,t-1},y_{i,t-1})\right), \nonumber
	\end{eqnarray}
	where
	\begin{eqnarray}
		f_k(y_{it}|x_{it}, \pi_{it}, z_{it}) &=& 
		f(y_{it}| x_{it}, \pi_{it},z_{it}; \beta_k, \mu) =
		\frac{1}{\mu} \phi\!\left( \frac{y_{it} - [1,x_{it},\pi_{it},z_{it}]' \beta_k}{\mu} \right), \nonumber\\
		\tau_{ki} = \tau_k(\overline{x}_{i},\overline{\pi}_i;\eta) &=& \frac{\exp([1,\overline{x}_i,\overline{\pi}_i]'\eta_k)}{1+\sum_{l=1}^{K-1}{\exp([1,\overline{x}_i,\overline{\pi}_i]' \eta_l)}}, \nonumber
	\end{eqnarray}
	with the normalization $\eta_K=0$. We have experimented with various choices of conditioning variables in $\tau_{ki}$, including initial assets $z_{i1}$, but in the main text and appendix we report estimates based on average log income and mean beliefs and an intercept.  In our application we allow for the $\beta$ coefficients of mean beliefs, log family income, and the intercept to vary with the latent type. 
	
	\paragraph{EM algorithm.}
	We maximize the likelihood by solving
	\begin{eqnarray}
		\max_{\beta_1,\dots,\beta_K,\ \mu,\ \eta_1,\dots,\eta_{K-1}}
		\sum_{i=1}^N w_i  \log \left(\sum_{k=1}^K{\tau_k(\overline{x}_i,\overline{\pi}_i; \eta)
			\prod_{t=1}^{2} f(y_{it}|x_{it}, \pi_{it}, z_{it}; \beta_k, \mu)}\right) \nonumber
	\end{eqnarray}
	using the EM algorithm. In the E-step, we calculate the posterior type probabilities given the parameters estimated in the previous iteration,
	\begin{eqnarray}
		\widehat{p}_{ki}^{(s)} &=& \frac{\tau_k(\overline{x}_i, \overline{\pi}_i; \widehat\eta^{(s-1)}) 
			\prod_{t=1}^T{f(y_{it}|x_{it},\pi_{it},z_{it}; \widehat\beta_k^{(s-1)}, \widehat\mu^{(s-1)})}}{
			\sum_{l=1}^K{\pi_l(\overline{x}_i, \overline{\pi}_i; \widehat\eta^{(s-1)}) 
				\prod_{t=1}^T{f(y_{it}|x_{it},\pi_{it},z_{it}; \widehat\beta_l^{(s-1)}, \widehat\mu^{(s-1)})}}}. \nonumber
	\end{eqnarray}
	\noindent In the M-step, we update the parameters taking the posterior probabilities as given. 
	\begin{eqnarray}
		&& \arg\max_{\{\beta_k,\}_{k=1}^K,\mu} \sum_iw_i\sum_k\widehat{p}_{ki}^{(s)}\sum_t  	
		\log(f(y_{it}|x_{it},\pi_{it},z_{it}; \beta_k, \mu))\nonumber\\
		&& \arg\max_{\{\eta_k\}_{k=1}^{K-1}} \sum_iw_i\sum_k   \widehat{p}_{ki}^{(s)} \log(\tau_k(\overline{x}_i,\overline{\pi}_i; \eta)) . \nonumber 
	\end{eqnarray}
	These equations correspond to estimating a linear model and a multinomial logit model in an extended data set, where each observation appears $K$ times and is weighted additionally by the posterior probabilities estimated in the E-step (and the survey weights). 
	
	%%%%%%%%%%%%%%%%%%%%%%%%%%%%%%%%%%%%%%%%%%%%%%
	\paragraph{Implementation details.}
	We run the algorithm for 2,000 iterations and use a numerical tolerance of $1e^{-7}$. To account for the presence of local maxima, we draw 100 starting values, randomly chosen in a neighborhood around the pooled OLS estimate and another 100 values randomly chosen in a neighborhood of the estimate based on first differences. In each case, we allow for different starting values for the type-specific coefficients in $\beta_k$, while fixing the homogeneous coefficients in $\beta$ and $\mu$ to their pooled OLS or first-differenced values. We set type coefficients $\eta$ equal to zero for the starting values, implying homogeneous type probabilities. The search leads to stable results for $K=2$ and $K=3$, but seems more unstable for $K=4$, which motivates our focus on two and three types.
	
	%%%%%%%%%%%%%%%%%%%%%%%%%%%%%%%%%%%%%%%%%%%%%%
	\paragraph{Standard errors.}
	Standard errors are based on the estimated asymptotic variance, calculated as the inverse Hessian of the partial likelihood (\citealp{cox1975partial}). 
	
	%%%%%%%%%%%%%%%%%%%%%%%%%%%%%%%%%%%%%%%%%%%%%%
	\paragraph{Decomposition.}
	When studying counterfactual effects in the cross-sectional sample, we fix most of the parameters to those estimated using the panel sample. However, to account for composition differences in unobserved types between the panel and cross-sectional samples, we re-estimate the type coefficients $\zeta$ in the (larger) cross-sectional sample. That is, we solve,
	\begin{eqnarray}
		\max_{\zeta_1,\dots,\zeta_{K-1}}\sum_{i=1}^N \sum_{t=1}^{T_i} w_i  \log\left(
		\sum_{k=1}^K {\tau}_k(x_{it}, \pi_{it};\zeta)  f(y_{it}|x_{it}, \pi_{it}, z_{it}; \widehat\beta_k, \widehat\mu)
		\right).\nonumber
	\end{eqnarray}
	This a concave problem which we solve using the EM algorithm. For estimating the standard errors of these quantities, we use the bootstrap. To avoid re-estimating the finite-mixture model in each replication, we instead draw parameter values from their estimated (Gaussian) asymptotic distribution. Then, we bootstrap the remaining steps (i.e., the re-estimation of type probabilities in the cross section, and the calculation of average effects) by re-sampling at the household level.}

%\begin{eqnarray}
%	TAPE(\delta) &=& \frac{1}{(\sum_{i,t} w_{it})}\sum_{i=1}^N \sum_{t=1}^{T_i}\sum_{k=1}^K w_{it} \cdot {\tau}_k(x_{it},\pi_{it};\zeta)\left[\phi_k(x_{it}^{(\delta)}, \pi_{it}^{(\delta)}, z_{it}) - \phi_k(x_{it}, \pi_{it}, z_{it}) \right] \nonumber
%\end{eqnarray}

%%%%%%%%%%%%%%%%%%%%%%%%%%%%%%%%%%%%%%%%%%%%%%%%%%%%%%%%%%%%%%%%%%%%%%%%%%%%%%%%%%%%%%%%%%%%%%%%
%%%%%%%%%%%%%%%%%%%%%%%%%%%%%%%%%%%%%%%%%%%%%%%%%%%%%%%%%%%%%%%%%%%%%%%%%%%%%%%%%%%%%%%%%%%%%%%%
%%%%%%%%%%%%%%%%%%%%%%%%%%%%%%%%%%%%%%%%%%%%%%%%%%%%%%%%%%%%%%%%%%%%%%%%%%%%%%%%%%%%%%%%%%%%%%%%
\setcounter{figure}{0}\renewcommand{\thefigure}{\thesection\arabic{figure}}
\setcounter{table}{0}\renewcommand{\thetable}{\thesection\arabic{table}}
\section{Tax counterfactuals: details about estimation\label{app_counter}}

In this section of the appendix we detail the calculations of tax counterfactuals and present additional empirical estimates.

%%%%%%%%%%%%%%%%%%%%%%%%%%%%%%%%%%%%%%%%%%%%%%%%%%%%%%%%%%%%%%%%%%%%%%%%%%%%%%%%%%%%%%%%%%%%%%%%
\subsection{Tax schedule}

We assume the tax schedule takes the parametric form $T(\widetilde{w}_r)=\widetilde{w}_r-\lambda \widetilde{w}_r^{1-\tau}$, where $\widetilde{w}_r$ denotes gross income in multiples of its population average, as in \citet{benabou2002tax}. This parametric form can be re-written as a similar function that depends on gross income $\widetilde{w}$, with the same parameter $\tau$ but a different parameter $\widetilde\lambda$.\footnote{Specifically, $\widetilde\lambda=\lambda K^\tau$, for $K$ the average gross income in the population.} For the baseline level of the tax, we rely on the estimates obtained by \citet{holter2019tax} for Italy, averaged over family characteristics in our sample: $\lambda_0=0.94$ and $\tau_0=0.196$. 
%
%As there is a one-to-one mapping between $\lambda$ and $\widetilde\lambda$, sometimes we refer to one of them and sometimes to the other. 

Let $\lambda_1$ and $\tau_1$ denote the parameters defining the tax schedule under a counterfactual scenario. We assume the tax schedule applies to gross family income, and that each individual pays taxes proportionally to their contribution in the family, $r_{it}$, a proportion we assume does not change in counterfactual scenarios. {Hence, the baseline and counterfactual tax parameters determine the function $\delta(\cdot)$. Let $x^{(\delta)}_{it}$ denote log family income and ($\mu^{(\delta)}_{it}$,$\sigma^{2(\delta)}_{it}$) denote the parameters of income beliefs under a counterfactual scenario. Let ($x_{it}$, $\mu_{it}$, $\sigma^2_{it}$)} denote their baseline values, observed in sample. In this case, {equation (\ref{eq_delta_change3}) implies
	\begin{eqnarray}
		\delta(x) &=& \big(x-log(\widetilde{\lambda}_0)\big)\bigg(\frac{1-\tau_1}{1-\tau_0}\bigg) + log(\widetilde\lambda_1), \nonumber\\
		x^{(\delta)}_{it} &=& \delta(x_{it}), \nonumber\\
		\mu^{(\delta)}_{it} &=& \mu_{it} \frac{\frac{1-\tau_1}{1-\tau_0}+\xi}{1+\xi} + \frac{1}{1+\xi}\bigg[log(\widetilde\lambda_1)-log(\widetilde\lambda_0)\frac{1-\tau_1}{1-\tau_0}+log(r_{it})\frac{\tau_1-\tau_0}{1-\tau_0}\bigg], \nonumber\\
		\sigma^{2(\delta)}_{it} &=& \sigma^2_{it}\frac{\frac{(1-\tau_1)^2}{(1-\tau_0)^2}+\xi}{1+\xi} + \xi(\mu^{(\delta)}_{it}-\mu_{it})^2.\nonumber
\end{eqnarray}}%
Given a counterfactual tax schedule ($\lambda_1$, $\tau_1$), we use these values to compute average partial effects. {Our baseline results correspond to the case of full pass-through, that is, $\xi=0$.}

We consider three counterfactual scenarios. In the \textit{transitory tax}  increase and \textit{permanent tax} increase counterfactuals, we set $\lambda_1=\lambda_0-0.1$ and $\tau_1=\tau_0$. In the \textit{regressivity} counterfactual, we set $\tau_1=0.142$, the progressivity parameter of the tax system in France according to \citet{holter2019tax}, and set $\lambda_1$ such that the tax change is revenue neutral.\footnote{Assuming that family gross income is log-normally distributed with parameters $\mu_{\widetilde{w}}$ and $\sigma^2_{\widetilde{w}}$, a change in the parameters of the tax system is revenue neutral if 
	\begin{eqnarray}
		\log(\widetilde\lambda_1) - \log(\widetilde\lambda_0) = \frac{1}{2}\sigma^2_{\widetilde{w}}\bigg[(1-\tau_0)^2-(1-\tau_1)^2\bigg]+\mu_{\widetilde{w}}(\tau_1-\tau_0). \nonumber
	\end{eqnarray}
	Furthermore, $\mu_{\widetilde{w}}=(\mu_x-log(\widetilde\lambda_0))/(1-\tau_0)$ and $\sigma_{\widetilde{w}}=\sigma_x/(1-\tau_0)$, where $\mu_x$ and $\sigma_x^2$ are the mean and variance of the log of disposable family income, which we estimate from the SHIW.}

%%%%%%%%%%%%%%%%%%%%%%%%%%%%%%%%%%%%%%%%%%%%%%%%%%%%%%%%%%%%%%%%%%%%%%%%%%%%%%%%%%%%%%%%%%%%%%%%
\subsection{Double Lasso estimation\label{App_dlasso}} 

In this subsection we describe how we estimate the consumption function using the double Lasso method introduced by \citet{belloni2014inference}. Consider the equation,
\begin{eqnarray}
	y_{it} &=& a'\Psi(s_{it}) + \beta_k k_{it} + \alpha_i  + \varepsilon_{it}, \label{eq_lasso_reparam}
\end{eqnarray}
where $\Psi(s_{it})$ includes polynomial functions of the main covariates (age, log income, log assets, and the income beliefs' means and variances), and $k_{it}$ includes the other demographic controls. Under this specification, an average partial effect corresponding to a counterfactual of interest is given by 
\begin{eqnarray}
	a'\bigg(\frac{1}{nT}\sum_{i,t}{(\Psi(\widetilde{s}_{it})-\Psi(s_{it}))}\bigg)  \nonumber
\end{eqnarray}
where $s_{it}$ are the main covariates under the baseline, and $\widetilde{s}_{it}$ are the main covariates under the counterfactual. 

Letting
$$v=\frac{1}{nT}\sum_{i,t}{(\Psi(\widetilde{s}_{it})-\Psi(s_{it}))},$$
we first reparameterize the polynomials so that the average partial effect of interest coincides with the coefficient of the first regressor. To that end, we construct an invertible matrix $A$ whose first column is equal to $v$.\footnote{For example, we set $A=[v \quad \iota_2 ..\quad \iota_L]$, where $\iota_{\ell}$ are the canonical vectors in $\mathbb{R}^L$ and $L=\dim \Psi$, provided such a matrix $A$ is invertible.} Then, we rewrite (\ref{eq_lasso_reparam}) using the reparameterized polynomials $\widetilde\Psi(s_{it})=A^{-1}\Psi(s_{it})$, and obtain
\begin{eqnarray}
	y_{it} &=& \left(A'a\right)'\widetilde\Psi(s_{it}) + \beta_k k_{it}+ \alpha_i  + \varepsilon_{it}. \label{eq_lasso_reparam2}
\end{eqnarray}
Note that the coefficient of the first covariate in (\ref{eq_lasso_reparam2}) is equal to $a'v$, which is the average partial effect of interest.

To estimate $a'v$, we apply the double Lasso estimator to (\ref{eq_lasso_reparam2}). To account for household fixed effects, we take first differences. We always include (i.e., we do not penalize) the following regressors: the first order polynomials (age, log income, log assets, and the beliefs' means and variances), as well as the variables in $k_{it}$ (existence of a spouse, marital status, family size, number of children 0-5, 6-13, 14-17 years old in the household, number of children outside the household, number of income earners in the household, and a wave indicator). 

The double Lasso method is implemented in two steps. In a first step, we apply the Lasso to regress the first element in $\widetilde\Psi(s_{it})$ on its second to last elements and $k_{it}$, in first differences. In the second step, we again apply the Lasso to regress $y_{it}$ on the second to last elements of $\widetilde\Psi(s_{it})$ and $k_{it}$, in first differences. In both steps, we choose the penalty parameters by 10-fold cross-validation (\citealp{chetverikov2021cross}). Lastly, we regress $y_{it}$ on the first element in $\widetilde\Psi(s_{it})$ and all the controls selected in the two Lasso steps, again in first differences. We account for estimation uncertainty (in particular, for the fact that $v$ is estimated) by computing bootstrapped standard errors.

%%%%%%%%%%%%%%%%%%%%%%%%%%%%%%%%%%%%%%%%%%%%%%%%%%%%%%%%%%%%%%%%%%%%%%%%%%%%%%%%%%%%%%%%%%%%%%%%%%%%%%%%%%%%%%%%%
\subsection{Empirical estimates}

In Table \ref{tab_counter_types} we report average partial effects based on the finite mixture specification for $K=2$ and $K=3$, respectively, aggregated across types. In Tables \ref{tab_counter_types_K2} and \ref{tab_counter_types_K3} we report APE estimates by types, for $K=2$ and $K=3$, respectively. In panels A and B of Table \ref{tab_counter} we report APE based on OLS estimates of the consumption function and APE based on the double Lasso. We show these in graphical form in Figures \ref{fig_counter_ols} and \ref{fig_counter_lasso}, respectively. Lastly, in panel C of Table \ref{tab_counter} we report APE based on correlated-random effects estimates.

%In unreported results, we also similar patterns for APE based on OLS estimates from different samples, ein Table \ref{Table_cons_by_wave}.

%%%%%%%%%%%%%%%%%%%%%%%%%%%%%%%%%%%%%%%%%%%%%%%%%%%%%%%%%%%%%%%%%%%%%%%%%%%%%%%%%%%%%%%%%%%%%%%%
%%%%%%%%%%%%%%%%%%%%%%%%%%%%%%%%%%%%%%%%%%%%%%%%%%%%%%%%%%%%%%%%%%%%%%%%%%%%%%%%%%%%%%%%%%%%%%%%
%%%%%%%%%%%%%%%%%%%%%%%%%%%%%%%%%%%%%%%%%%%%%%%%%%%%%%%%%%%%%%%%%%%%%%%%%%%%%%%%%%%%%%%%%%%%%%%%
\section{Finite mixture model without beliefs\label{app_no_beliefs}}	
To compare our results with those obtained when beliefs are ignored, we estimate a finite mixture model excluding belief-related variables (from both the main equation and the type equation), for $K=2$ and $K=3$ types. We present the results in Table \ref{Table_nobel} and Figures \ref{fig_nobel} and \ref{fig_nobel_type}. When excluding belief-related variables, there is by definition no dynamic APE.

%%%%%%%%%%%%%%%%%%%%%%%%%%%%%%%%%%%%%%%%%%%%%%%%%%%%%%%%%%%%%%%%%%%%%%%%%%%%%%%%%%%%%%%%%%%%%%%%
%%%%%%%%%%%%%%%%%%%%%%%%%%%%%%%%%%%%%%%%%%%%%%%%%%%%%%%%%%%%%%%%%%%%%%%%%%%%%%%%%%%%%%%%%%%%%%%%
%%%%%%%%%%%%%%%%%%%%%%%%%%%%%%%%%%%%%%%%%%%%%%%%%%%%%%%%%%%%%%%%%%%%%%%%%%%%%%%%%%%%%%%%%%%%%%%%
{
\section{Correlated random effects specification\label{app_CRE}}
As an additional robustness check, we consider a case with continuous instead of discrete unobserved heterogeneity.

Denoting as $X_{it}$ all the covariates in the model (including the belief-related variables, all of them being assumed to be strictly exogenous), we assume that $y_{it}=X_{it}'\beta_i+u_{it},$ where the slopes $\beta_i$ are individual-specific with $\mathbb{E}[\beta_i\,|\, X_{i1},...,X_{iT}]=\Gamma \overline{X}_i,$ and $\overline{X}_i=\frac{1}{T}\sum_{t=1}^TX_{it}$. Then, $\mathbb{E}[y_{it}\,|\,  X_{i1},...,X_{iT}]=\left(\overline{X}_i\otimes X_{it}\right)'\limfunc{vec}(\Gamma).$ Hence, $\gamma=\limfunc{vec}(\Gamma)$ can be estimated consistently by OLS on interactions between elements of $X_{it}$ and $ \overline{X}_i$.

Average partial effects are then obtained easily. For example, a change $X_{it,k}\mapsto X_{it,k}+\delta$ implies the response $\delta\beta_{i,k}$, and the average response for individuals with characteristics $\overline{X}_i$ is $\delta e_k'\Gamma\overline{X}_i,$ where $e_k$ is the $k$-th canonical vector in $\mathbb{R}^{\dim X}$.

In our implementation we allow for the same specification of the $\phi_i$ function as in our main estimates, where only the coefficients for income, mean beliefs and the intercept are allowed to be heterogeneous. However, here we do not assume that heterogeneity is discrete, and instead estimate the parameters $\Gamma$ that describe the conditional mean of the continuous $\beta_i$ coefficients, which we specify as a linear function of household-specific averages of these same three variables (i.e., income, mean beliefs, and an intercept). In Table \ref{Table_CRE} we report the estimated $\Gamma$ coefficients. 
	
}

%%%%%%%%%%%%%%%%%%%%%%%%%%%%%%%%%%%%%%%%%%%%%%%%%%%%%%%%%%%%%%%%%%%%%%%%%%%%%%%%%%%%%%%%%%%%%%%%
%%%%%%%%%%%%%%%%%%%%%%%%%%%%%%%%%%%%%%%%%%%%%%%%%%%%%%%%%%%%%%%%%%%%%%%%%%%%%%%%%%%%%%%%%%%%%%%%
%%%%%%%%%%%%%%%%%%%%%%%%%%%%%%%%%%%%%%%%%%%%%%%%%%%%%%%%%%%%%%%%%%%%%%%%%%%%%%%%%%%%%%%%%%%%%%%%
{
\section{Monte Carlo Simulation \label{app_mcarlo}}

As an additional check, we perform a small Monte Carlo experiment to assess the magnitude of the bias when slope heterogeneity is ignored. We simulate data from a finite mixture model, with parameters set to our baseline estimates, and estimate average partial effects from a specification that allows only for heterogeneous intercepts.

For a given $K$, we repeat $B=10,000$ times the following three steps. First, we simulate the data, that is, we simulate individual types from the multinomial specification and simulate type-specific consumption in period one; then, we modify period-two assets using the budget constraint and the simulated period-one consumption;\footnote{That is, letting $x$ denote the original value and $\widetilde{x}$ denote the modified one, we define period-two assets as $\widetilde{z}_{i2} = z_{i2} - (\widetilde{c}_{i1}-c_{i1})$.} and, finally, we simulate type-specific consumption in period two using the simulated value of period-two assets. Next, we run regressions of log consumption on income and beliefs in first differences, following the specification in column (5) of Table \ref{Table_cons}. Finally, we use these coefficients to estimate average partial effects in the cross-sectional simulated data as explained in Section \ref{app_counter}.

The average partial effects (averaged across simulations) are presented in Table \ref{tab_counter_mcarlo} and Figure \ref{fig_decompo_mcarlo}. Even tough these estimates ignore slope heterogeneity when estimating the consumption function, the average partial effects are similar to the true ones predicted under the data generating process (compare with Table \ref{tab_counter_types} and Figure \ref{fig_decompo_cons_types}).
}

%%%%%%%%%%%%%%%%%%%%%%%%%%%%%%%%%%%%%%%%%%%%%%%%%%%%%%%%%%%%%%%%%%%%%%%%%%%%%%%%%%%%%%%%%%%%%%%%
%%%%%%%%%%%%%%%%%%%%%%%%%%%%%%%%%%%%%%%%%%%%%%%%%%%%%%%%%%%%%%%%%%%%%%%%%%%%%%%%%%%%%%%%%%%%%%%%
%%%%%%%%%%%%%%%%%%%%%%%%%%%%%%%%%%%%%%%%%%%%%%%%%%%%%%%%%%%%%%%%%%%%%%%%%%%%%%%%%%%%%%%%%%%%%%%%
\setcounter{figure}{0}\renewcommand{\thefigure}{\thesection\arabic{figure}}
\setcounter{table}{0}\renewcommand{\thetable}{\thesection\arabic{table}}
\section{Measurement error\label{app_merror}}

In this section of the appendix we describe how we correct for measurement error in the beliefs responses. We focus on the 1989--1991 waves. In those two waves, individuals are asked to distribute 100 balls into 12 bins, corresponding to different intervals of beliefs about log income growth. Assuming log income growth beliefs to be normally distributed, a simple model of the responses is that individuals draw 100 i.i.d. values from their normal belief distributions ${\cal{N}}(\mu_g,\sigma_g^2)$, and put those in the bins. 

However, this model may not provide a good approximation to the response process of individuals when answering the questions in the SHIW. In fact, it is possible that respondents are only able to imagine a smaller number $M<100$ of ``income growth scenarios'', corresponding to events that they expect might happen in the next year, such as a promotion or a demotion, a job change, etc. To provide empirical support for this possibility, we predict, for each respondent, the number of non-empty bins reported by the respondent under the model, for various values of $M$. The estimates in Table \ref{Table_merror_nbins} show that taking $M=100$ implies that, on average, respondents should report 3.6 non-empty bins, while in the data this number is only 1.7. The table also shows that taking smaller values of $M$ provides a better approximation to the distribution of the number of non-empty bins across individuals. 

Given this motivation, we consider a model with $M<100$ draws, where we vary $M$.\footnote{In the model of measurement error that we propose, $M$ is constant across individuals. An alternative model would let $M_i$ vary across individuals. \citet{manski2010rounding} exploit repeated responses by the same individual to infer individual types of measurement error in responses.} Given this model, we propose a correction for measurement error and apply it to revisit the log consumption regression estimates of Table \ref{Table_cons_by_wave}. Our approach is based on a ``small-$\sigma$'' approximation (e.g., \citealp{evdokimov2022simple}). Since, for a given $M$ value, the model of measurement error is parametric, the correction can be implemented using a simple parametric bootstrap method, which we now describe.\footnote{Since the measurement error model is parametric, one could alternatively rely on an exact approach for deconvolving the measurement error, without the need for an approximation. An advantage of the specific approach that we implement here is its simplicity.} {Note that the parametric form of measurement error that we postulate in this exercise is non-classical.}    

We consider the specification of the consumption regression in column (3) of Table \ref{Table_cons_by_wave}, which only accounts for mean beliefs. We draw $S=1,000$ samples where, for each respondent, we draw $M$ observations from a ${\cal{N}}(\widehat{\mu}_g,\widehat{\sigma}_g^2)$, for $\widehat{\mu}_g$ and $\widehat{\sigma}_g^2$ our original estimates of $\mu_g$ and $\sigma_g^2$, respectively. This gives us simulated responses $\widehat{p}_j^{(s)}$, for each sample $s$, from which we estimate $\mu_g$ and $\sigma_g$ and, based on those, the regression coefficients for consumption, exactly in the same way as we did to obtain the estimates in Table \ref{Table_cons_by_wave}.\footnote{In particular, we still consider a likelihood model with 100 trials and an uninformative prior.} Let $\widehat\beta^{(s)}$ denote the estimated coefficients in this last regression. We then construct the bootstrapped bias-corrected counterpart to the original coefficients $\widehat\beta^{\rm OLS}$ as $$\widehat\beta^{\rm BC}=2\widehat\beta^{\rm OLS}-\frac{1}{S}\sum_{s=1}^S\widehat{\beta}^{(s)}.$$
We repeat this exercise for values of $M$ between $1$ and $100$. 

\begin{figure}[h!]
	\begin{center}
		\caption{Bias-corrected coefficients of mean beliefs and log income}
		\label{Figure_merror_sim}
		\begin{tabular}{cc}
			(a) $\beta$ for mean income beliefs & (b) $\beta$ for current log income \\
			\includegraphics[width=8cm]{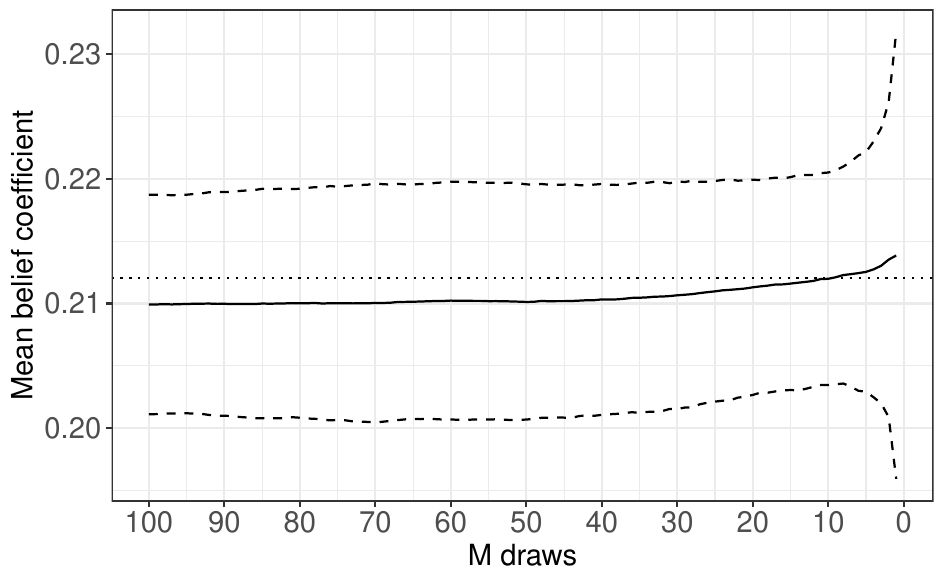}& 
			\includegraphics[width=8cm]{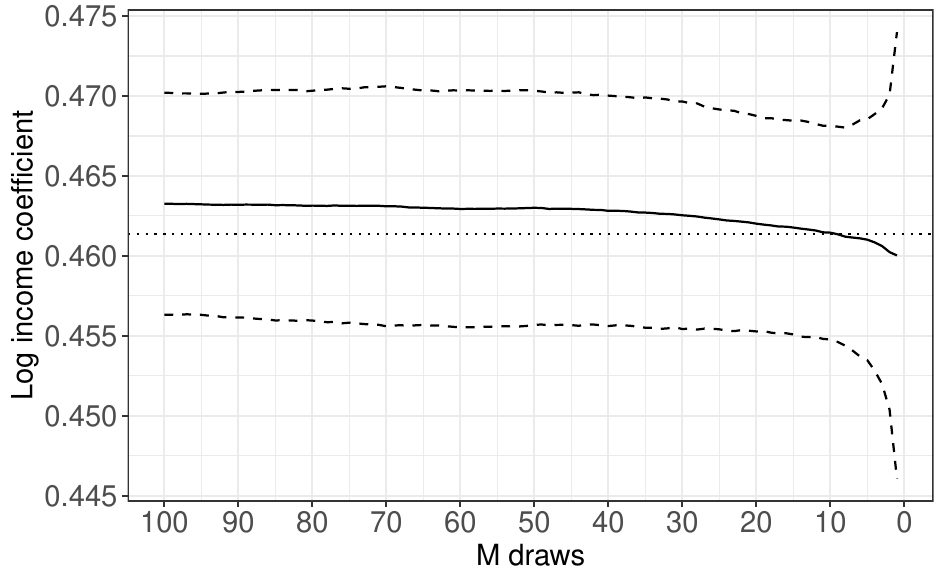} 
		\end{tabular}
	\end{center}
	
	{\footnotesize \textit{Notes: SHIW, 1989--1991, sample from column (3) in Table \ref{Table_cons_by_wave}. The horizontal dotted lines show the corresponding elements of $\widehat\beta^{\rm OLS}$ from column (3) in Table \ref{Table_cons_by_wave}. The solid lines show $\widehat\beta^{\rm BC}$, and the dashed lines add a band of plus or minus twice the standard deviation of $\widehat\beta^{(s)}$ across simulations. $1,000$ simulations.
	}}
\end{figure}

In Figure \ref{Figure_merror_sim} we report the bias-corrected estimator $\widehat\beta^{\rm BC}$ for two of the regression parameters: the coefficient of the mean income beliefs, and the coefficient of current log income. We report the results for different values of $M$. The figure shows that the results are fairly robust to this form of measurement error, with $\widehat\beta^{\rm BC}$ and $\widehat\beta^{\rm OLS}$ being close to each other irrespective of $M$. In addition, the variability induced by this form of measurement error, as captured by the dashed lines in the figure, appears moderate.

While this sensitivity analysis exercise is reassuring, it is important to acknowledge that it relies on a specific model of measurement error, and our ability to entertain other models is limited by the short panel dimension available in the SHIW.

Lastly, a possible source of measurement error specific to the SHIW, and not captured by the model we have just outlined, relates to the timing of the expectations questions. As pointed out by \citet{pistaferri2001superior}, since income and consumption refer to the previous calendar year, yet expectations are asked a few months after the end of the year, one needs to assume that individuals do not update their information sets during these few months.\footnote{Alternatively, one could instead follow a structural approach and specify a complete structural model of consumption choices and belief formation. \citet{stoltenberg2022consumption} propose such an approach and find that income beliefs, corrected for the timing discrepancy within the structure of their model (which assumes rational expectations), have larger effects on consumption than the original beliefs.}

%%%%%%%%%%%%%%%%%%%%%%%%%%%%%%%%%%%%%%%%%%%%%%%%%%%%%%%%%%%%%%%%%%%%%%%%%%%%%%%%%%%%%%%%%%%%%%%%%%%%%%%%%%%%%%%%%
%%%%%%%%%%%%%%%%%%%%%%%%%%%%%%%%%%%%%%%%%%%%%%%%%%%%%%%%%%%%%%%%%%%%%%%%%%%%%%%%%%%%%%%%%%%%%%%%%%%%%%%%%%%%%%%%%
%%%%%%%%%%%%%%%%%%%%%%%%%%%%%%%%%%%%%%%%%%%%%%%%%%%%%%%%%%%%%%%%%%%%%%%%%%%%%%%%%%%%%%%%%%%%%%%%%%%%%%%%%%%%%%%%%
\section{Simulated tax counterfactuals {in a structural model}\label{app_struct}}

In this section of the appendix we present the details of the calibration that we used to produce Table \ref{Table_sim_decom}, and report additional output from the simulation.

%%%%%%%%%%%%%%%%%%%%%%%%%%%%%%%%%%%%%%%%%%%%%%%%%%%%%%%%%%%%%%%%%%%%%%%%%%%%%%%%%%%%%%%%%%%%%%%%%%%%%%%%%%%%%%%%%%%%%%%%%%%
\subsection{Model }

The model closely follows \citet{Kaplan_Violante_2010}, with some differences. Agents live for $T$ periods, and work until age $T_{\rm ret}$, where both $T$ and $T_{\rm ret}$ are exogenous and fixed. \textit{Ex ante} identical households maximize expected life-time utility 
\begin{eqnarray*}
	\mathbb{E}_{0}\left[\sum_{t=1}^T\beta^{t-1}u(c_{it})\right].
\end{eqnarray*}

During working years $1\leq t\leq T_{\rm ret}$, agents receive after-tax labor income $w_{it}=\exp(x_{it})$, the log of which is the sum of a deterministic experience profile $\kappa_t$, a permanent component $\eta_{it}$, and a transitory component $\varepsilon_{it}$,
\begin{eqnarray*}
	x_{it} &=& \kappa_t + \eta_{it} + \varepsilon_{it}, \\
	\eta_{it}&=& \eta_{i,t-1} + v_{it},
\end{eqnarray*}
where $\eta_{i1}$ is drawn from an initial normal distribution with mean zero and variance $\sigma^2_{\eta_1}$. The shocks
$\varepsilon_{it}$ and $v_{it}$ have zero mean, are independent at all leads and lags, and are normally distributed with variances $\sigma^2_{\varepsilon}$ and $\sigma^2_{v}$, respectively. 

We define gross labor income as $\widetilde{w}_{it}=G(w_{it})$, where $G$ is the inverse of (one minus) the tax 
function
\begin{eqnarray*}
	\tau(\widetilde{w}_{it}) = \widetilde{w}_{it}-\widetilde\lambda\widetilde{w}_{it}^{1-\tau}. 
\end{eqnarray*}
After retirement, agents receive after-tax social security transfers $w_{it}^{\rm ss}$, which are a function of average individual gross income over the last few years of their working life, 
\begin{eqnarray*}
	w_{it}^{\rm ss} = P\bigg(\frac{1}{T_{\rm ret}-T_{\rm cont}}\sum_{t=T_{\rm cont}}^{T_{\rm ret}-1}{\widetilde{w}_{it}}\bigg). 
\end{eqnarray*}

Lastly, throughout their lifetime, households can save (but not borrow) through a single risk-free, one-period bond whose constant return is $1 + r$, and they face a period-to-period budget constraint
\begin{eqnarray*}
	z_{i,t+1} &=& (1 + r)z_{it} + w_{it} - c_{it} \quad \mbox{ if } t<T_{\rm ret} \nonumber\\
	z_{i,t+1} &=& (1 + r)z_{it} + w_{it}^{\rm ss} - c_{it} \quad \mbox{ if } t\geq T_{\rm ret}.
\end{eqnarray*}

We consider two cases:
\begin{itemize}
	\item A case with rational expectations, where individuals observe $\eta_{it}$ each period, and beliefs about after-tax log income next period are normally distributed with
	\begin{eqnarray*}
		\mathbb{E}_{t}(x_{i,t+1}) &=& \kappa_{t+1}+\eta_{it},\\
		{\limfunc{Var}}_t(x_{i,t+1}) &=& \sigma^2_{v}+\sigma^2_\varepsilon .
	\end{eqnarray*}
	
	\item A case with adaptive expectations, where beliefs about after-tax log income next period are normally distributed with
	\begin{eqnarray*}
		\mathbb{E}_{t}(x_{i,t+1}) &=& \kappa_{t+1}+\left(\mathbb{E}_{t-1}(x_{it})-\kappa_t\right) + \Gamma\cdot(x_{it}-\mathbb{E}_{t-1}(x_{it})) + u_{it}, \quad u_{it}\sim {\cal{N}}(0,V_u),\\
		{\limfunc{Var}}_t(x_{i,t+1}) &=& \sigma^2_{v}+\sigma^2_\varepsilon,
	\end{eqnarray*}
	where $\Gamma$ is a constant, $u_{it}$ are independent of all other shocks in the model, and initial mean beliefs are given by $\mathbb{E}_{1}(x_{i2})=\kappa_2+\eta_{i1}$. 
\end{itemize}
%
%\noindent In both cases, $$\log( X_{i,t+1}) = \kappa_{t+1} + y_{i,t+1}  \sim {\cal{N}}(\mathbb{E}_{t}(y_{i,t+1})+\kappa_{t+1},	{\limfunc{Var}}_t(y_{i,t+1})),$$ 
%and 
%$$\mathbb{E}_{t}(X_{i,t+1}) =\exp\bigg(\mathbb{E}_{t}(y_{i,t+1})+\kappa_{t+1}+\frac{{\limfunc{Var}}_t(y_{i,t+1})}{2}\bigg).$$

%%%%%%%%%%%%%%%%%%%%%%%%%%%%%%%%%%%%%%%%%%%%%%%%%%%%%%%%%%%%%%%%%%%%%%%%%%%%%%%%%%%%%%%%%%%%%%%%%%%%%%%%%%%%%%%%%%%%%%%%%%%%
\subsection{Calibration}

We closely follow the calibration strategy in \citet{Kaplan_Violante_2010}.\\

\noindent \textbf{Demographics.} The model period is one year. Agents enter the labor market at age 25, retire at age 60, and die with certainty at age 95. So we set $T_{\rm ret} = 35$, and $T = 70$. \\

\noindent \textbf{Preferences.} The utility function is CRRA, $u(c) = c^{1-\gamma}/(1-\gamma)$, where
the risk aversion parameter is set to $\gamma = 2$.\\
\\
\textbf{Discount factor and interest rate.} The interest rate is $r=0.03$, and $\beta=1/(1+r)$.\\
\\
\textbf{Income process.} We use the deterministic age profile $\kappa_t$ from \citet{Kaplan_Violante_2010}. For the stochastic components of the income process, we set $\sigma^2_{\eta_1}=0.15$, $\sigma^2_{v}=0.01$, and $\sigma^2_{\varepsilon}=0.05$.\\
\\
\textbf{Initial wealth and borrowing limit.} Households' initial assets are set to 0 and there is no borrowing possible.\\
\\
\textbf{Tax system.} We use parameters derived from \citet{holter2019tax}, $\widetilde\lambda=3.826$, $\tau=0.137$.\\
\\
\textbf{Social security benefits.} Social security benefits are a function of average individual gross earnings between the ages of 50 and 60, $w_{it}^{\rm ss}= P\bigg(\frac{1}{T_{\rm ret}-T_{\rm cont}}\sum_{t=T_{\rm cont}}^{T_{\rm ret}-1}{\widetilde{w}_{it}}\bigg)$, where $T_{\rm cont}=25$. Pre-tax benefits are equal to 90\% of average past earnings up to a given bend point, 32\% from this first bend point to a second bend point, and 15\% beyond that. The two bend points are set at, respectively, 0.18 and 1.10 times cross-sectional average gross earnings. Benefits are then scaled proportionately so that a worker earning average wages between ages 50 and 60 is entitled to a pre-tax replacement rate of 45\%. There is also a cap on pre-tax earnings contributing to pensions (cap of 2.2) and only 85\% of pre-tax pensions are taxed.\\
\\
\textbf{Adaptive beliefs.} We take $\Gamma=0.5$ and $V_u=0.2$. \\

There are two main differences between our calibration and the one from \citet{Kaplan_Violante_2010}, besides including the adaptive expectations case and using a different tax function. First, pensions depend on contributions made between ages 50 and 60, so the history of past income is not a relevant state variable before age 50. Second, we do not consider random mortality during retirement years.

%%%%%%%%%%%%%%%%%%%%%%%%%%%%%%%%%%%%%%%%%%%%%%%%%%%%%%%%%%%%%%%%%%%%%%%%%%%%%%%%%%%%%%%%%%%%%%%%%%%%%%%%%%%%%%%%%%%%%%%%%%%%
\subsection{Additional simulation results}

In this subsection we report results based on the calibrated structural model. 

\begin{figure}[tbp]
	\begin{center}
		\caption{Policy rules by type of expectations and age}
		\label{fig_sim_policy_rules}
		\begin{tabular}{ccc}
			\multicolumn{3}{c}{A. Rational expectations}\\
			(a) 26 years old & (b) 35 years old & (c) 45 years old\\
			\includegraphics[width=0.33\linewidth]{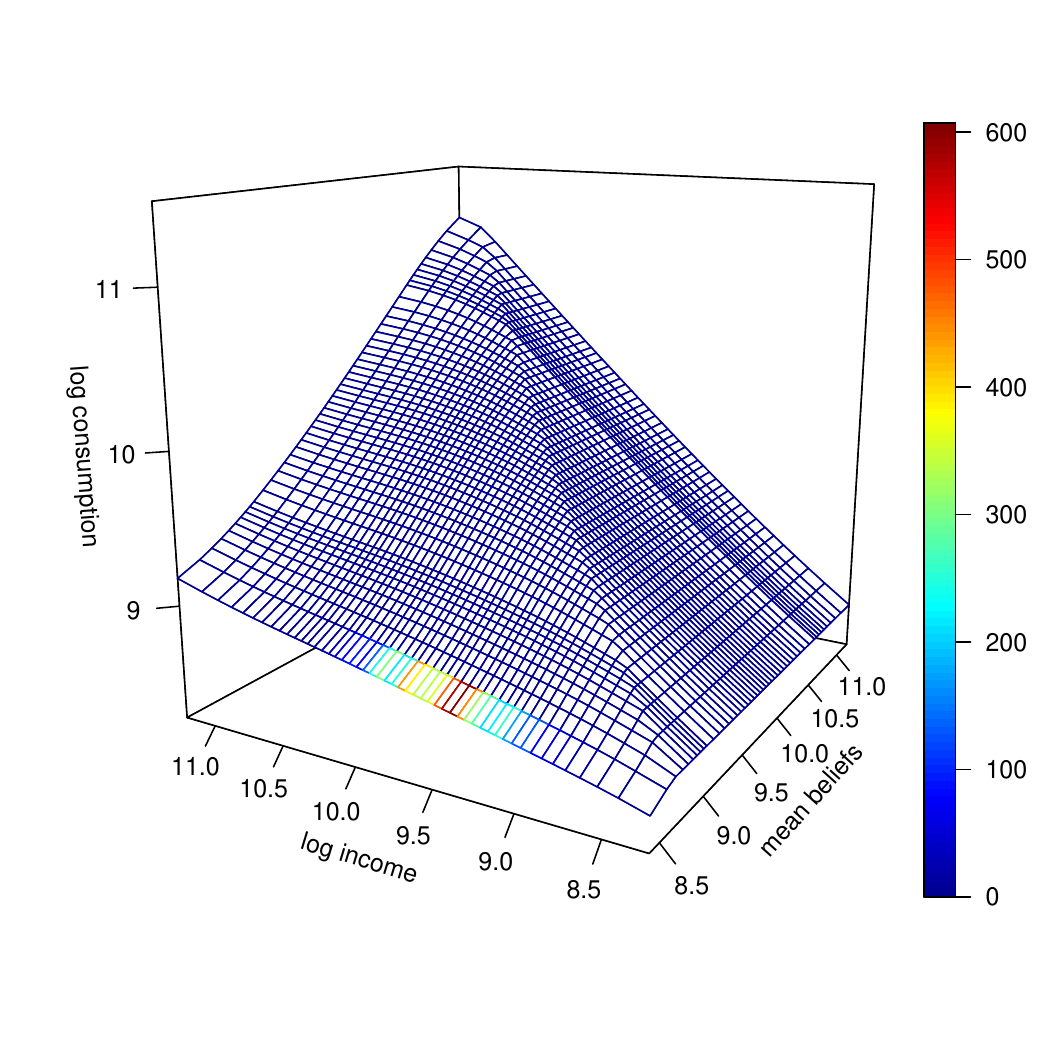} & 
			\includegraphics[width=0.33\linewidth]{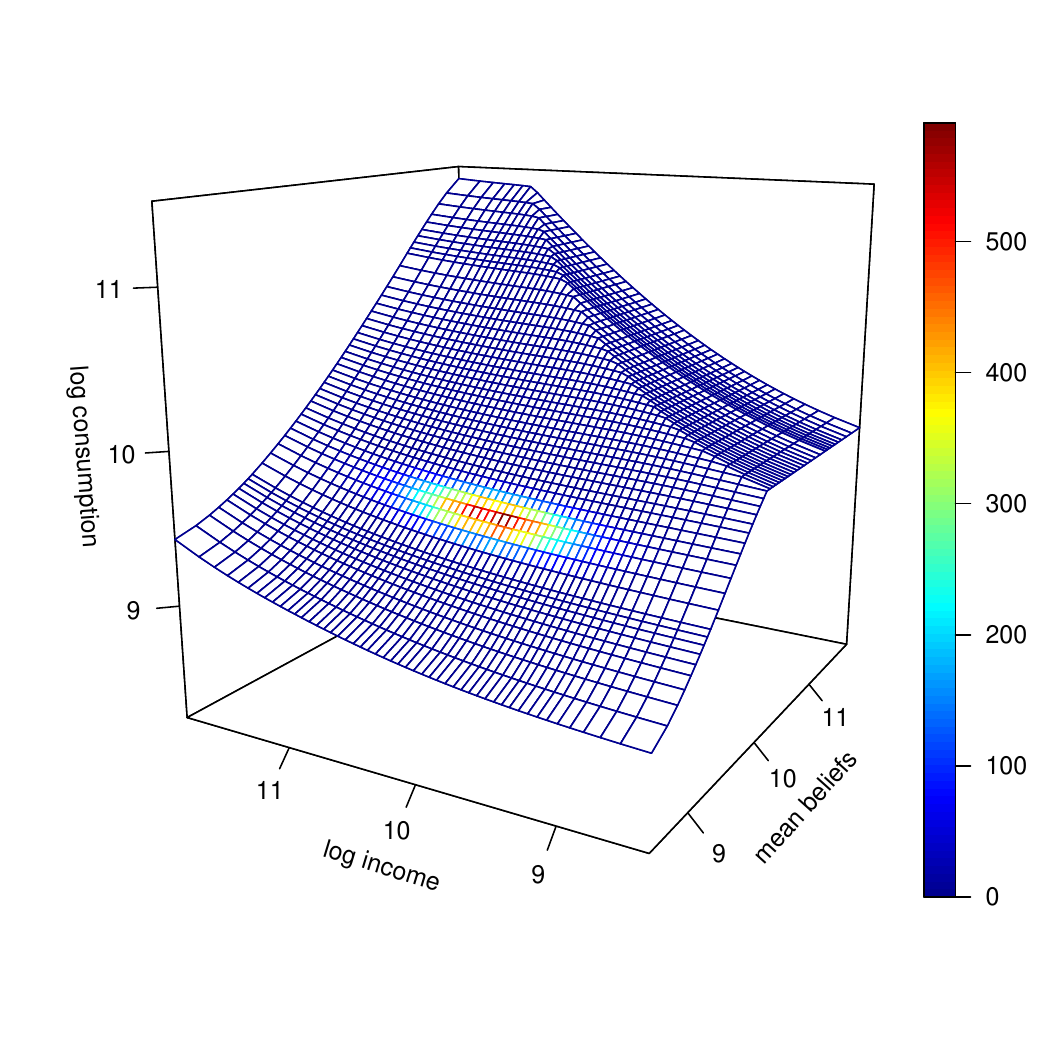} & 
			\includegraphics[width=0.33\linewidth]{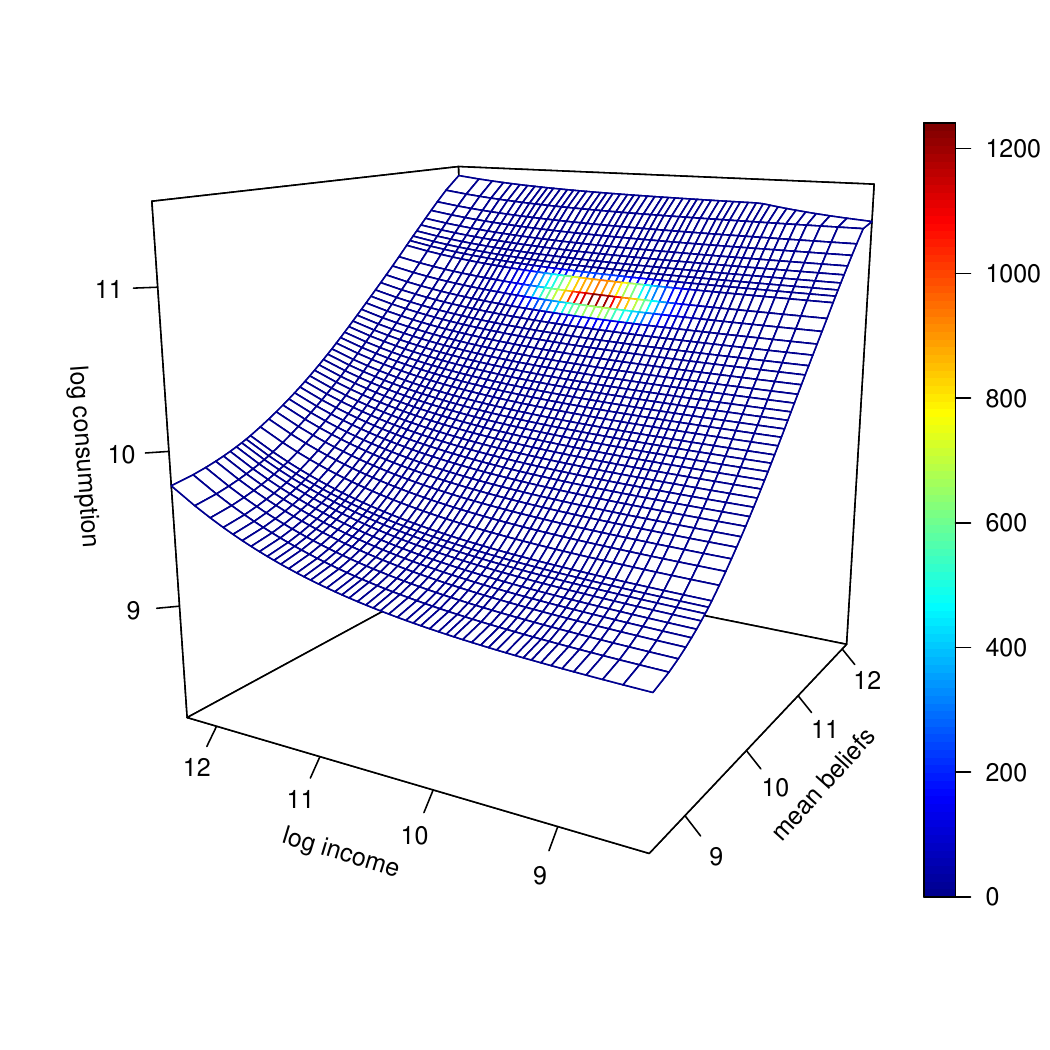} \\
			\multicolumn{3}{c}{B. Adaptive expectations}\\
			(a) 26 years old & (b) 35 years old & (c) 45 years old\\
			\includegraphics[width=0.33\linewidth]{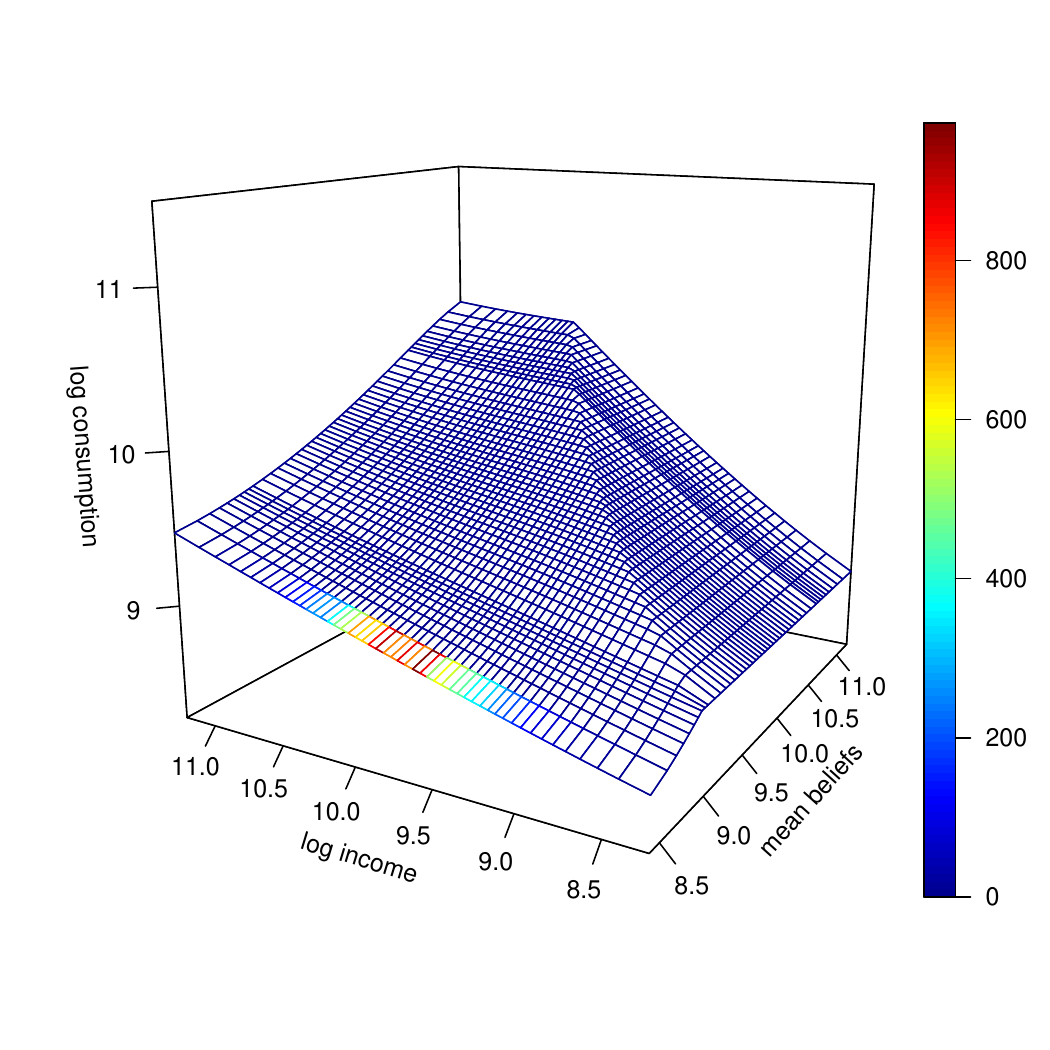} & 
			\includegraphics[width=0.33\linewidth]{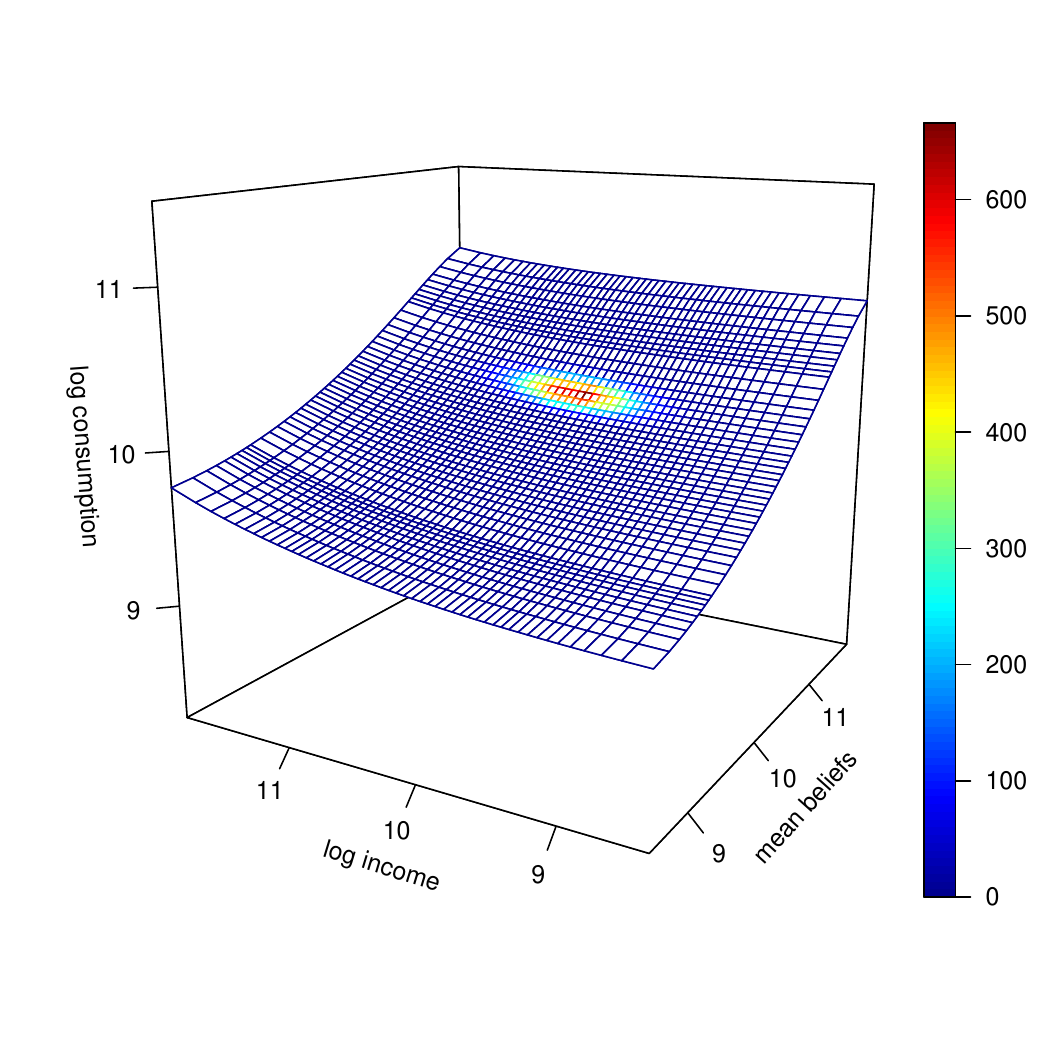} & 
			\includegraphics[width=0.33\linewidth]{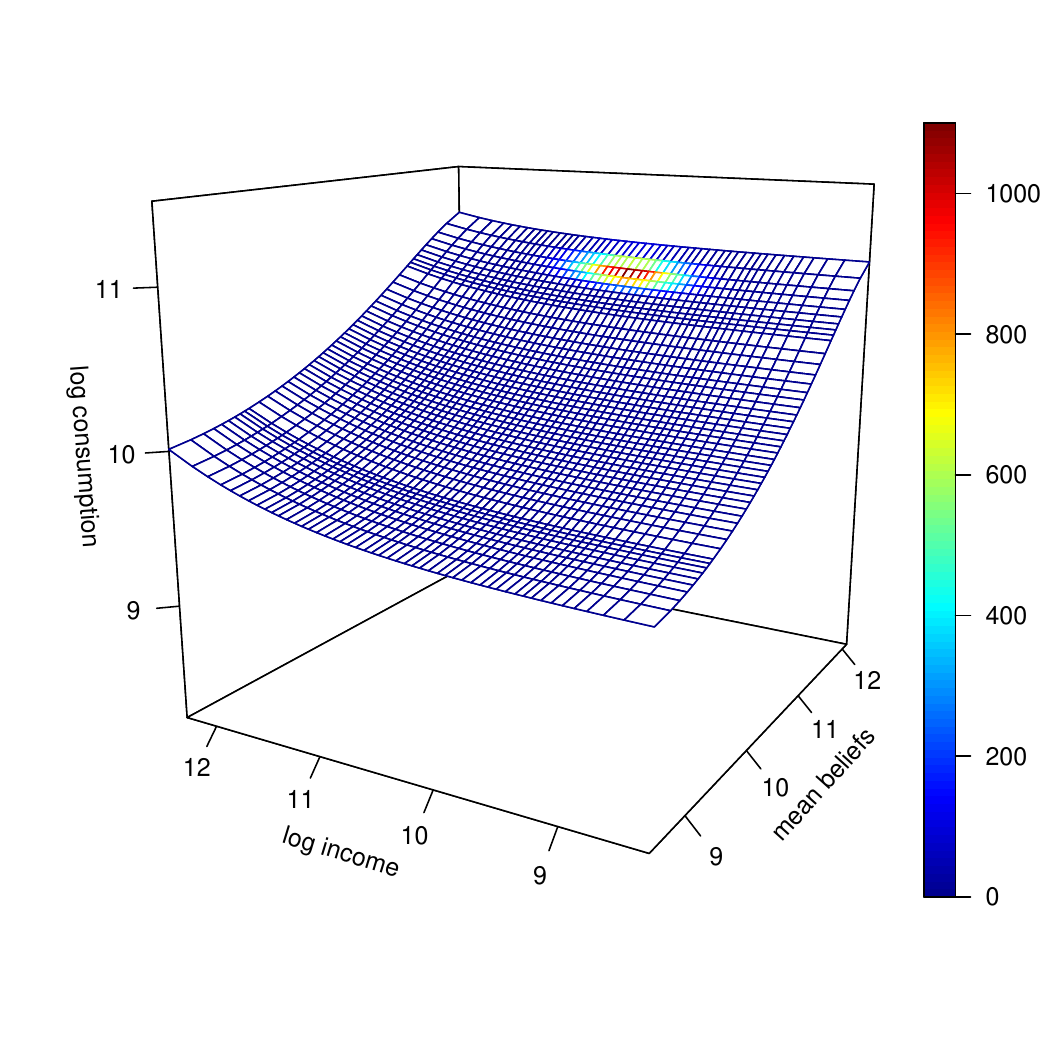} \\
		\end{tabular}
	\end{center}
	\footnotesize{\textit{Notes: The top panel plots policy rules under rational expectations and the bottom panel plots policy rules under adaptive expectations. The horizontal axes show log income and mean beliefs, and the vertical axis shows log consumption. In each figure, assets are fixed at the median value among simulated cases with positive assets. The colors represent the number of observations in the corresponding simulated data set.}} 
\end{figure}

\begin{figure}[tbp]
	\begin{center}
		\caption{Simulation results, rational versus adaptive expectations}
		\label{fig_sim}
		\begin{tabular}{cc}
			\multicolumn{2}{c}{A. Consumption}\\
			(a) Mean & (b) Variance\\
			\includegraphics[width=0.5\linewidth]{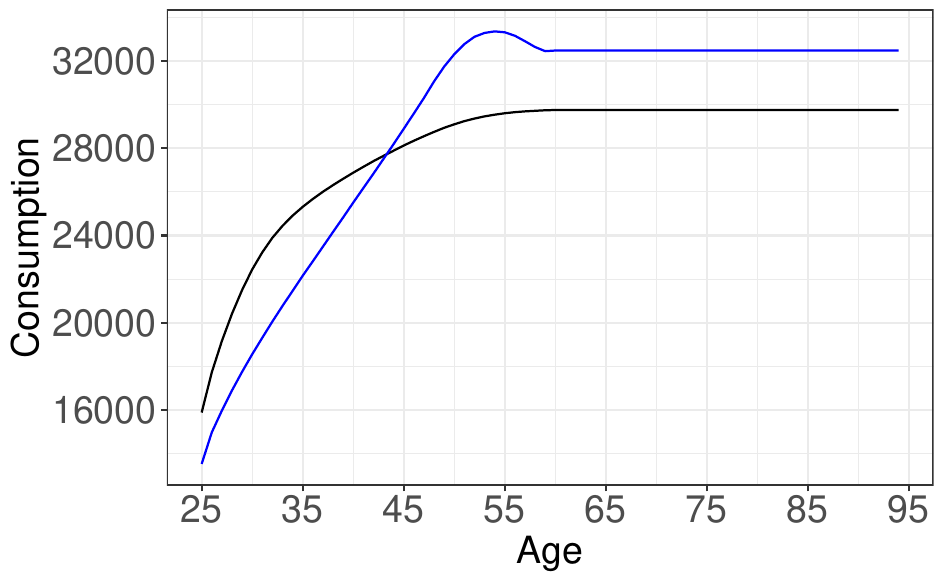}  & 
			\includegraphics[width=0.5\linewidth]{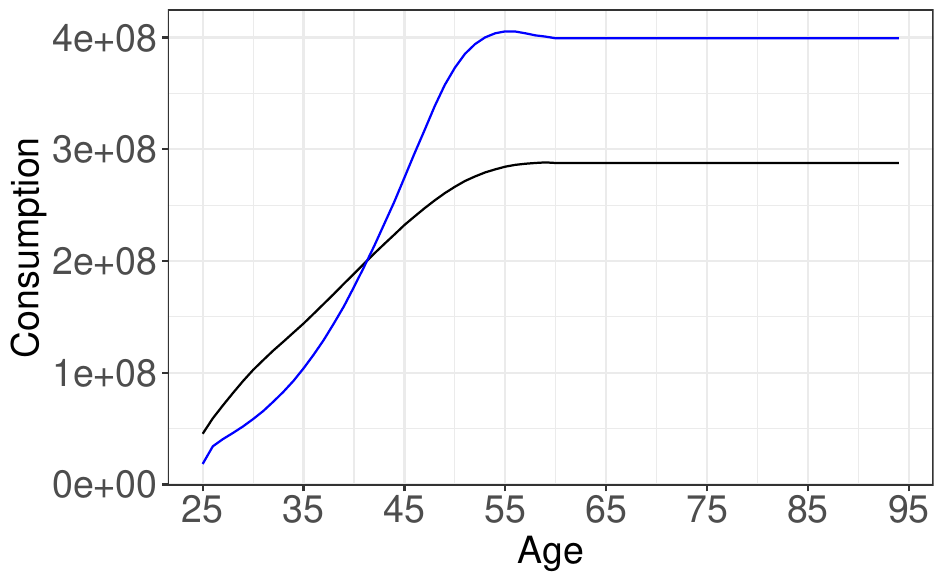}  \\
			\multicolumn{2}{c}{B. Assets}\\
			(a) Mean & (b) Variance\\
			\includegraphics[width=0.5\linewidth]{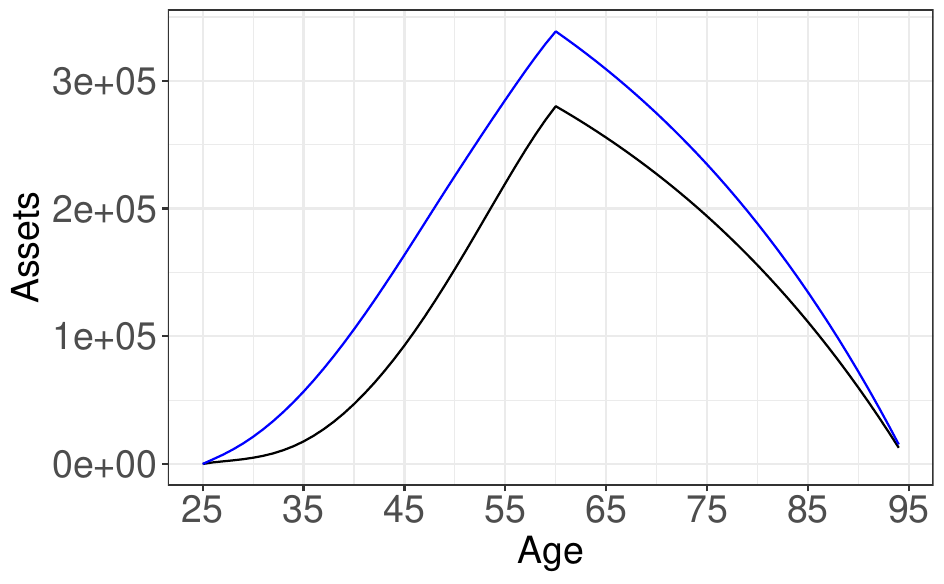}  & 
			\includegraphics[width=0.5\linewidth]{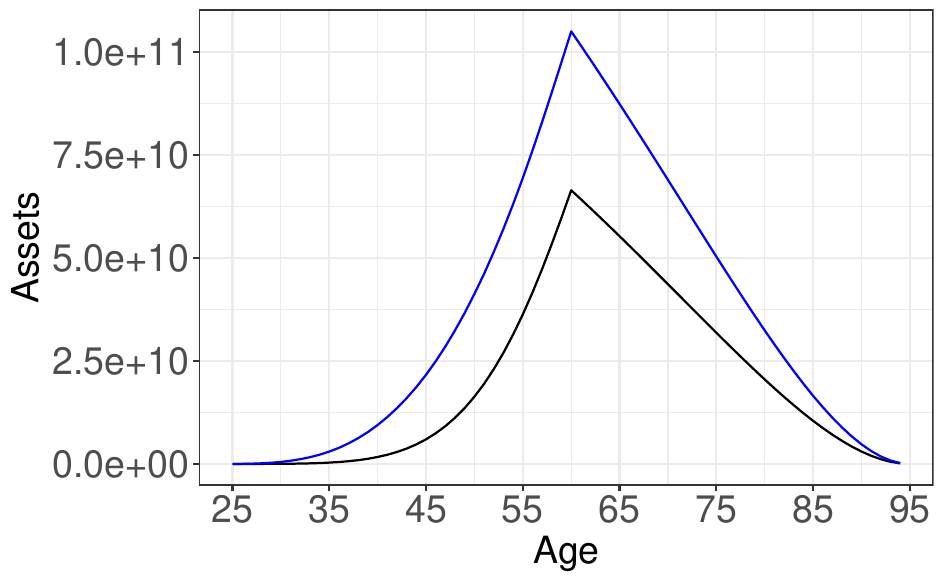}  \\
			\multicolumn{2}{c}{B. Income}\\
			(a) Mean & (b) Variance\\
			\includegraphics[width=0.5\linewidth]{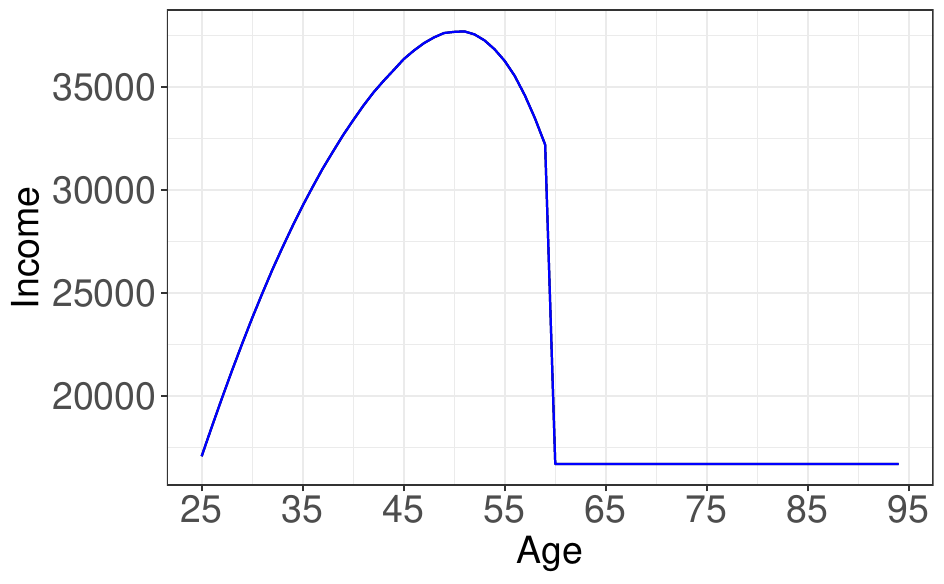}  & 
			\includegraphics[width=0.5\linewidth]{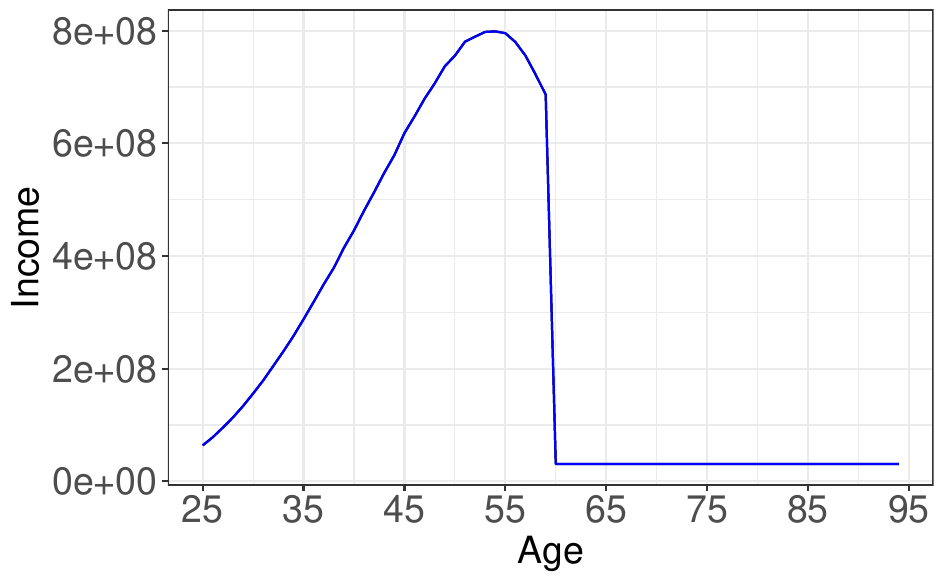}  \\
		\end{tabular}
	\end{center}
	{\footnotesize \textit{Notes: Simulations results based on the structural model. Black lines show results under rational expectations, blue lines show results under adaptive expectations.}} 
\end{figure}

In Table \ref{Table_sim_decom_age} we report structural counterfactual effects {and average partial effects} of a permanent 10\% income tax, as in Table \ref{Table_sim_decom}, for three different ages: 26, 35, and 45. We see that, under rational expectations (left panel), the contemporaneous effect of the tax is higher for the young than for older households, while the dynamic impact is lower. This reflects the fact that households start their working life without assets, and that they cannot borrow. The average partial effects reproduce the structural policy effects well. In the case of adaptive expectations (right panel) there is less variation by age, and while a linear specification tends to produce too high a contemporaneous effect for the old, the quadratic and spline specifications agree well with the structural predictions. For completeness, in Figures \ref{fig_sim_policy_rules} and \ref{fig_sim} we plot the policy rules and the mean and variance profiles of consumption, assets and income under the model.

%%%%%%%%%%%%%%%%%%%%%%%%%%%%%%%%%%%%%%%%%%%%%%%%%%%%%%%%%%%%%%%%%%%%%%%%%%%%%%%%%%%%%%%%%%%%%%%%%%%%%%%%%
%%%%%%%%%%%%%%%%%%%%%%%%%%%%%%%%%%%%%%%%%%%%%%%%%%%%%%%%%%%%%%%%%%%%%%%%%%%%%%%%%%%%%%%%%%%%%%%%%%%%%%%%%
%%%%%%%%%%%%%%%%%%%%%%%%%%%%%%%%%%%%%%%%%%%%%%%%%%%%%%%%%%%%%%%%%%%%%%%%%%%%%%%%%%%%%%%%%%%%%%%%%%%%%%%%%

\clearpage

\section{Appendix tables and figures}

%%%%%%%%%%%%%%%%%%%%%%%%%%%%%%%%%%%%%%%%%%%%%%%%%%%%%%%%%%%%%%%%%%%%%%%%%%%%%%%%%%%%%%%%%%%%%%%%%%%%
\begin{table}[h!]
	\begin{center}
		\caption{Descriptive statistics on income expectations questions 1989--1991}
		\label{Table_desc_expq_89}
		\adjustbox{max width=\textwidth}{
			\begin{tabular}{lrrrrrrrr}\hline\hline
				& \multicolumn{4}{c}{Cross-sectional sample} & \multicolumn{4}{c}{Panel sample} \\  \cmidrule(lr){2-5}  \cmidrule(lr){6-9} 
				                    &         Obs&         P25&        Mean&         P75&         Obs&         P25&        Mean&         P75\\
\hline
Income growth $>25\%$&       5,486&           0&        0.79&           0&       1,096&           0&        0.63&           0\\
Income growth $20-25\%$&       5,486&           0&        0.85&           0&       1,096&           0&        1.18&           0\\
Income growth $15-20\%$&       5,486&           0&        1.80&           0&       1,096&           0&        1.09&           0\\
Income growth $13-15\%$&       5,486&           0&        2.72&           0&       1,096&           0&        2.92&           0\\
Income growth $10-13\%$&       5,486&           0&        5.50&           0&       1,096&           0&        4.85&           0\\
Income growth $8-10\%$&       5,486&           0&        8.22&           0&       1,096&           0&        8.50&           0\\
Income growth $7-8\%$&       5,486&           0&        6.78&           0&       1,096&           0&        7.99&           0\\
Income growth $6-7\%$&       5,486&           0&        7.70&           0&       1,096&           0&        9.01&           0\\
Income growth $5-6\%$&       5,486&           0&       12.18&           0&       1,096&           0&       13.15&           5\\
Income growth $3-5\%$&       5,486&           0&       20.49&          30&       1,096&           0&       20.16&          30\\
Income growth $0-3\%$&       5,486&           0&       29.24&          80&       1,096&           0&       28.13&          70\\
Income growth $<0\%$&       5,486&           0&        3.72&           0&       1,096&           0&        2.39&           0\\
Income growth - by how much if $<0\%$&         163&           3&       10.05&          10&          15&           1&       12.18&          12\\ \hline\hline
%Number of bins with positive prob.&       5,486&           1&        1.65&           2&       1,096&           1&        1.83&           2\\
 
		\end{tabular}}
	\end{center}
	
	%\vspace{0.1cm}
	\footnotesize{\textit{Notes: Descriptive statistics are weighted using the survey's weights. }}
\end{table}

%%%%%%%%%%%%%%%%%%%%%%%%%%%%%%%%%%%%%%%%%%%%%%%%%%%%%%%%%%%%%%%%%%%%%%%%%%%%%%%%%%%%%%%%%%%%%%%%%%%%
\begin{table}[tbp]
	\begin{center}
		\caption{Descriptive statistics on income expectations questions 1995--1998}
		\label{Table_desc_expq_95}
		\adjustbox{max width=\textwidth}{
			\begin{tabular}{lrrrrrrrr}\hline\hline
				& \multicolumn{4}{c}{Cross-sectional sample} & \multicolumn{4}{c}{Panel sample} \\  \cmidrule(lr){2-5}  \cmidrule(lr){6-9} 
				                    &         Obs&         P25&        Mean&         P75&         Obs&         P25&        Mean&         P75\\
\hline
Minimum amount expected to earn&       2,310&    13,515.1&    18,401.7&    20,503.5&         550&    14,645.4&    18,866.1&    21,968.1\\
Maximum amount expected to earn&       2,310&    16,109.9&    21,363.3&    23,798.7&         550&    16,893.8&    21,551.2&    24,897.1\\
Prob. of earning less than half&       2,302&       40.00&       50.73&       70.00&         548&       30.00&       50.75&       70.00\\\hline\hline

		\end{tabular}}
	\end{center}
	
	%\vspace{0.1cm}
	\footnotesize{\textit{Notes: Amounts are in 2010 euros. Descriptive statistics are weighted using the survey's weights.}}
\end{table}

%%%%%%%%%%%%%%%%%%%%%%%%%%%%%%%%%%%%%%%%%%%%%%%%%%%%%%%%%%%%%%%%%%%%%%%%%%%%%%%%%%%%%%%%%%%%%%%%%%%%
\begin{table}[tbp]
	\begin{center}
		\caption{Descriptive statistics}
		\label{Table_desc_bel}
		\adjustbox{max width=\textwidth}{
			\begin{tabular}{lrrrrrrrr} \hline\hline
				& \multicolumn{4}{c}{Cross-sectional sample} & \multicolumn{4}{c}{Panel sample} \\  \cmidrule(lr){2-5}  \cmidrule(lr){6-9} 
				                    &         Obs&         P25&        Mean&         P75&         Obs&         P25&        Mean&         P75\\
\hline
Log family consumption&       7,796&        9.78&       10.05&       10.31&       1,646&        9.78&       10.07&       10.33\\
Log family assets   &       7,496&       10.03&       11.04&       12.18&       1,587&       10.33&       11.21&       12.28\\
Log family income   &       7,795&       10.03&       10.39&       10.74&       1,645&       10.07&       10.43&       10.79\\
Log individual income&       7,791&        9.69&        9.87&       10.07&       1,644&        9.73&        9.91&       10.11\\
Mean expected log income&       7,796&        9.72&        9.92&       10.13&       1,646&        9.75&        9.96&       10.16\\
SD expected log income&       7,796&       0.005&       0.015&       0.017&       1,646&       0.005&       0.015&       0.017\\\hline\hline
 
		\end{tabular}}
	\end{center}
	
	%\vspace{0.1cm}
	\footnotesize{\textit{Notes: Amounts are in 2010 euros. Descriptive statistics are weighted using the survey's weights. Individual income excludes property income and income from transfers. Individual-level variables (i.e., income and income expectations) corresponds to the household head.}}
\end{table}

%%%%%%%%%%%%%%%%%%%%%%%%%%%%%%%%%%%%%%%%%%%%%%%%%
\begin{table}[h!]
	\begin{center}
		\caption{Predictive power of income beliefs }
		\label{Table_income}
		\adjustbox{max width=\linewidth}{
			\begin{tabular}{lrrrrrrrrr}\hline\hline
				&\multicolumn{4}{c}{$\log(w_{i,t+1})$} & \multicolumn{4}{c}{$\log(w_{i,t+1})-\log(w_{it})$} \\ \cmidrule(lr){2-5} \cmidrule(lr){6-9}  
				                &\multicolumn{1}{c}{(1)}&\multicolumn{1}{c}{(2)}&\multicolumn{1}{c}{(3)}&\multicolumn{1}{c}{(4)}&\multicolumn{1}{c}{(5)}&\multicolumn{1}{c}{(6)}&\multicolumn{1}{c}{(7)}&\multicolumn{1}{c}{(8)}\\
\midrule
Mean expected log income&         &    0.596&         &    0.367&         &         &         &         \\
                &         &  (0.036)&         &  (0.082)&         &         &         &         \\
\addlinespace
Mean expected change in log income&         &         &         &         &         &    0.659&         &    0.367\\
                &         &         &         &         &         &  (0.116)&         &  (0.082)\\
\addlinespace
Log individual income&         &         &    0.566&    0.239&         &         &   -0.434&   -0.394\\
                &         &         &  (0.041)&  (0.083)&         &         &  (0.041)&  (0.038)\\
\midrule
Sample          &1989-1998&1989-1998&1989-1998&1989-1998&1989-1998&1989-1998&1989-1998&1989-1998\\
Controls        &      Yes&      Yes&      Yes&      Yes&      Yes&      Yes&      Yes&      Yes\\
N observations  &    2,994&    2,994&    2,994&    2,994&    2,994&    2,994&    2,994&    2,994\\
R-squared       &    0.290&    0.466&    0.460&    0.470&    0.047&    0.098&    0.196&    0.211\\\hline\hline

		\end{tabular}}
	\end{center}
	
	{\footnotesize{\textit{Notes: SHIW, 1989--1991 and 1995--1998. Regression for household heads. Controls include age and age squared, gender, education, indicator of spouse, marital status, family size, number of children 0-5, 6-13, 14-17 years old in the household, number of children outside the household, area, number of income earners in the household, and a wave indicator. Regression estimates are weighted using survey weights. Standard errors (shown in parentheses) are clustered at the household level.}}}
\end{table}

%%%%%%%%%%%%%%%%%%%%%%%%%%%%%%%%%%%%%%%%%%%%%%%%%%%%%%%%%%%%%%%%%%%%%

\begin{table}[h!]
	\begin{center}
		\caption{Regression estimates: robustness checks}
		\label{Table_cons_rob}
		\adjustbox{max width=\linewidth}{
			\resizebox{\linewidth}{!}{	\begin{tabular}{lrrrrrrrr}\hline\hline
					                &\multicolumn{1}{c}{(1)}&\multicolumn{1}{c}{(2)}&\multicolumn{1}{c}{(3)}&\multicolumn{1}{c}{(4)}&\multicolumn{1}{c}{(5)}&\multicolumn{1}{c}{(6)}&\multicolumn{1}{c}{(7)}&\multicolumn{1}{c}{(8)}\\
\midrule
Mean expected log income head&    0.235&    0.229&    0.237&    0.230&    0.235&    0.229&    0.245&    0.242\\
                &  (0.095)&  (0.093)&  (0.095)&  (0.094)&  (0.095)&  (0.093)&  (0.095)&  (0.093)\\
\addlinespace
(Mean expect. log income head)$\cdot$(Log family income)&         &    0.106&         &    0.105&         &    0.104&         &    0.103\\
                &         &  (0.061)&         &  (0.061)&         &  (0.061)&         &  (0.062)\\
\addlinespace
Mean expected log income spouse&         &         &         &         &         &         &    0.018&   -0.022\\
                &         &         &         &         &         &         &  (0.054)&  (0.064)\\
\addlinespace
(Mean expect. log income spouse)$\cdot$(Log family income)&         &         &         &         &         &         &         &    0.011\\
                &         &         &         &         &         &         &         &  (0.009)\\
\addlinespace
Log family income &    0.438&    0.438&    0.438&    0.438&    0.439&    0.439&    0.428&    0.439\\
                &  (0.091)&  (0.090)&  (0.090)&  (0.089)&  (0.089)&  (0.089)&  (0.091)&  (0.091)\\
\addlinespace
Log family assets&    0.016&    0.017&    0.018&    0.019&    0.018&    0.019&    0.018&    0.020\\
                &  (0.024)&  (0.024)&  (0.023)&  (0.023)&  (0.023)&  (0.023)&  (0.023)&  (0.023)\\
\midrule
Household fixed effect&      Yes&      Yes&      Yes&      Yes&      Yes&      Yes&      Yes&      Yes\\
Controls        &      Yes&      Yes&      Yes&      Yes&      Yes&      Yes&      Yes&      Yes\\
Distribution assumption&Disc - Triang&Disc - Triang&Log-normal&Log-normal&Log-normal&Log-normal&Log-normal&Log-normal\\
M draws         &         &         &       10&       10&       50&       50&      100&      100\\
N observations  &    1,514&    1,514&    1,536&    1,536&    1,536&    1,536&    1,536&    1,536\\
N households    &      757&      757&      768&      768&      768&      768&      768&      768\\
R-squared       &     0.26&     0.26&     0.26&     0.26&     0.26&     0.26&     0.26&     0.26\\
Pvalue F beliefs head&     0.01&     0.01&     0.01&     0.02&     0.01&     0.02&     0.01&     0.01\\
Pvalue F beliefs spouse&         &         &         &         &         &         &     0.74&     0.45\\
Pvalue F beliefs head and spouse&         &         &         &         &         &         &     0.04&     0.04\\ \hline\hline

		\end{tabular}}}
	\end{center}
	
	%\vspace{0.1cm}
	\footnotesize{\textit{Notes: SHIW, regression of log consumption for household heads. In columns (1) and (2) we assume a different distribution of beliefs (discrete distribution in waves 1989--1991 and triangular distribution in waves 1995--1998). In columns (3) to (6) we
			vary the number $M$ of draws used in estimation. In columns (7) and (8), we add spouse's beliefs (for spouses that are employees and have beliefs questions, and 0 for everyone else). The expectations variables and log family income are centered around the weighted average in the sample. Controls include age and age squared, existence of a spouse, marital status, family size, number of children 0-5, 6-13, 14-17 years old in the household, number of children outside the household, number of income earners in the household, and a wave indicator. In columns (7) and (8), we also control for a categorical variable indicating spousal situation (no spouse, spouse is homemaker, spouse is employee with beliefs questions, spouse is employee without beliefs questions, other). Regression estimates are weighted using survey weights. Standard errors (shown in parentheses) are clustered at the household level.}}
\end{table}

%%%%%%%%%%%%%%%%%%%%%%%%%%%%%%%%%%%%%%%%%%%%%%%%%%%%%%%%%%%%
%\clearpage
\begin{table}[h!]
	\begin{center}
		\caption{Regression estimates by wave}
		\label{Table_cons_by_wave}
		\adjustbox{max width=\linewidth}{
			\resizebox{\linewidth}{!}{	\begin{tabular}{lrrrrrr}\hline\hline
					                &\multicolumn{1}{c}{(1)}&\multicolumn{1}{c}{(2)}&\multicolumn{1}{c}{(3)}&\multicolumn{1}{c}{(4)}&\multicolumn{1}{c}{(5)}&\multicolumn{1}{c}{(6)}\\
\midrule
Mean expected log income &    0.235&    0.229&    0.212&    0.242&    0.323&    0.342\\
                &  (0.094)&  (0.093)&  (0.110)&  (0.108)&  (0.171)&  (0.172)\\
\addlinespace
(Mean expect. log income)$\cdot$(Log family income)&         &    0.104&         &    0.113&         &   -0.125\\
                &         &  (0.061)&         &  (0.060)&         &  (0.177)\\
\addlinespace
Log family income &    0.439&    0.439&    0.461&    0.442&    0.277&    0.264\\
                &  (0.089)&  (0.089)&  (0.101)&  (0.100)&  (0.169)&  (0.168)\\
\addlinespace
Log family assets&    0.018&    0.019&    0.046&    0.048&   -0.063&   -0.060\\
                &  (0.023)&  (0.023)&  (0.027)&  (0.026)&  (0.039)&  (0.039)\\
\midrule
Sample          &1989-1998&1989-1998&1989-1991&1989-1991&1995-1998&1995-1998\\
Household fixed effect&      Yes&      Yes&      Yes&      Yes&      Yes&      Yes\\
Controls        &      Yes&      Yes&      Yes&      Yes&      Yes&      Yes\\
N observations  &    1,536&    1,536&      962&      962&      512&      512\\
N households    &      768&      768&      481&      481&      256&      256\\
R-squared       &     0.26&     0.26&     0.35&     0.37&     0.16&     0.17\\
Pvalue F beliefs&     0.01&     0.02&     0.05&     0.03&     0.06&     0.14\\\hline\hline

		\end{tabular}}}
	\end{center}
	
	%\vspace{0.1cm}
	\footnotesize{\textit{Notes: SHIW, regression of log consumption for household heads. The expectations variables and log family income are centered around the weighted average in the sample. Controls include age and age squared, existence of a spouse, marital status, family size, number of children 0-5, 6-13, 14-17 years old in the household, number of children outside the household, number of income earners in the household, and a wave indicator. When available, we also control for other expectations variables: columns (3) and (4) also control for mean expected inflation, and columns (5) and (6) also control for the beliefs about the probability of being employed next year. Regression estimates are weighted using survey weights. Standard errors (shown in parentheses) are clustered at the household level.}}
\end{table}

%%%%%%%%%%%%%%%%%%%%%%%%%%%%%%%%%%%%%%%%%%%%%%%%%%%%%%%%%%%%
%\clearpage
\begin{table}[h!]
	\begin{center}
		\caption{Regression estimates: robustness to assets}
		\label{Table_cons_assets}
		\adjustbox{max width=\linewidth}{
			\resizebox{\linewidth}{!}{	\begin{tabular}{lrrrrrrrrr}\hline\hline
					                &\multicolumn{1}{c}{(1)}&\multicolumn{1}{c}{(2)}&\multicolumn{1}{c}{(3)}&\multicolumn{1}{c}{(4)}&\multicolumn{1}{c}{(5)}&\multicolumn{1}{c}{(6)}&\multicolumn{1}{c}{(7)}&\multicolumn{1}{c}{(8)}\\
\midrule
Mean expected log income&    0.245&    0.238&    0.167&    0.159&    0.191&    0.186&    0.223&    0.216\\
                &  (0.097)&  (0.095)&  (0.107)&  (0.106)&  (0.091)&  (0.089)&  (0.096)&  (0.095)\\
\addlinespace
(Mean expect. log income)$\cdot$(Log family income)&         &    0.095&         &    0.093&         &    0.038&         &    0.102\\
                &         &  (0.061)&         &  (0.062)&         &  (0.068)&         &  (0.060)\\
\addlinespace
Log family income &    0.410&    0.413&    0.642&    0.648&    0.494&    0.499&    0.475&    0.476\\
                &  (0.097)&  (0.097)&  (0.144)&  (0.144)&  (0.096)&  (0.095)&  (0.097)&  (0.096)\\
\addlinespace
Log family assets&    0.033&    0.032&   -0.084&   -0.087&         &         &         &         \\
                &  (0.032)&  (0.032)&  (0.055)&  (0.054)&         &         &         &         \\
\addlinespace
(Log family assets)$^2$&    0.007&    0.006&         &         &         &         &         &         \\
                &  (0.006)&  (0.006)&         &         &         &         &         &         \\
\addlinespace
Log (family assets - savings)&         &         &         &         &    0.051&    0.050&         &         \\
                &         &         &         &         &  (0.022)&  (0.022)&         &         \\
\midrule
Household fixed effect&      Yes&      Yes&      Yes&      Yes&      Yes&      Yes&      Yes&      Yes\\
Controls        &      Yes&      Yes&      Yes&      Yes&      Yes&      Yes&      Yes&      Yes\\
IV              &       No&       No&      Yes&      Yes&       No&       No&       No&       No\\
N observations  &    1,536&    1,536&    1,536&    1,536&    1,404&    1,404&    1,536&    1,536\\
N households    &      768&      768&      768&      768&      702&      702&      768&      768\\
R-squared       &     0.26&     0.26&        .&        .&     0.33&     0.33&     0.26&     0.26\\
Pvalue F beliefs&     0.01&     0.02&     0.12&     0.13&     0.04&     0.11&     0.02&     0.02\\
Pvalue first stage&         &         &     0.00&     0.00&         &         &         &         \\\hline\hline

		\end{tabular}}}
	\end{center}
	
	%\vspace{0.1cm}
	\footnotesize{\textit{Notes: SHIW, regression of log consumption for household heads. In columns (1) and (2) we control for log assets squared. In columns (3) and (4) we instrument the difference of log family assets by first-period assets and income. In columns (5) and (6) we replace end-of-year family assets by end-of-year family assets minus savings during the year. Lastly, in columns (7) and (8) we do not include any controls for assets. The expectations variables and log family income are centered around the weighted average in the sample. Controls include age and age squared, existence of a spouse, marital status, family size, number of children 0-5, 6-13, 14-17 years old in the household, number of children outside the household, number of income earners in the household, and a wave indicator. Regression estimates are weighted using survey weights. Standard errors (shown in parentheses) are clustered at the household level.}}
\end{table}

%%%%%%%%%%%%%%%%%%%%%%%%%%%%%%%%%%%%%%%%%%%%%%%%%%%%%%%%%%%%%%%%%%%%
\begin{table}[h!]
	\begin{center}
		\caption{Average partial effects estimates}
		\label{tab_counter_types}
		\adjustbox{max width=\linewidth}{
			\resizebox{0.9\linewidth}{!}{	
				\begin{tabular}{lccccccccc}\hline\hline
					\multirow{2}{*}{Quintile} & \multicolumn{3}{c}{\textit{Transitory tax} counterfactual} & \multicolumn{3}{c}{\textit{Permanent tax } counterfactual} & \multicolumn{3}{c}{\textit{Regressivity} counterfactual} \\  \cmidrule(lr){2-4}  \cmidrule(lr){5-7} \cmidrule(lr){8-10}
					& CAPE & DAPE & TAPE & CAPE & DAPE & TAPE & CAPE & DAPE & TAPE \\ \hline
					\multicolumn{10}{c}{A. $K=2$ types} \\ \hline
					% latex table generated in R 4.4.3 by xtable 1.8-4 package
% Sun Oct 12 08:19:21 2025
1 & -0.0709 &  0.0000 & -0.0709 & -0.0709 & -0.0068 & -0.0777 & -0.0414 & -0.0032 & -0.0446 \\ 
   & (0.0041) & (0.0000) & (0.0041) & (0.0041) & (0.0045) & (0.0048) & (0.0025) & (0.0028) & (0.0030) \\ 
  2 & -0.0712 &  0.0000 & -0.0712 & -0.0712 & -0.0102 & -0.0814 & -0.0234 & -0.0039 & -0.0274 \\ 
   & (0.0037) & (0.0000) & (0.0037) & (0.0037) & (0.0041) & (0.0033) & (0.0013) & (0.0016) & (0.0013) \\ 
  3 & -0.0718 &  0.0000 & -0.0718 & -0.0718 & -0.0119 & -0.0837 & -0.0110 & -0.0025 & -0.0135 \\ 
   & (0.0035) & (0.0000) & (0.0035) & (0.0035) & (0.0040) & (0.0031) & (0.0007) & (0.0009) & (0.0008) \\ 
  4 & -0.0722 &  0.0000 & -0.0722 & -0.0722 & -0.0135 & -0.0858 &  0.0007 & -0.0009 & -0.0002 \\ 
   & (0.0034) & (0.0000) & (0.0034) & (0.0034) & (0.0043) & (0.0032) & (0.0005) & (0.0003) & (0.0007) \\ 
  5 & -0.0727 &  0.0000 & -0.0727 & -0.0727 & -0.0167 & -0.0894 &  0.0186 &  0.0027 &  0.0214 \\ 
   & (0.0036) & (0.0000) & (0.0036) & (0.0036) & (0.0051) & (0.0042) & (0.0010) & (0.0010) & (0.0013) \\ 
  Total & -0.0718 &  0.0000 & -0.0718 & -0.0718 & -0.0118 & -0.0836 & -0.0113 & -0.0016 & -0.0129 \\ 
   & (0.0035) & (0.0000) & (0.0035) & (0.0035) & (0.0040) & (0.0031) & (0.0008) & (0.0010) & (0.0010)
 \\\hline
					\multicolumn{10}{c}{B. $K=3$ types} \\ \hline
					% latex table generated in R 4.4.3 by xtable 1.8-4 package
% Sun Oct 12 08:19:21 2025
1 & -0.0674 &  0.0000 & -0.0674 & -0.0674 & -0.0096 & -0.0770 & -0.0393 & -0.0039 & -0.0433 \\ 
   & (0.0055) & (0.0000) & (0.0055) & (0.0055) & (0.0093) & (0.0088) & (0.0033) & (0.0059) & (0.0058) \\ 
  2 & -0.0651 &  0.0000 & -0.0651 & -0.0651 & -0.0193 & -0.0844 & -0.0214 & -0.0073 & -0.0287 \\ 
   & (0.0048) & (0.0000) & (0.0048) & (0.0048) & (0.0062) & (0.0046) & (0.0017) & (0.0023) & (0.0017) \\ 
  3 & -0.0676 &  0.0000 & -0.0676 & -0.0676 & -0.0213 & -0.0889 & -0.0103 & -0.0047 & -0.0150 \\ 
   & (0.0044) & (0.0000) & (0.0044) & (0.0044) & (0.0066) & (0.0060) & (0.0008) & (0.0015) & (0.0014) \\ 
  4 & -0.0692 &  0.0000 & -0.0692 & -0.0692 & -0.0238 & -0.0930 &  0.0006 & -0.0015 & -0.0009 \\ 
   & (0.0042) & (0.0000) & (0.0042) & (0.0042) & (0.0094) & (0.0094) & (0.0005) & (0.0007) & (0.0009) \\ 
  5 & -0.0675 &  0.0000 & -0.0675 & -0.0675 & -0.0325 & -0.1000 &  0.0171 &  0.0057 &  0.0228 \\ 
   & (0.0047) & (0.0000) & (0.0047) & (0.0047) & (0.0139) & (0.0148) & (0.0015) & (0.0023) & (0.0031) \\ 
  Total & -0.0674 &  0.0000 & -0.0674 & -0.0674 & -0.0213 & -0.0887 & -0.0107 & -0.0023 & -0.0130 \\ 
   & (0.0044) & (0.0000) & (0.0044) & (0.0044) & (0.0064) & (0.0058) & (0.0010) & (0.0016) & (0.0017) 
 \\ \hline \hline
		\end{tabular}}}
	\end{center}
	
	\footnotesize{\textit{Notes: SHIW, 1989–1991 and 1995–1998, cross-sectional sample. Estimates based on a parametric model with finite types: two types in the top panel, and three types in the bottom panel. Standard errors are based on 100 bootstrap replications.}}
\end{table}

\clearpage
%%%%%%%%%%%%%%%%%%%%%%%%%%%%%%%%%%%%%%%%%%%%%%%%%%%%%%%%%%%%%%%%%%%%%%%%%%%%
\begin{table}[h!]
	\begin{center}
		\caption{Average partial effects estimates by type, $K=2$}
		\label{tab_counter_types_K2}
		\adjustbox{max width=\linewidth}{
			\resizebox{\linewidth}{!}{	
				\begin{tabular}{lccccccccccccc}\hline\hline                 
					& \multicolumn{2}{c}{\textit{Transitory tax} counterfactual} & \multicolumn{2}{c}{\textit{Permanent tax } counterfactual} & \multicolumn{2}{c}{\textit{Regressivity} counterfactual} \\  \cmidrule(lr){2-3}  \cmidrule(lr){4-5} \cmidrule(lr){6-7}   
					& $k=1$ & $k=2$& $k=1$ & $k=2$& $k=1$ & $k=2$ \\\hline
					% latex table generated in R 4.4.3 by xtable 1.8-4 package
% Tue Oct 14 19:30:16 2025
CAPE & -0.0915 & -0.0599 & -0.0915 & -0.0599 & -0.0142 & -0.0091 \\ 
   & (0.0062) & (0.0041) & (0.0062) & (0.0041) & (0.0011) & (0.0008) \\ 
  DAPE &  0.0000 &  0.0000 & -0.0009 & -0.0183 &  0.0008 & -0.0033 \\ 
   & (0.0000) & (0.0000) & (0.0065) & (0.0047) & (0.0015) & (0.0011) \\ 
  TAPE & -0.0915 & -0.0599 & -0.0924 & -0.0783 & -0.0134 & -0.0124 \\ 
   & (0.0062) & (0.0041) & (0.0055) & (0.0032) & (0.0013) & (0.0011) \\ 
  Type proportion &  0.3750 &  0.6250 &  0.3750 &  0.6250 &  0.3750 &  0.6250 \\ 
   & (0.0656) & (0.0656) & (0.0656) & (0.0656) & (0.0656) & (0.0656) 
 \\\hline\hline
		\end{tabular}}}
	\end{center}
	
	\vspace{0.1cm}
	\footnotesize{\textit{Notes: See the notes to Appendix Table \ref{tab_counter_types}. Estimates based on a parametric model with two types. Standard errors are based on 100 bootstrap replications.}}
\end{table}

%%%%%%%%%%%%%%%%%%%%%%%%%%%%%%%%%%%%%%%%%%%%%%%%%%%%%%%%%%%%%%%%%%%%%%%%%%%%
\begin{table}[h!]
	\begin{center}
		\caption{Average partial effects estimates by type, $K=3$}
		\label{tab_counter_types_K3}
		\adjustbox{max width=\linewidth}{
			\resizebox{\linewidth}{!}{	
				\begin{tabular}{lccccccccccccc}\hline\hline                 
					& \multicolumn{3}{c}{\textit{Transitory tax} counterfactual} & \multicolumn{3}{c}{\textit{Permanent tax } counterfactual} & \multicolumn{3}{c}{\textit{Regressivity} counterfactual} \\  \cmidrule(lr){2-4}  \cmidrule(lr){5-7} \cmidrule(lr){8-10}   
					& $k=1$ & $k=2$ & $k=3$& $k=1$ & $k=2$ & $k=3$& $k=1$ & $k=2$ & $k=3$ \\\hline
					% latex table generated in R 4.4.3 by xtable 1.8-4 package
% Tue Oct 14 19:30:16 2025
CAPE & -0.0143 & -0.0611 & -0.0955 & -0.0143 & -0.0611 & -0.0955 & -0.0013 & -0.0088 & -0.0144 \\ 
   & (0.0165) & (0.0047) & (0.0139) & (0.0165) & (0.0047) & (0.0139) & (0.0022) & (0.0011) & (0.0014) \\ 
  DAPE &  0.0000 &  0.0000 &  0.0000 & -0.0641 & -0.0239 & -0.0007 & -0.0131 & -0.0038 &  0.0016 \\ 
   & (0.0000) & (0.0000) & (0.0000) & (0.0329) & (0.0057) & (0.0123) & (0.0092) & (0.0022) & (0.0018) \\ 
  TAPE & -0.0143 & -0.0611 & -0.0955 & -0.0784 & -0.0850 & -0.0962 & -0.0145 & -0.0126 & -0.0127 \\ 
   & (0.0165) & (0.0047) & (0.0139) & (0.0194) & (0.0041) & (0.0231) & (0.0080) & (0.0028) & (0.0022) \\ 
  Type proportion &  0.1891 &  0.3703 &  0.4406 &  0.1891 &  0.3703 &  0.4406 &  0.1891 &  0.3703 &  0.4406 \\ 
   & (0.0815) & (0.1177) & (0.1456) & (0.0815) & (0.1177) & (0.1456) & (0.0815) & (0.1177) & (0.1456) 
 \\\hline\hline
		\end{tabular}}}
	\end{center}
	
	\vspace{0.1cm}
	\footnotesize{\textit{Notes: See the notes to Appendix Table \ref{tab_counter_types}. Estimates based on a parametric model with three types. Standard errors are based on 100 bootstrap replications.}}
\end{table}

%%%%%%%%%%%%%%%%%%%%%%%%%%%%%%%%%%%%%%%%%%%%%%%%%%%%%%%%%%%%%%%%%%%%%

\begin{table}[h!]
	\begin{center}
		\caption{{Estimates of the log consumption function in a model without beliefs}}
		\label{Table_nobel}
		\adjustbox{max width=\linewidth}{
			\resizebox{0.8\linewidth}{!}{	
				\begin{tabular}{lccccccccccccc}\hline\hline
					& \multicolumn{2}{c}{$K=2$ types} & \multicolumn{3}{c}{$K=3$ types} \\\cmidrule(lr){2-3}\cmidrule(lr){4-6}
					& $k=1$ & $k=2$& $k=1$ & $k=2$ & $k=3$ \\\hline
					Intercept & 10.495 & 10.346 & 10.402 & 10.345 & 10.549 \\ 
   & (0.132) & (0.129) & (0.130) & (0.135) & (0.133) \\ 
  Log family income &  0.861 &  0.643 &  0.606 &  0.619 &  0.834 \\ 
   & (0.047) & (0.024) & (0.028) & (0.067) & (0.049) \\ 
  Log family assets &  0.008 &  0.008 &  0.008 &  0.008 &  0.008 \\ 
   & (0.005) & (0.005) & (0.005) & (0.005) & (0.005) 
%  SD of error term ($\sigma$) &  0.214 &  0.214 &  0.213 &  0.213 &  0.213 \\ 
%   & (0.004) & (0.004) & (0.004) & (0.004) & (0.004) 
%   \hline
 \\\hline\hline
		\end{tabular}}}
	\end{center}
	\vspace{0.1cm}
	\footnotesize{\textit{Notes: SHIW, 1989--1991 and 1995--1998. Estimates of the conditional mean of log consumption in the finite mixture model with types $k\in\{1,...,K\}$. Same list of covariates as in our baseline specification, excluding belief-related variables. Estimates are weighted using survey weights. Analytical standard errors are shown in parentheses.}}
\end{table}

%%%%%%%%%%%%%%%%%%%%%%%%%%%%%%%%%%%%%%%%%%%%%%%%%%%%%%%%%%%%%%%%%%%%%
\begin{table}[h!]
	\begin{center}
		\caption{{Estimates of the log consumption function's coefficients based on a correlated random-effects specification}}
		\label{Table_CRE}
		\adjustbox{max width=\linewidth}{
			\resizebox{0.8\linewidth}{!}{	
				\begin{tabular}{lccccccccccccc}\hline\hline
					Variable $X_{it,k}$ & $\overline{X}_{i,1}$ & $\overline{X}_{i,2}$ & $\overline{X}_{i,3}$ \\\hline
					Intercept $X_{it,1}$ & 10.283 & -0.113 &  0.131 \\ 
   & (0.221) & (0.099) & (0.090) \\ 
  Mean expected log income $X_{it,2}$ &  0.206 & -0.006 &  0.257 \\ 
   & (0.093) & (0.092) & (0.286) \\ 
  Log family income $X_{it,3}$ &  0.525 & -0.190 &  0.016 \\ 
   & (0.091) & (0.182) & (0.061) \\ 
  (Mean expected log income)$\cdot$(Log family income) &  0.008 &  &  \\ 
   & (0.139) &  &  \\ 
  Var expected log income &  3.609 &  &  \\ 
   & (1.430) &  &  \\ 
  (Var expected log income)$\cdot$(Log family income) & -0.249 &  &  \\ 
   & (2.353) &  &  \\ 
  Log family assets &  0.013 &  &  \\ 
   & (0.007) &  & 
 \\\hline\hline
				\end{tabular}}}
	\end{center}
	\vspace{0.1cm}
	\footnotesize{\textit{Notes: See Appendix \ref{app_CRE} for a description of the correlated random-effects approach. Each row presents the estimated coefficients of $\mathbbm{E}(\beta_{i,k}|X_{i1}\ldots X_{it})$, which are assumed to be homogeneous except for the coefficients corresponding to the intercept $X_{it,1}$, the mean expected log income $X_{it,2}$, and log family income $X_{it,3}$. The bottom four rows  show the estimates of the homogeneous coefficients in the model. The top three rows show the estimates of the means of the heterogeneous coefficients, which correspond to the intercept, mean beliefs, and income, and are functions of the household-specific averages of these same three variables. Standard errors clustered at the household level are shown in parentheses. }}
\end{table}

%%%%%%%%%%%%%%%%%%%%%%%%%%%%%%%%%%%%%%%%%%%%%%%%%
\begin{table}[h!]
	\begin{center}
		\caption{Average partial effects estimates, other specifications}
		%Decomposition results \textcolor{red}{based on ols estimates} by quintile of family income}
	\label{tab_counter}
	\adjustbox{max width=\linewidth}{
		\resizebox{0.78\linewidth}{!}{	
			\begin{tabular}{lccccccccc}\hline\hline
				\multirow{2}{*}{Quintile} & \multicolumn{3}{c}{\textit{Transitory tax} counterfactual} & \multicolumn{3}{c}{\textit{Permanent tax } counterfactual} & \multicolumn{3}{c}{\textit{Regressivity} counterfactual} \\  \cmidrule(lr){2-4}  \cmidrule(lr){5-7} \cmidrule(lr){8-10}
				& CAPE & DAPE & TAPE & CAPE & DAPE & TAPE & CAPE & DAPE & TAPE \\ \hline
				\multicolumn{10}{c}{A. OLS estimates} \\ \hline
				1           &     -0.0449&      0.0000&     -0.0449&     -0.0449&     -0.0160&     -0.0608&     -0.0257&     -0.0097&     -0.0355\\
            &    (0.0105)&    (0.0000)&    (0.0105)&    (0.0105)&    (0.0119)&    (0.0118)&    (0.0063)&    (0.0077)&    (0.0079)\\
2           &     -0.0482&      0.0000&     -0.0482&     -0.0482&     -0.0209&     -0.0691&     -0.0158&     -0.0088&     -0.0246\\
            &    (0.0102)&    (0.0000)&    (0.0102)&    (0.0102)&    (0.0108)&    (0.0091)&    (0.0034)&    (0.0043)&    (0.0035)\\
3           &     -0.0489&      0.0000&     -0.0489&     -0.0489&     -0.0242&     -0.0731&     -0.0075&     -0.0057&     -0.0132\\
            &    (0.0102)&    (0.0000)&    (0.0102)&    (0.0102)&    (0.0105)&    (0.0086)&    (0.0016)&    (0.0024)&    (0.0019)\\
4           &     -0.0498&      0.0000&     -0.0498&     -0.0498&     -0.0274&     -0.0771&      0.0005&     -0.0023&     -0.0018\\
            &    (0.0103)&    (0.0000)&    (0.0103)&    (0.0103)&    (0.0106)&    (0.0088)&    (0.0004)&    (0.0009)&    (0.0011)\\
5           &     -0.0528&      0.0000&     -0.0528&     -0.0528&     -0.0321&     -0.0849&      0.0138&      0.0047&      0.0185\\
            &    (0.0105)&    (0.0000)&    (0.0105)&    (0.0105)&    (0.0114)&    (0.0108)&    (0.0028)&    (0.0018)&    (0.0027)\\
Total       &     -0.0489&      0.0000&     -0.0489&     -0.0489&     -0.0241&     -0.0730&     -0.0070&     -0.0044&     -0.0113\\
            &    (0.0102)&    (0.0000)&    (0.0102)&    (0.0102)&    (0.0105)&    (0.0086)&    (0.0018)&    (0.0027)&    (0.0026)
 \\\hline
				\multicolumn{10}{c}{B. Double Lasso estimates} \\ \hline
				% latex table generated in R 4.1.1 by xtable 1.8-4 package
% Mon Sep 11 15:33:39 2023
1 & -0.0371 &  0.0000 & -0.0371 & -0.0371 & -0.0102 & -0.0473 & -0.0207 & -0.0091 & -0.0298 \\ 
   & (0.0264) & (0.0000) & (0.0264) & (0.0264) & (0.0205) & (0.0259) & (0.0174) & (0.0308) & (0.0333) \\ 
  2 & -0.0438 &  0.0000 & -0.0438 & -0.0438 & -0.0250 & -0.0688 & -0.0138 & -0.0111 & -0.0249 \\ 
   & (0.0153) & (0.0000) & (0.0153) & (0.0153) & (0.0159) & (0.0162) & (0.0052) & (0.0221) & (0.0225) \\ 
  3 & -0.0455 &  0.0000 & -0.0455 & -0.0455 & -0.0277 & -0.0733 & -0.0064 & -0.0063 & -0.0127 \\ 
   & (0.0127) & (0.0000) & (0.0127) & (0.0127) & (0.0116) & (0.0105) & (0.0018) & (0.0097) & (0.0096) \\ 
  4 & -0.0452 &  0.0000 & -0.0452 & -0.0452 & -0.0276 & -0.0728 &  0.0008 & -0.0022 & -0.0013 \\ 
   & (0.0146) & (0.0000) & (0.0146) & (0.0146) & (0.0113) & (0.0125) & (0.0004) & (0.0165) & (0.0165) \\ 
  5 & -0.0494 &  0.0000 & -0.0494 & -0.0494 & -0.0262 & -0.0756 &  0.0135 &  0.0035 &  0.0170 \\ 
   & (0.0179) & (0.0000) & (0.0179) & (0.0179) & (0.0126) & (0.0165) & (0.0061) & (0.0759) & (0.0760) \\ 
  Total & -0.0452 &  0.0000 & -0.0452 & -0.0452 & -0.0276 & -0.0729 & -0.0052 & -0.0056 & -0.0108 \\ 
   & (0.0129) & (0.0000) & (0.0129) & (0.0129) & (0.0126) & (0.0113) & (0.0041) & (0.0186) & (0.0189) 
 \\ \hline 
				\multicolumn{10}{c}{C. Correlated random-effects estimates} \\ \hline
				1 & -0.0653 &  0.0000 & -0.0653 & -0.0653 & -0.0017 & -0.0670 & -0.0389 &  0.0007 & -0.0382 \\ 
   & (0.0122) & (0.0000) & (0.0122) & (0.0122) & (0.0127) & (0.0105) & (0.0079) & (0.0083) & (0.0067) \\ 
  2 & -0.0604 &  0.0000 & -0.0604 & -0.0604 & -0.0138 & -0.0743 & -0.0199 & -0.0051 & -0.0250 \\ 
   & (0.0097) & (0.0000) & (0.0097) & (0.0097) & (0.0093) & (0.0080) & (0.0032) & (0.0036) & (0.0029) \\ 
  3 & -0.0597 &  0.0000 & -0.0597 & -0.0597 & -0.0221 & -0.0818 & -0.0091 & -0.0045 & -0.0136 \\ 
   & (0.0097) & (0.0000) & (0.0097) & (0.0097) & (0.0106) & (0.0086) & (0.0015) & (0.0022) & (0.0017) \\ 
  4 & -0.0588 &  0.0000 & -0.0588 & -0.0588 & -0.0299 & -0.0887 &  0.0005 & -0.0019 & -0.0014 \\ 
   & (0.0102) & (0.0000) & (0.0102) & (0.0102) & (0.0139) & (0.0106) & (0.0004) & (0.0010) & (0.0011) \\ 
  5 & -0.0543 &  0.0000 & -0.0543 & -0.0543 & -0.0416 & -0.0959 &  0.0134 &  0.0066 &  0.0200 \\ 
   & (0.0140) & (0.0000) & (0.0140) & (0.0140) & (0.0203) & (0.0127) & (0.0041) & (0.0031) & (0.0026) \\ 
  Total & -0.0597 &  0.0000 & -0.0597 & -0.0597 & -0.0218 & -0.0815 & -0.0108 & -0.0008 & -0.0117 \\ 
   & (0.0097) & (0.0000) & (0.0097) & (0.0097) & (0.0105) & (0.0085) & (0.0023) & (0.0026) & (0.0022) 
 \\ \hline \hline
	\end{tabular}}}
\end{center}

\footnotesize{\textit{Notes: SHIW, 1989--1991 and 1995--1998, cross-sectional sample. In the top panel we report results based on OLS estimates, see column (5) in Table \ref{Table_cons}. In the middle panel we report estimates based on the double/debiased Lasso, for a dictionary including interactions and power of the covariates up to the third order. In the bottom panel we report estimates based on a correlated random-effects specification described in Appendix \ref{app_CRE}. Standard errors are based on 1,000 bootstrap replications.}}
\end{table}

%%%%%%%%%%%%%%%%%%%%%%%%%%%%%%%%%%%%%%%%%%%%%%%%%

\begin{table}[h!]
	\begin{center}
		\caption{{Average partial effects based on OLS, Monte Carlo simulation}}
		\label{tab_counter_mcarlo}
		\adjustbox{max width=\linewidth}{
			\resizebox{0.9\linewidth}{!}{	
				\begin{tabular}{lccccccccc}\hline\hline
					\multirow{2}{*}{Quintile} & \multicolumn{3}{c}{\textit{Transitory tax} counterfactual} & \multicolumn{3}{c}{\textit{Permanent tax } counterfactual} & \multicolumn{3}{c}{\textit{Regressivity} counterfactual} \\  \cmidrule(lr){2-4}  \cmidrule(lr){5-7} \cmidrule(lr){8-10}
					& CAPE & DAPE & TAPE & CAPE & DAPE & TAPE & CAPE & DAPE & TAPE \\ \hline
					\multicolumn{10}{c}{DGP with $K=2$} \\ \hline
					% latex table generated in R 4.5.2 by xtable 1.8-4 package
% Wed Apr  8 11:42:00 2026
1 & -0.0590 &  0.0000 & -0.0590 & -0.0590 & -0.0107 & -0.0698 & -0.0342 & -0.0058 & -0.0400 \\ 
   & (0.0097) & (0.0000) & (0.0097) & (0.0097) & (0.0106) & (0.0128) & (0.0058) & (0.0067) & (0.0082) \\ 
  2 & -0.0611 &  0.0000 & -0.0611 & -0.0611 & -0.0137 & -0.0747 & -0.0201 & -0.0054 & -0.0254 \\ 
   & (0.0091) & (0.0000) & (0.0091) & (0.0091) & (0.0087) & (0.0086) & (0.0030) & (0.0034) & (0.0031) \\ 
  3 & -0.0615 &  0.0000 & -0.0615 & -0.0615 & -0.0157 & -0.0772 & -0.0094 & -0.0034 & -0.0128 \\ 
   & (0.0091) & (0.0000) & (0.0091) & (0.0091) & (0.0082) & (0.0078) & (0.0014) & (0.0018) & (0.0015) \\ 
  4 & -0.0620 &  0.0000 & -0.0620 & -0.0620 & -0.0175 & -0.0795 &  0.0006 & -0.0012 & -0.0006 \\ 
   & (0.0091) & (0.0000) & (0.0091) & (0.0091) & (0.0085) & (0.0081) & (0.0001) & (0.0007) & (0.0007) \\ 
  5 & -0.0638 &  0.0000 & -0.0638 & -0.0638 & -0.0204 & -0.0842 &  0.0165 &  0.0032 &  0.0197 \\ 
   & (0.0094) & (0.0000) & (0.0094) & (0.0094) & (0.0100) & (0.0114) & (0.0025) & (0.0017) & (0.0028) \\ 
  Total & -0.0615 &  0.0000 & -0.0615 & -0.0615 & -0.0156 & -0.0771 & -0.0093 & -0.0025 & -0.0118 \\ 
   & (0.0091) & (0.0000) & (0.0091) & (0.0091) & (0.0082) & (0.0078) & (0.0016) & (0.0023) & (0.0025) 
 \\ \hline
					\multicolumn{10}{c}{DGP with $K=3$} \\ \hline
					% latex table generated in R 4.5.2 by xtable 1.8-4 package
% Wed Apr  8 11:41:52 2026
1 & -0.0544 &  0.0000 & -0.0544 & -0.0544 & -0.0155 & -0.0699 & -0.0313 & -0.0087 & -0.0400 \\ 
   & (0.0096) & (0.0000) & (0.0096) & (0.0096) & (0.0106) & (0.0127) & (0.0058) & (0.0067) & (0.0081) \\ 
  2 & -0.0579 &  0.0000 & -0.0579 & -0.0579 & -0.0204 & -0.0783 & -0.0190 & -0.0082 & -0.0272 \\ 
   & (0.0091) & (0.0000) & (0.0091) & (0.0091) & (0.0088) & (0.0086) & (0.0030) & (0.0034) & (0.0031) \\ 
  3 & -0.0585 &  0.0000 & -0.0585 & -0.0585 & -0.0239 & -0.0824 & -0.0090 & -0.0053 & -0.0143 \\ 
   & (0.0091) & (0.0000) & (0.0091) & (0.0091) & (0.0082) & (0.0077) & (0.0014) & (0.0018) & (0.0015) \\ 
  4 & -0.0594 &  0.0000 & -0.0594 & -0.0594 & -0.0270 & -0.0865 &  0.0006 & -0.0020 & -0.0014 \\ 
   & (0.0091) & (0.0000) & (0.0091) & (0.0091) & (0.0084) & (0.0080) & (0.0001) & (0.0007) & (0.0007) \\ 
  5 & -0.0625 &  0.0000 & -0.0625 & -0.0625 & -0.0318 & -0.0943 &  0.0163 &  0.0049 &  0.0212 \\ 
   & (0.0094) & (0.0000) & (0.0094) & (0.0094) & (0.0099) & (0.0112) & (0.0025) & (0.0016) & (0.0028) \\ 
  Total & -0.0586 &  0.0000 & -0.0586 & -0.0586 & -0.0237 & -0.0823 & -0.0085 & -0.0038 & -0.0123 \\ 
   & (0.0091) & (0.0000) & (0.0091) & (0.0091) & (0.0082) & (0.0077) & (0.0016) & (0.0023) & (0.0025) \\   \hline \hline

		\end{tabular}}}
	\end{center}
	
	\footnotesize{\textit{Notes: Average estimates of APEs in simulated data across 10,000 replications, based on OLS in first differences. The underlying data generating process is a finite mixture	model with $K=2$ types in the top panel and $K=3$ types in the bottom panel. Standard deviations across simulations are shown in parentheses.}}
\end{table}

%%%%%%%%%%%%%%%%%%%%%%%%%%%%%%%%%%%%%%%%%%%%%%%%%%%%%%%%%%%%%%%%%
%\clearpage
\begin{table}[h!]
	\begin{center}
		\caption{Predicted distribution of number of bins by number of draws $M$}
		\label{Table_merror_nbins}
		\begin{tabular}{lccccccccccccc}\hline \hline
			& \multicolumn{12}{c}{Number of bins with non-zero frequencies} \\ \cmidrule(lr){2-13}
			% latex table generated in R 4.1.1 by xtable 1.8-4 package
% Thu Jun 08 09:56:23 2023
 & 1 & 2 & 3 & 4 & 5 & 6 & 7 & 8 & 9 & 10 & 11 & 12 & Mean \\ 
  \hline
Data & 0.59 & 0.24 & 0.09 & 0.05 & 0.01 & 0.01 & 0.00 & 0.00 & 0.00 & 0.00 & 0.00 & 0.00 & 1.75 \\ 
  $M=1$ & 1.00 & 0.00 & 0.00 & 0.00 & 0.00 & 0.00 & 0.00 & 0.00 & 0.00 & 0.00 & 0.00 & 0.00 & 1.00 \\ 
  $M=2$ & 0.68 & 0.32 & 0.00 & 0.00 & 0.00 & 0.00 & 0.00 & 0.00 & 0.00 & 0.00 & 0.00 & 0.00 & 1.32 \\ 
  $M=3$ & 0.57 & 0.35 & 0.08 & 0.00 & 0.00 & 0.00 & 0.00 & 0.00 & 0.00 & 0.00 & 0.00 & 0.00 & 1.51 \\ 
  $M=4$ & 0.50 & 0.36 & 0.12 & 0.02 & 0.00 & 0.00 & 0.00 & 0.00 & 0.00 & 0.00 & 0.00 & 0.00 & 1.66 \\ 
  $M=5$ & 0.45 & 0.37 & 0.14 & 0.04 & 0.01 & 0.00 & 0.00 & 0.00 & 0.00 & 0.00 & 0.00 & 0.00 & 1.78 \\ 
  $M=6$ & 0.42 & 0.37 & 0.15 & 0.05 & 0.01 & 0.00 & 0.00 & 0.00 & 0.00 & 0.00 & 0.00 & 0.00 & 1.88 \\ 
  $M=7$ & 0.39 & 0.37 & 0.16 & 0.06 & 0.02 & 0.00 & 0.00 & 0.00 & 0.00 & 0.00 & 0.00 & 0.00 & 1.96 \\ 
  $M=8$ & 0.36 & 0.38 & 0.16 & 0.06 & 0.02 & 0.01 & 0.00 & 0.00 & 0.00 & 0.00 & 0.00 & 0.00 & 2.03 \\ 
  $M=9$ & 0.34 & 0.38 & 0.17 & 0.07 & 0.03 & 0.01 & 0.00 & 0.00 & 0.00 & 0.00 & 0.00 & 0.00 & 2.10 \\ 
  $M=10$ & 0.32 & 0.39 & 0.18 & 0.07 & 0.03 & 0.01 & 0.00 & 0.00 & 0.00 & 0.00 & 0.00 & 0.00 & 2.16 \\ 
  $M=20$ & 0.17 & 0.41 & 0.24 & 0.09 & 0.05 & 0.02 & 0.01 & 0.00 & 0.00 & 0.00 & 0.00 & 0.00 & 2.59 \\ 
  $M=30$ & 0.09 & 0.39 & 0.30 & 0.11 & 0.06 & 0.03 & 0.01 & 0.01 & 0.00 & 0.00 & 0.00 & 0.00 & 2.87 \\ 
  $M=40$ & 0.05 & 0.36 & 0.34 & 0.13 & 0.06 & 0.03 & 0.02 & 0.01 & 0.00 & 0.00 & 0.00 & 0.00 & 3.07 \\ 
  $M=50$ & 0.03 & 0.31 & 0.38 & 0.14 & 0.07 & 0.04 & 0.02 & 0.01 & 0.00 & 0.00 & 0.00 & 0.00 & 3.22 \\ 
  $M=60$ & 0.01 & 0.28 & 0.41 & 0.15 & 0.07 & 0.04 & 0.02 & 0.01 & 0.00 & 0.00 & 0.00 & 0.00 & 3.33 \\ 
  $M=70$ & 0.01 & 0.24 & 0.43 & 0.16 & 0.08 & 0.04 & 0.02 & 0.01 & 0.00 & 0.00 & 0.00 & 0.00 & 3.42 \\ 
  $M=80$ & 0.00 & 0.21 & 0.45 & 0.16 & 0.08 & 0.04 & 0.02 & 0.01 & 0.00 & 0.00 & 0.00 & 0.00 & 3.49 \\ 
  $M=90$ & 0.00 & 0.19 & 0.46 & 0.16 & 0.09 & 0.04 & 0.03 & 0.01 & 0.01 & 0.00 & 0.00 & 0.00 & 3.55 \\ 
  $M=100$ & 0.00 & 0.16 & 0.48 & 0.17 & 0.09 & 0.05 & 0.03 & 0.01 & 0.01 & 0.00 & 0.00 & 0.00 & 3.61 \\ 
   \hline\hline
 
		\end{tabular}
	\end{center}
	{\footnotesize \textit{Notes: SHIW, 1989--1991, sample from column (3) in Table \ref{Table_cons_by_wave}. Each row reports the simulated distribution of the number of non-empty bins in data simulated from a measurement error model with $M$ draws, averaged across observations and $S=1,000$ simulations.}}
\end{table}

\begin{table}[h!]
	\begin{center}
		\caption{Simulated tax counterfactuals under rational and adaptive expectations by age}
		\label{Table_sim_decom_age}
		\adjustbox{max width=\linewidth}{
			\resizebox{\linewidth}{!}{	
				\begin{tabular}{lccccccccc}\hline\hline
					& \multicolumn{8}{c}{Age 26} \\ \hline
					& \multicolumn{4}{c}{Rational expectations} & \multicolumn{4}{c}{Adaptive expectations} \\ \cmidrule(lr){2-5} \cmidrule(lr){6-9}
					& \multirow{2}{*}{Structural} & \multicolumn{3}{c}{APE} & \multirow{2}{*}{Structural} & \multicolumn{3}{c}{APE}  \\\cmidrule(lr){3-5}\cmidrule(lr){7-9}
					&  & Linear & Quadratic & Spline &  & Linear & Quadratic & Spline \\ \hline
					% latex table generated in R 4.1.1 by xtable 1.8-4 package
% Wed Feb 15 17:21:19 2023
%  \hline
CAPE & -0.0663 & -0.0599 & -0.0599 & -0.0599 & -0.0331 & -0.0318 & -0.0313 & -0.0315 \\ 
  DAPE & -0.0471 & -0.0550 & -0.0543 & -0.0540 & -0.0509 & -0.0536 & -0.0536 & -0.0535 \\ 
  TAPE & -0.1134 & -0.1149 & -0.1142 & -0.1139 & -0.0840 & -0.0854 & -0.0849 & -0.0850 %\\ 
%  Observations & 304,370 & 304,370 & 304,370 & 304,370 & 407,192 & 407,192 & 407,192 & 407,192 \\ 
%   \hline
 \\ \hline
					& \multicolumn{8}{c}{Age 35} \\ \hline
					& \multicolumn{4}{c}{Rational expectations} & \multicolumn{4}{c}{Adaptive expectations} \\ \cmidrule(lr){2-5} \cmidrule(lr){6-9}
					& \multirow{2}{*}{Structural} & \multicolumn{3}{c}{APE} & \multirow{2}{*}{Structural} & \multicolumn{3}{c}{APE}  \\\cmidrule(lr){3-5}\cmidrule(lr){7-9}
					&  & Linear & Quadratic & Spline &  & Linear & Quadratic & Spline \\ \hline
					% latex table generated in R 4.1.1 by xtable 1.8-4 package
% Wed Feb 15 17:21:21 2023
%  \hline
CAPE & -0.0110 & -0.0097 & -0.0097 & -0.0097 & -0.0111 & -0.0284 & -0.0149 & -0.0123 \\ 
  DAPE & -0.0921 & -0.0982 & -0.0948 & -0.0945 & -0.0507 & -0.0521 & -0.0519 & -0.0519 \\ 
  TAPE & -0.1031 & -0.1079 & -0.1044 & -0.1041 & -0.0618 & -0.0805 & -0.0668 & -0.0643 %\\ 
%  Observations & 481,660 & 481,660 & 481,660 & 481,660 & 484,714 & 484,714 & 484,714 & 484,714 \\ 
%   \hline
 \\ \hline
					& \multicolumn{8}{c}{Age 45} \\ \hline
					& \multicolumn{4}{c}{Rational expectations} & \multicolumn{4}{c}{Adaptive expectations} \\ \cmidrule(lr){2-5} \cmidrule(lr){6-9}
					& \multirow{2}{*}{Structural} & \multicolumn{3}{c}{APE} & \multirow{2}{*}{Structural} & \multicolumn{3}{c}{APE}  \\\cmidrule(lr){3-5}\cmidrule(lr){7-9}
					&  & Linear & Quadratic & Spline &  & Linear & Quadratic & Spline \\ \hline
					% latex table generated in R 4.1.1 by xtable 1.8-4 package
% Wed Feb 15 17:21:22 2023
%  \hline
CAPE & -0.0058 & -0.0062 & -0.0062 & -0.0061 & -0.0078 & -0.0337 & -0.0139 & -0.0084 \\ 
  DAPE & -0.0794 & -0.0877 & -0.0821 & -0.0805 & -0.0479 & -0.0508 & -0.0490 & -0.0491 \\ 
  TAPE & -0.0852 & -0.0939 & -0.0883 & -0.0866 & -0.0557 & -0.0846 & -0.0629 & -0.0575 %\\ 
%  Observations & 499,761 & 499,761 & 499,761 & 499,761 & 496,531 & 496,531 & 496,531 & 496,531 \\ 
%   \hline
 \\ \hline\hline 
		\end{tabular}}}
	\end{center}
	\footnotesize{\textit{Notes: See the notes to Table \ref{Table_sim_decom}. Results by age.}}
\end{table}

\clearpage

%%%%%%%%%%%%%%%%%%%%%%%%%%%%%%%%%%%%%%%%%%%%%%%%%%%%%%%%%%%%%%%%%%%%%
\begin{figure}[h!]
	\begin{center}
		\caption{{Average partial effects in a model without beliefs}}
		\label{fig_nobel}
		\begin{tabular}{ccc}
			\multicolumn{3}{c}{A. $K=2$ types} \\ 
			(a) Transitory tax & (b) Permanent tax & (c) Regressivity\\
			\includegraphics[width=5.35cm]{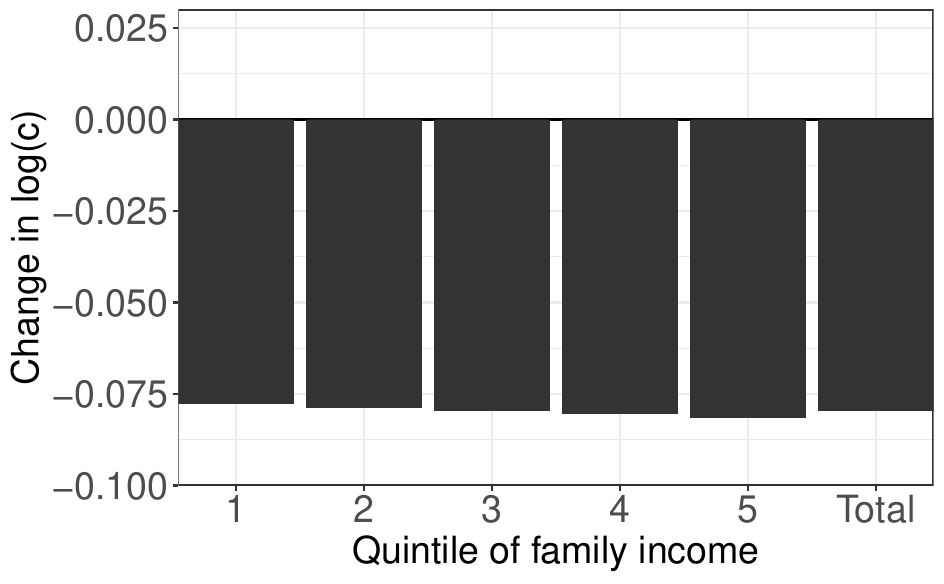}& 
			\includegraphics[width=5.35cm]{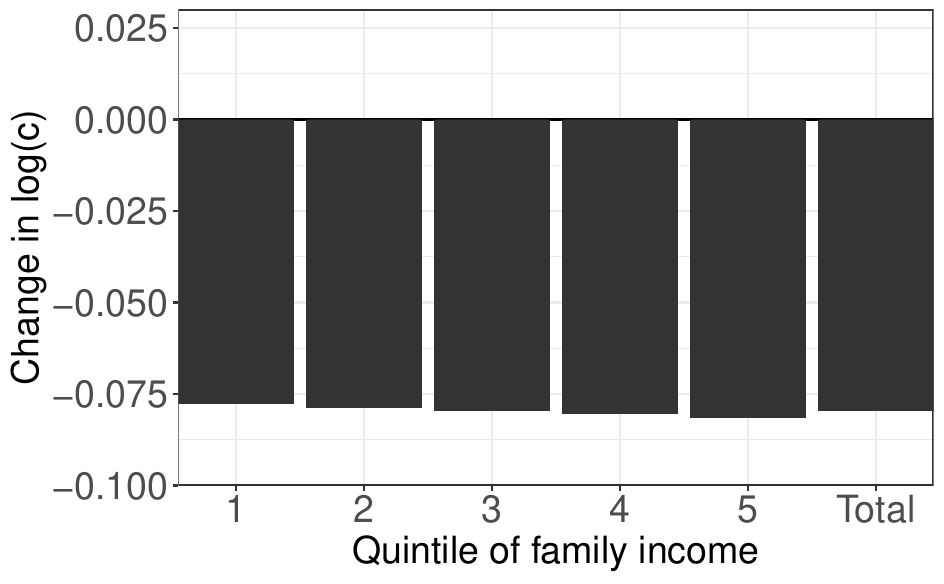}& 
			\includegraphics[width=5.35cm]{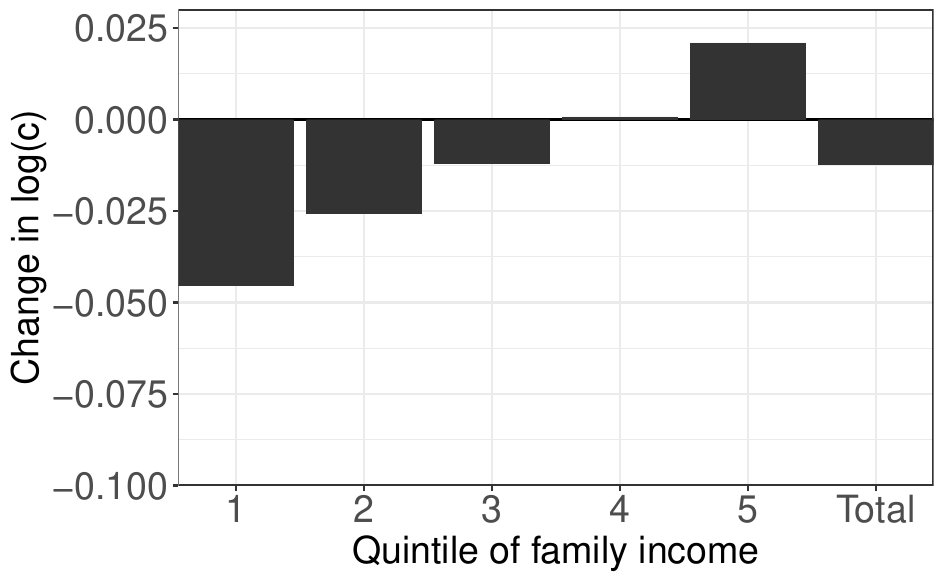}\\
			\multicolumn{3}{c}{B. $K=3$ types} \\ 
			(d) Transitory tax & (e) Permanent tax & (f) Regressivity\\
			\includegraphics[width=5.35cm]{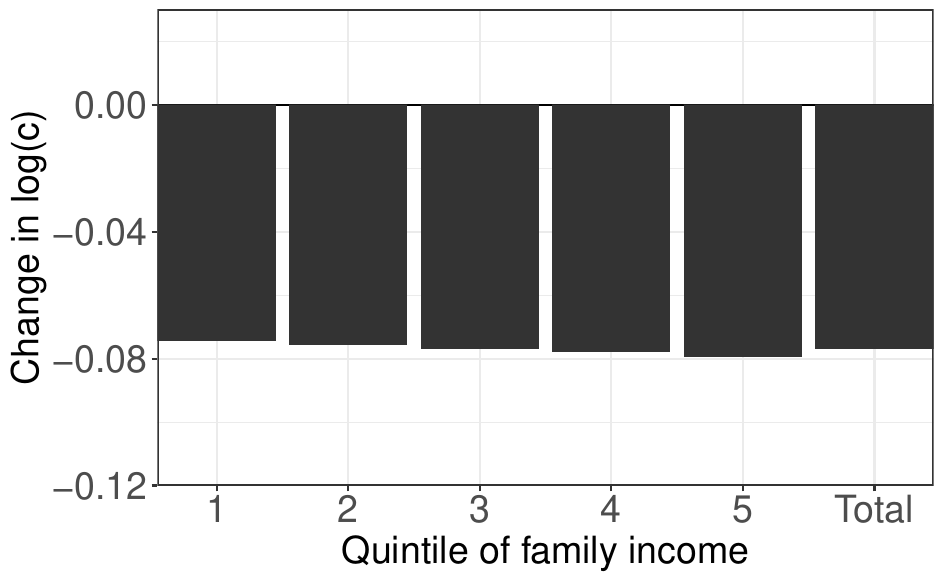}& 
			\includegraphics[width=5.35cm]{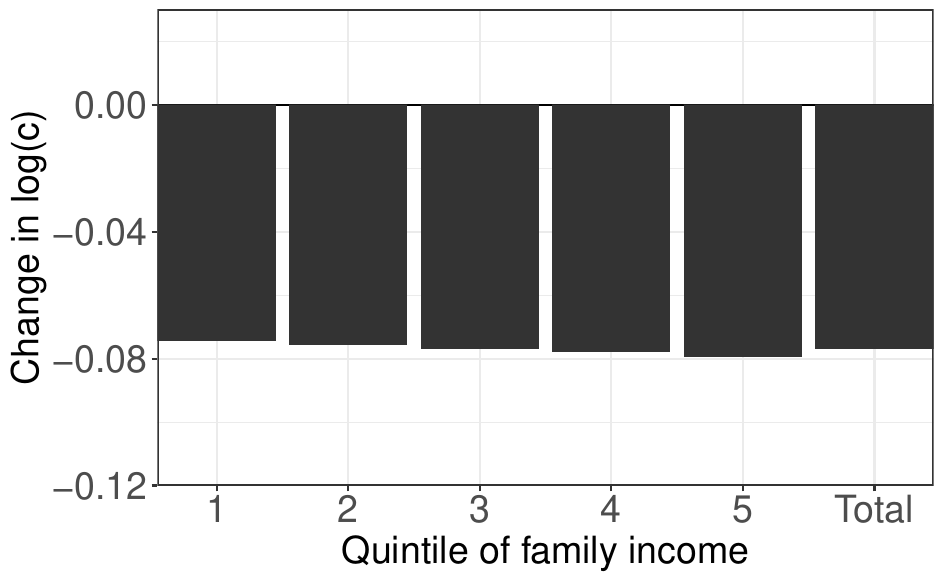}& 
			\includegraphics[width=5.35cm]{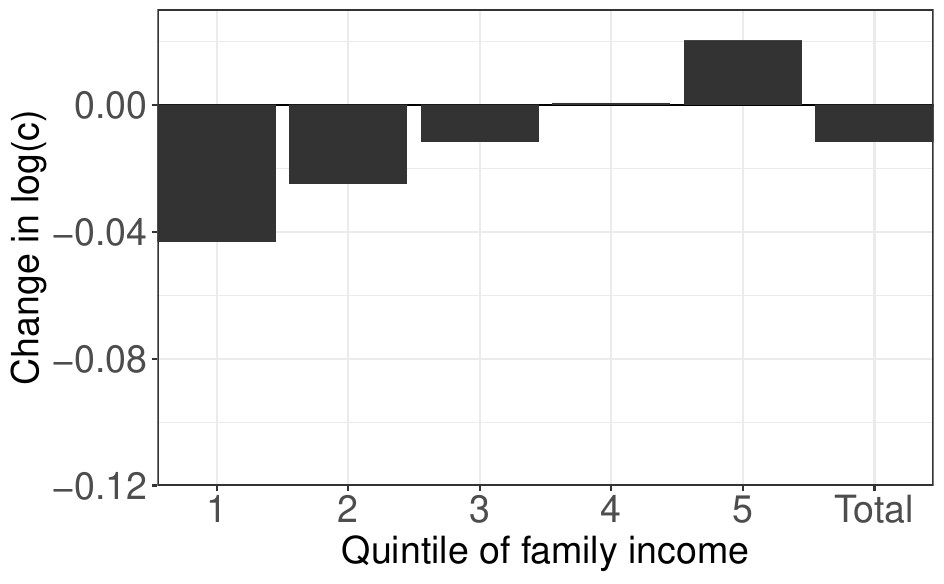}			
		\end{tabular}
	\end{center}
	
	{\footnotesize \textit{Notes: See the notes to Figure \ref{fig_decompo_cons_types}. Estimates based on a parametric finite-type model without beliefs.}}
	
\end{figure}

%%%%%%%%%%%%%%%%%%%%%%%%%%%%%%%%%%%%%%%%%%%%%%%%%%%%%%%%%%%%%%%%%%%%%
\begin{figure}[h!]
\begin{center}
	\caption{{Average partial effects by type in a model without beliefs}}
	\label{fig_nobel_type}
	\begin{tabular}{ccc}
		\multicolumn{3}{c}{A. $K=2$ types} \\ 
		(a) Transitory tax & (b) Permanent tax & (c) Regressivity\\
		\includegraphics[width=5.35cm]{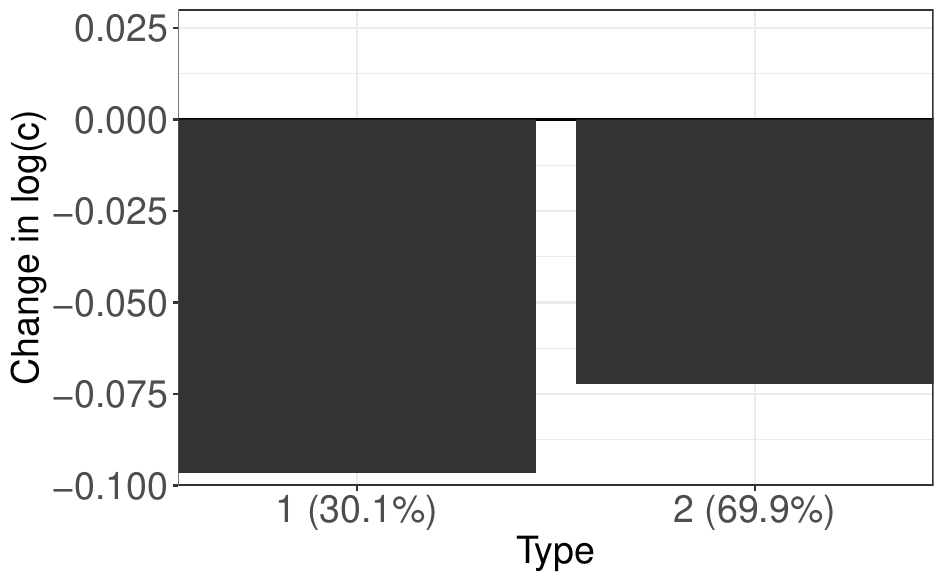}& 
		\includegraphics[width=5.35cm]{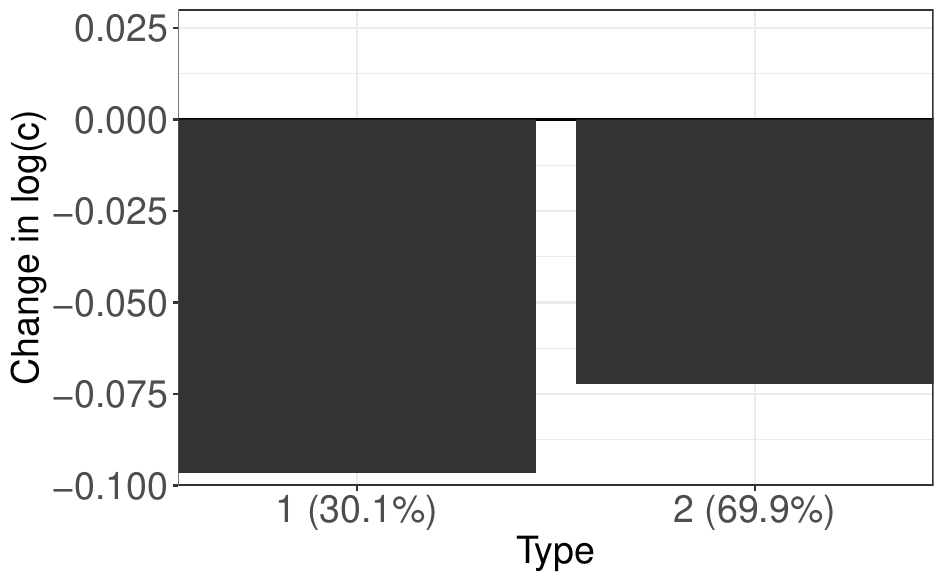}& 
		\includegraphics[width=5.35cm]{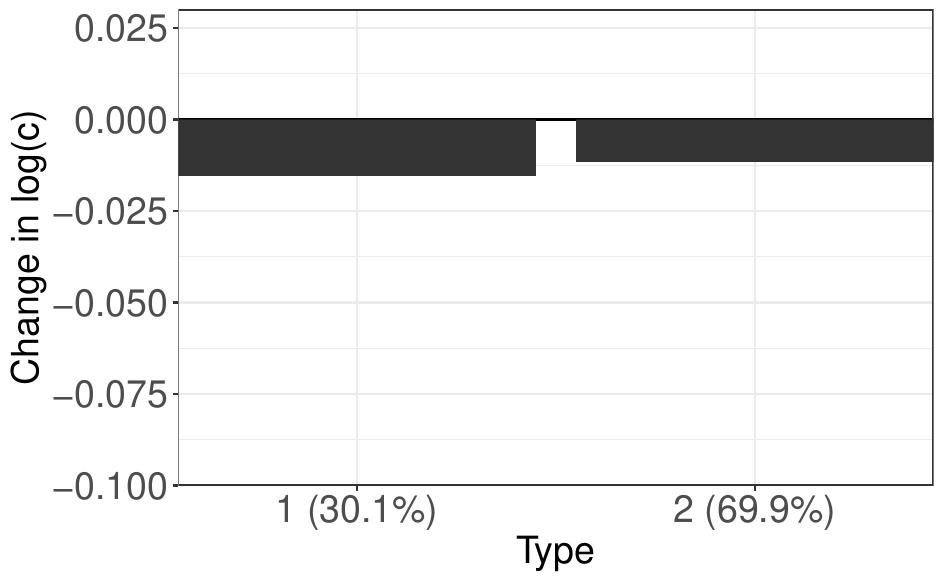}\\
		\multicolumn{3}{c}{B. $K=3$ types} \\ 
		(d) Transitory tax & (e) Permanent tax & (f) Regressivity\\
		\includegraphics[width=5.35cm]{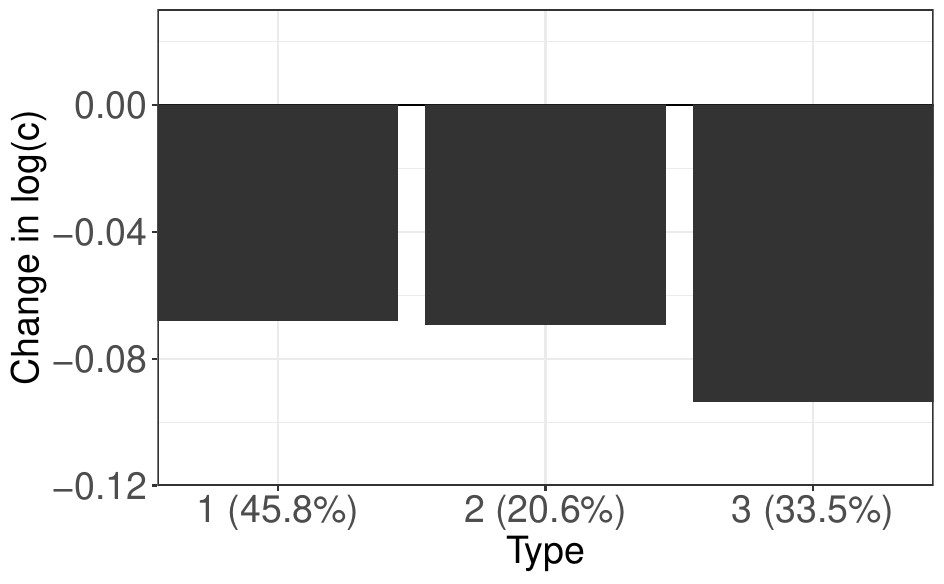}& 
		\includegraphics[width=5.35cm]{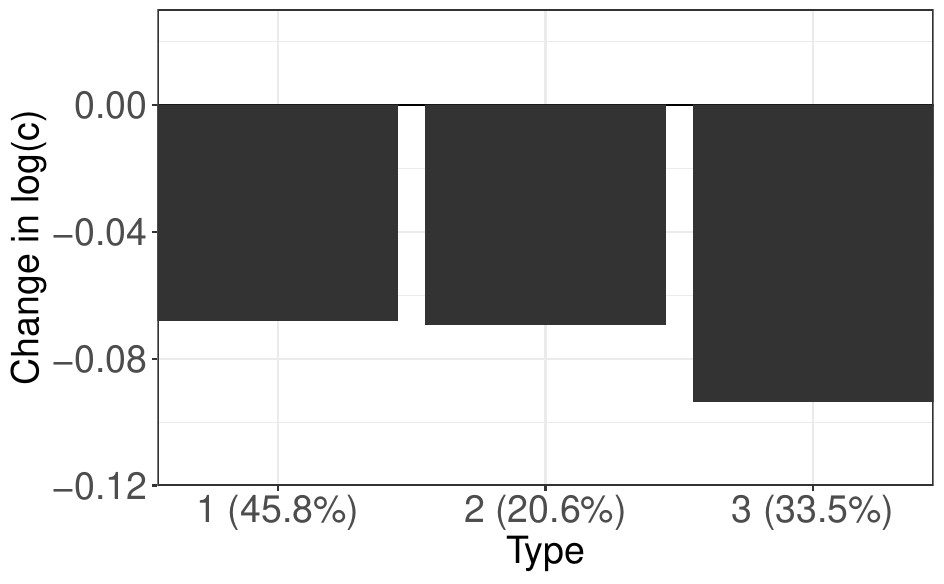}& 
		\includegraphics[width=5.35cm]{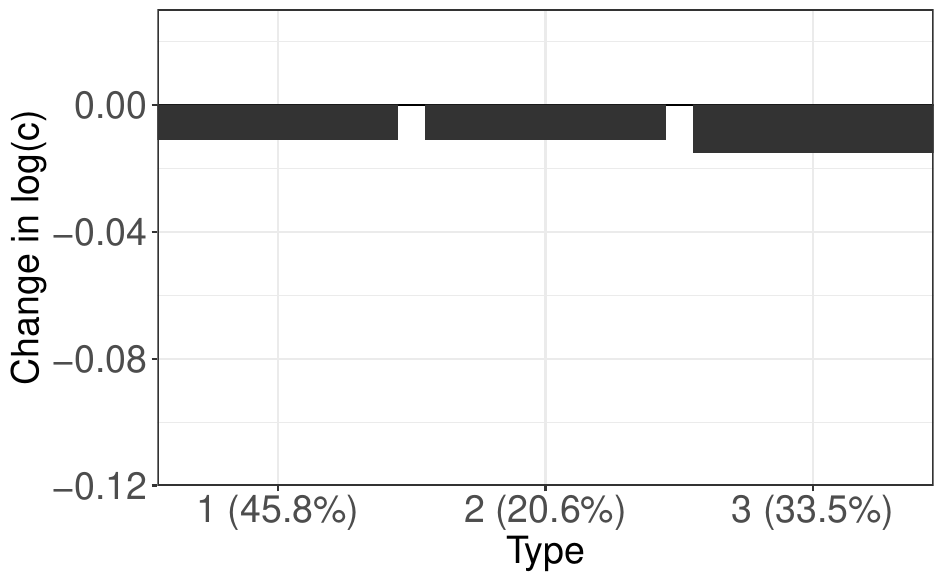}			
	\end{tabular}
\end{center}

{\footnotesize \textit{Notes: See the notes to Figure \ref{fig_decompo_cons_types}. Estimates based on a parametric finite-type model without beliefs. Type proportions in the cross-sectional sample are indicated on the x-axis.}}

\end{figure}

%%%%%%%%%%%%%%%%%%%%%%%%%%%%%%%%%%%%%%%%%%%%%%%%%
\begin{figure}[h!]
	\begin{center}
		\caption{Average partial effects estimates (OLS)}
		\label{fig_counter_ols}
		\begin{tabular}{ccc}
			\multicolumn{3}{c}{A. Mean beliefs only}\\
			(a) Transitory tax & (b) Permanent tax & (c) Regressivity\\
			\includegraphics[width=5.35cm]{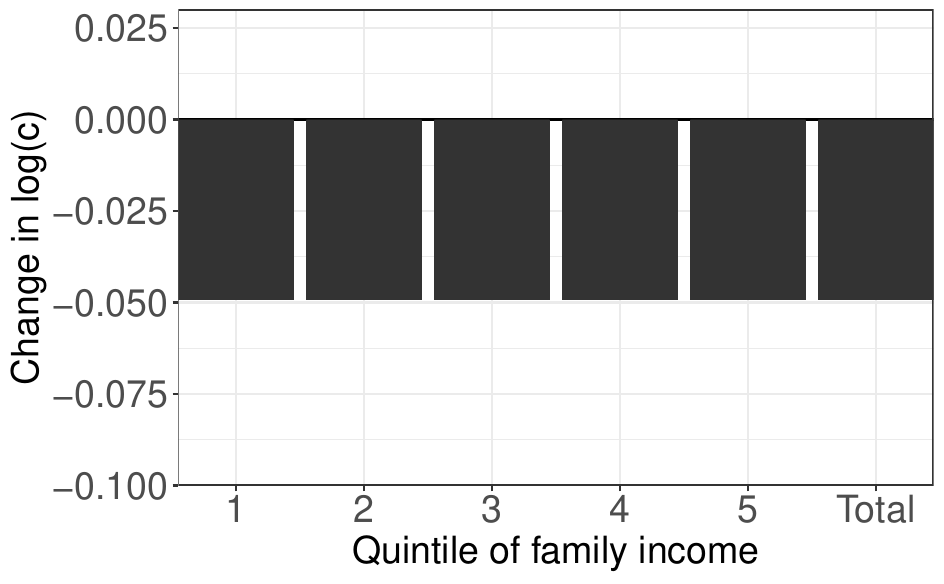}& 
			\includegraphics[width=5.35cm]{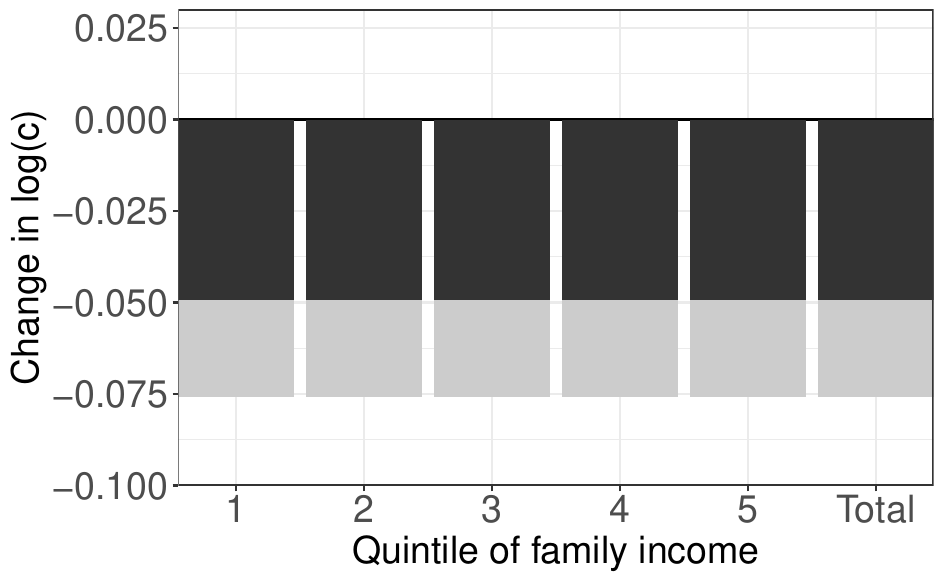}& 
			\includegraphics[width=5.35cm]{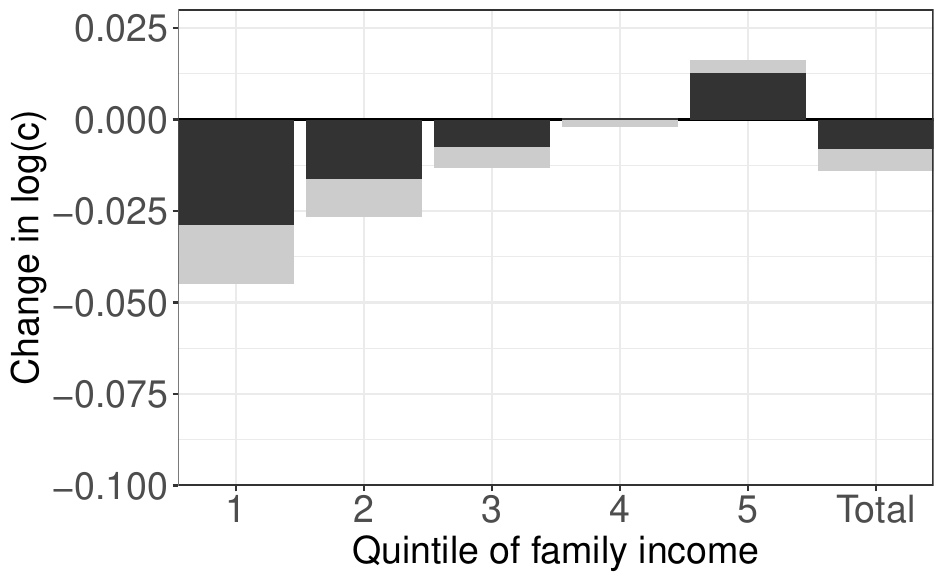}\\
			\multicolumn{3}{c}{B. Mean beliefs interacted with current log income}\\
			(d) Transitory tax & (e) Permanent tax & (f) Regressivity\\
			\includegraphics[width=5.35cm]{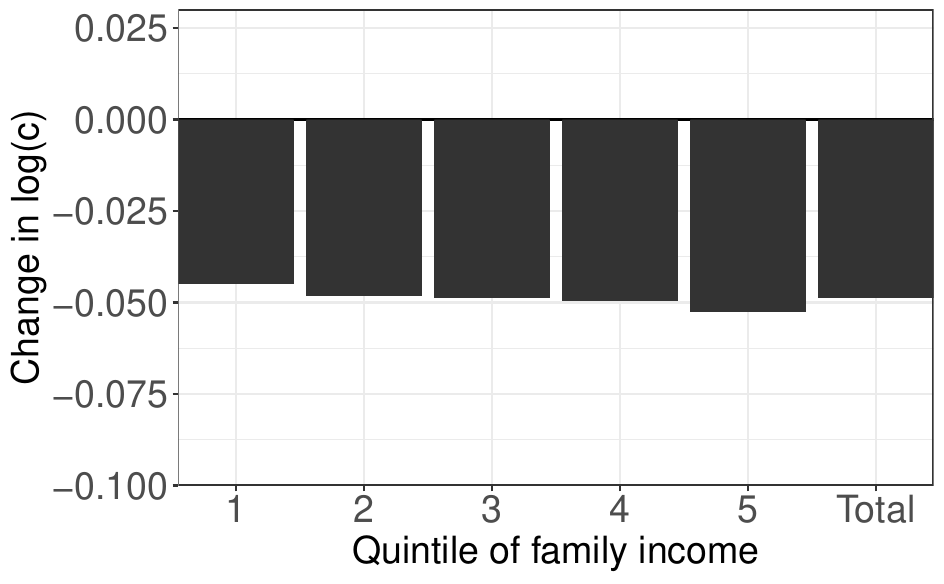}& 
			\includegraphics[width=5.35cm]{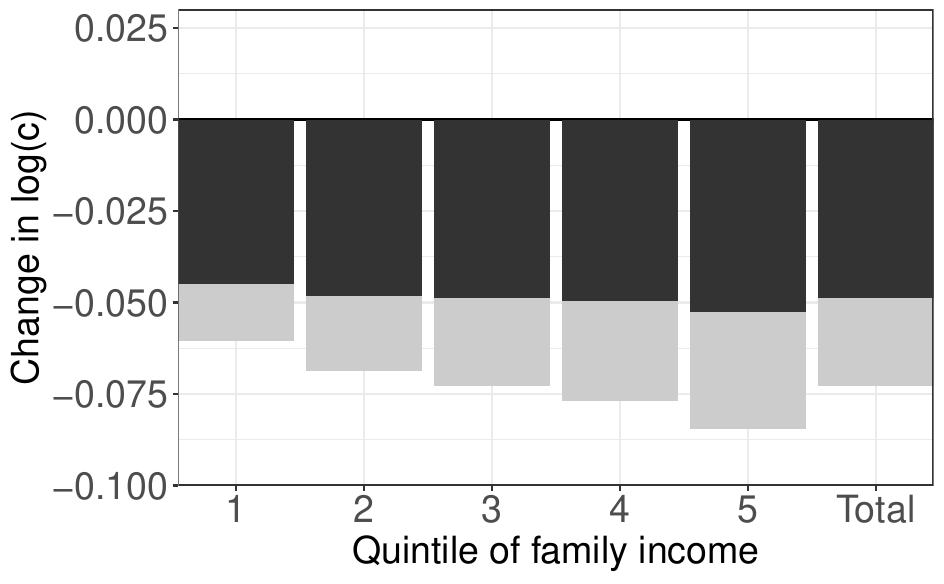}& 
			\includegraphics[width=5.35cm]{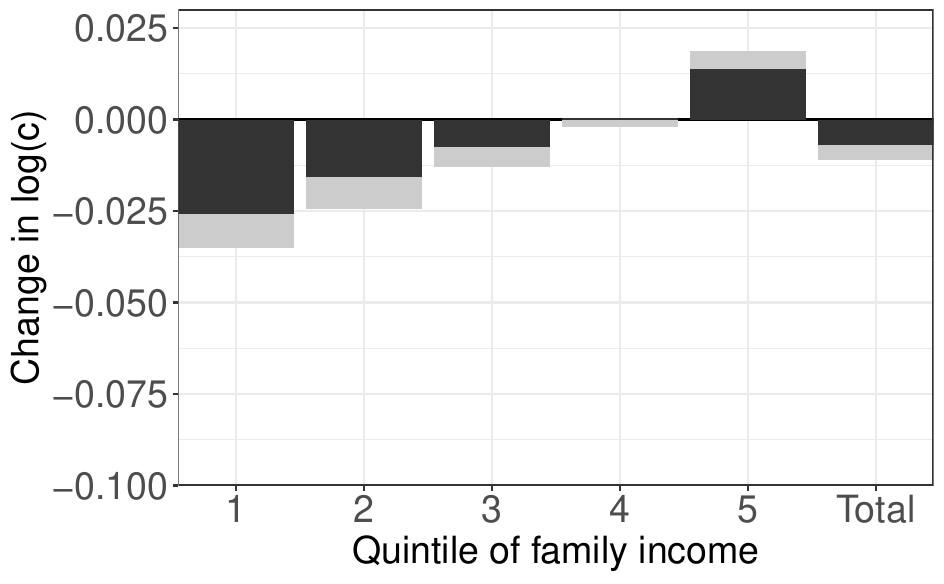}\\
			\multicolumn{3}{c}{C. Mean and variance of beliefs interacted with current log income}\\
			(g) Transitory tax & (h) Permanent tax & (i) Regressivity\\
			\includegraphics[width=5.35cm]{Plot_decomp_mean_var_i_count1.pdf}& 
			\includegraphics[width=5.35cm]{Plot_decomp_mean_var_i_count2.pdf}& 
			\includegraphics[width=5.35cm]{Plot_decomp_mean_var_i_count3.pdf}
		\end{tabular}
	\end{center}
	
	{\footnotesize \textit{Notes: SHIW, 1989--1991 and 1995--1998, cross-sectional sample. Black bars correspond to contemporaneous APE and grey bars correspond to dynamic APE. Total APE are the sums of CAPE and DAPE. The top panel is based on column (2) in Table \ref{Table_cons}, the middle panel on column (4), and the bottom panel on column (5).}}
\end{figure}

%%%%%%%%%%%%%%%%%%%%%%%%%%%%%%%%%%%%%%%%%%%%%%%%%
\begin{figure}[h!]
	\begin{center}
		\caption{Average partial effects estimates (Lasso)}
		\label{fig_counter_lasso}
		\begin{tabular}{ccc}
			\multicolumn{3}{c}{A. Double Lasso estimates, degree 2} \\ 
			(a) Transitory tax & (b) Permanent tax & (c) Regressivity\\
			\includegraphics[width=5.35cm]{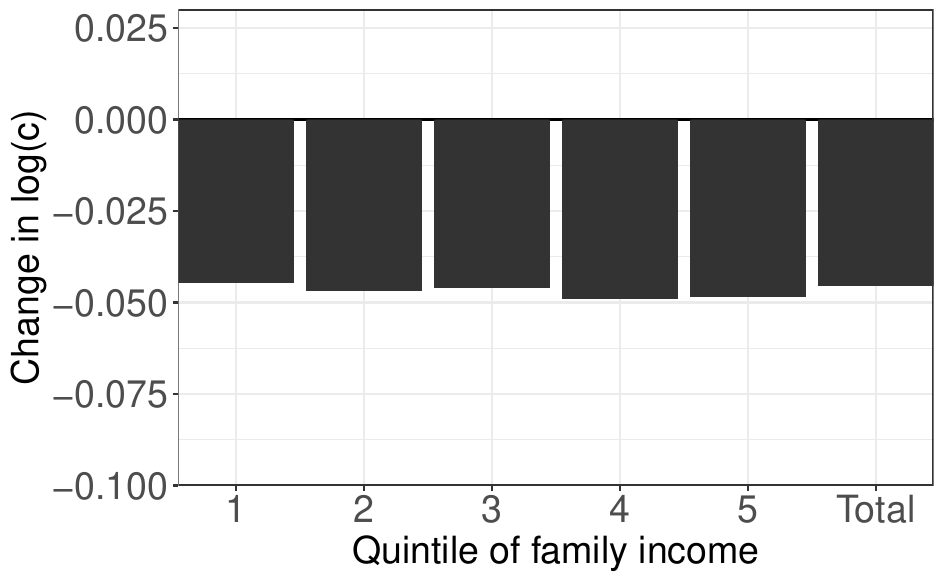}& 
			\includegraphics[width=5.35cm]{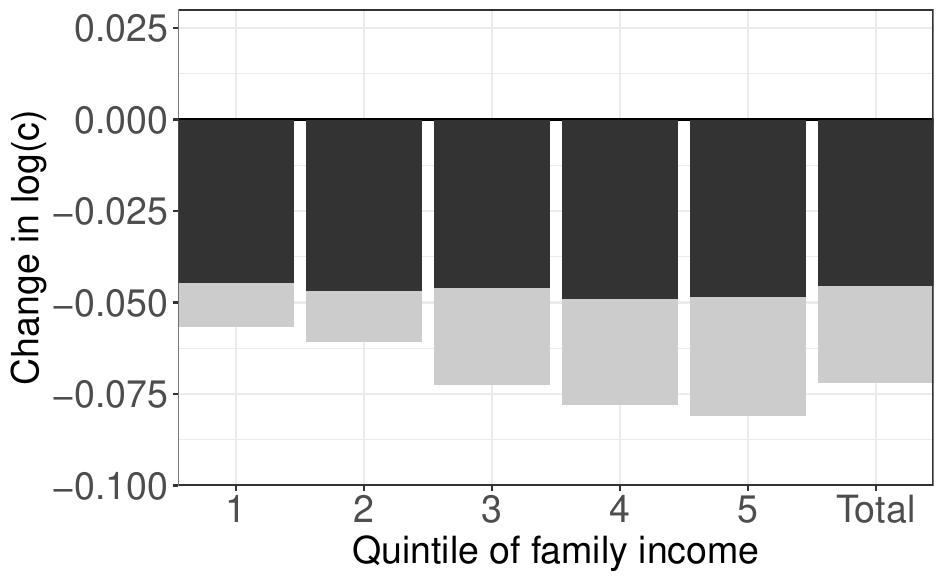}& 
			\includegraphics[width=5.35cm]{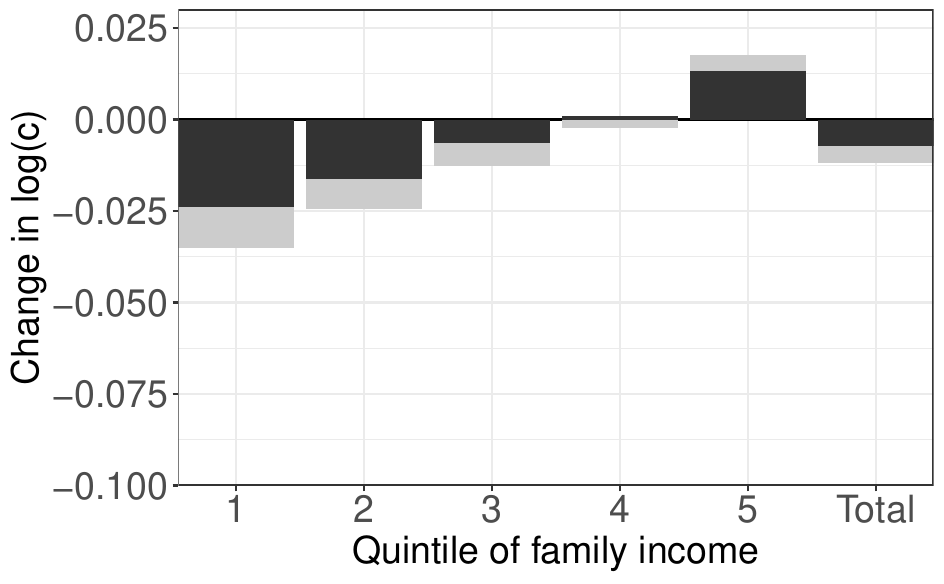}\\
			\multicolumn{3}{c}{B. Double Lasso estimates, degree 3} \\ 
			(d) Transitory tax & (e) Permanent tax & (f) Regressivity\\
			\includegraphics[width=5.35cm]{Plot_decomp_dlasso_dg3_count1.pdf}& 
			\includegraphics[width=5.35cm]{Plot_decomp_dlasso_dg3_count2.pdf}& 
			\includegraphics[width=5.35cm]{Plot_decomp_dlasso_dg3_count3.pdf}\\
			\multicolumn{3}{c}{C. Double Lasso estimates, degree 4} \\ 
			(g) Transitory tax & (h) Permanent tax & (i) Regressivity\\
			\includegraphics[width=5.35cm]{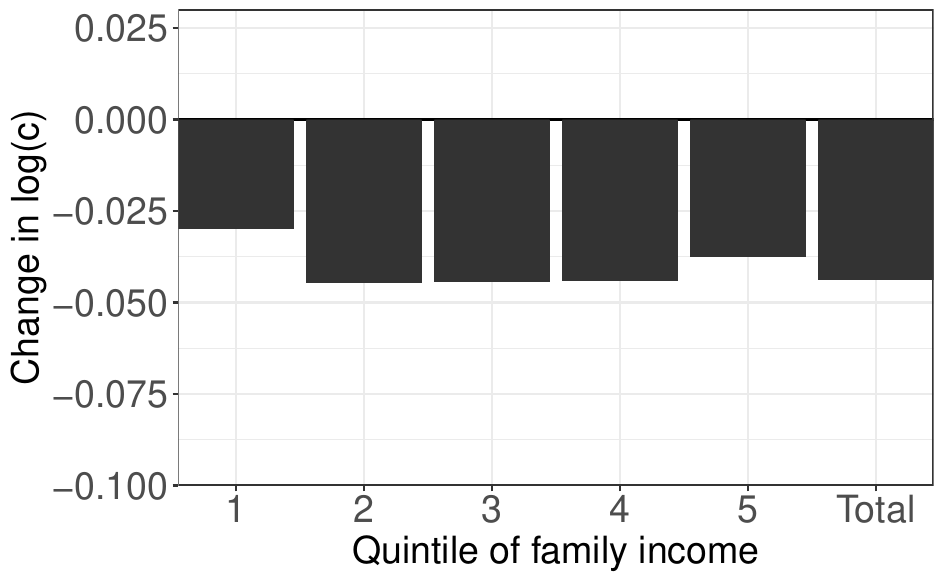}& 
			\includegraphics[width=5.35cm]{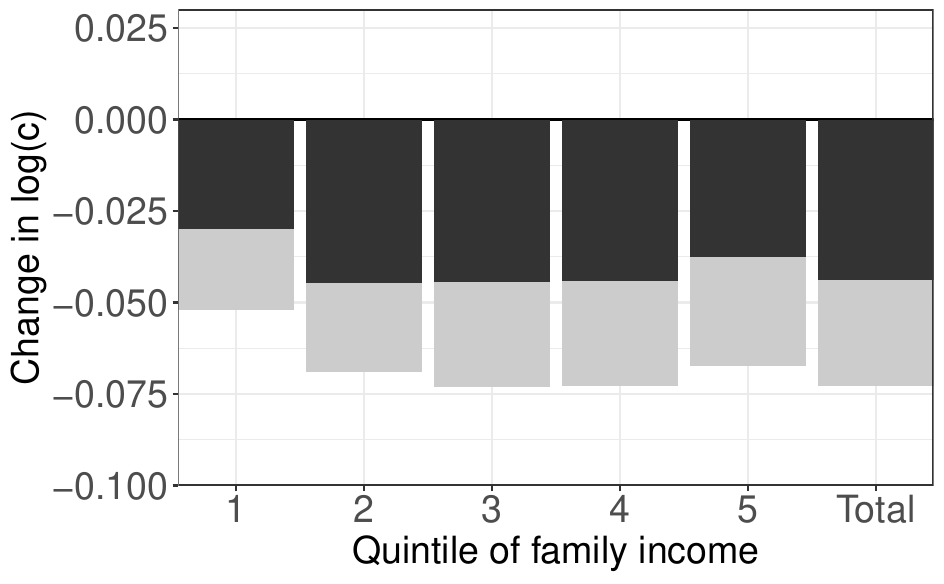}& 
			\includegraphics[width=5.35cm]{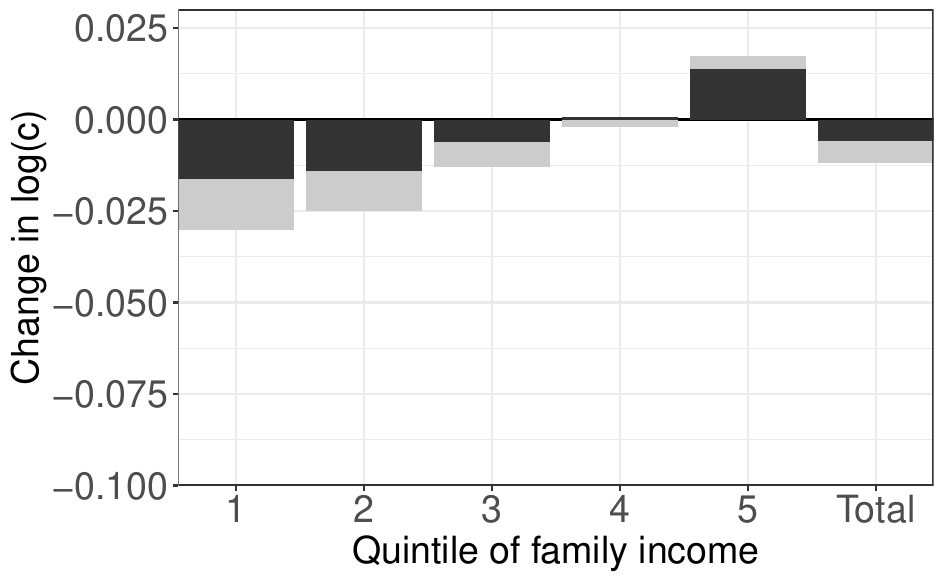}
		\end{tabular}
	\end{center}
	
	{\footnotesize \textit{Notes: SHIW, 1989--1991 and 1995--1998, cross-sectional sample. Black bars correspond to contemporaneous APE and grey bars correspond to dynamic APE. Total APE are the sums of CAPE and DAPE. Double Lasso estimates. The top panel is based on polynomials of degree 2, the middle panel on polynomials of degree 3, and the bottom panel on polynomials of degree 4.}}
\end{figure}

%%%%%%%%%%%%%%%%%%%%%%%%%%%%%%%%%%%%%%%%%%%%%%%%%%%%%%%%%%%%%%%%%

\begin{figure}[h!]
	\begin{center}
		\caption{{Average partial effects based on OLS, Monte Carlo simulation}}
		\label{fig_decompo_mcarlo}
		\begin{tabular}{ccc}
			\multicolumn{3}{c}{DGP with $K=2$} \\ 
			(a) Transitory tax & (b) Permanent tax & (c) Regressivity\\
			\includegraphics[width=5.35cm]{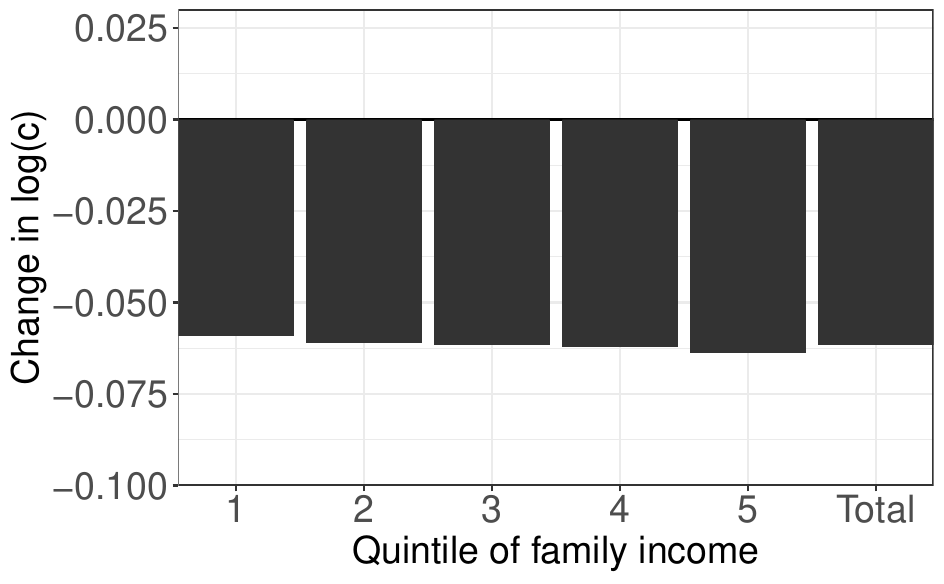}& 
			\includegraphics[width=5.35cm]{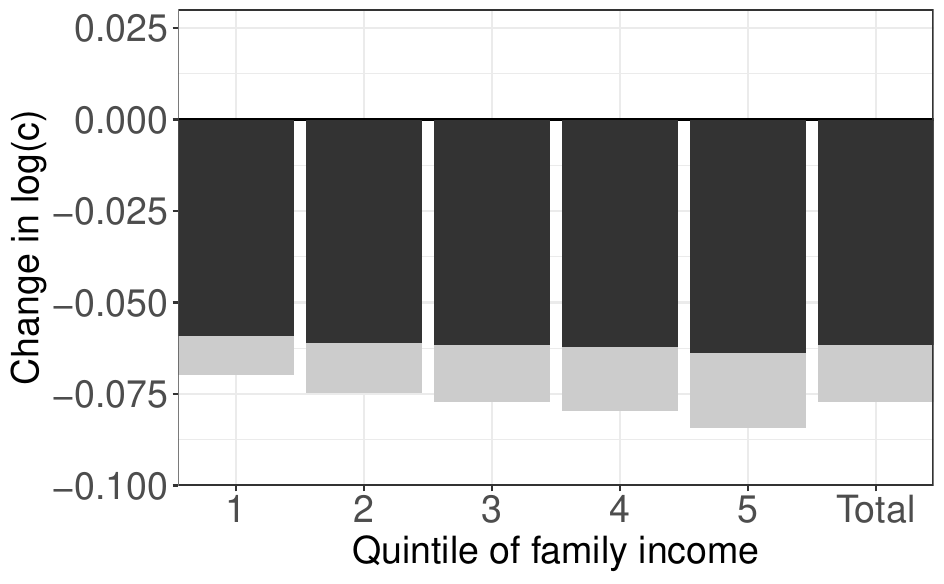}& 
			\includegraphics[width=5.35cm]{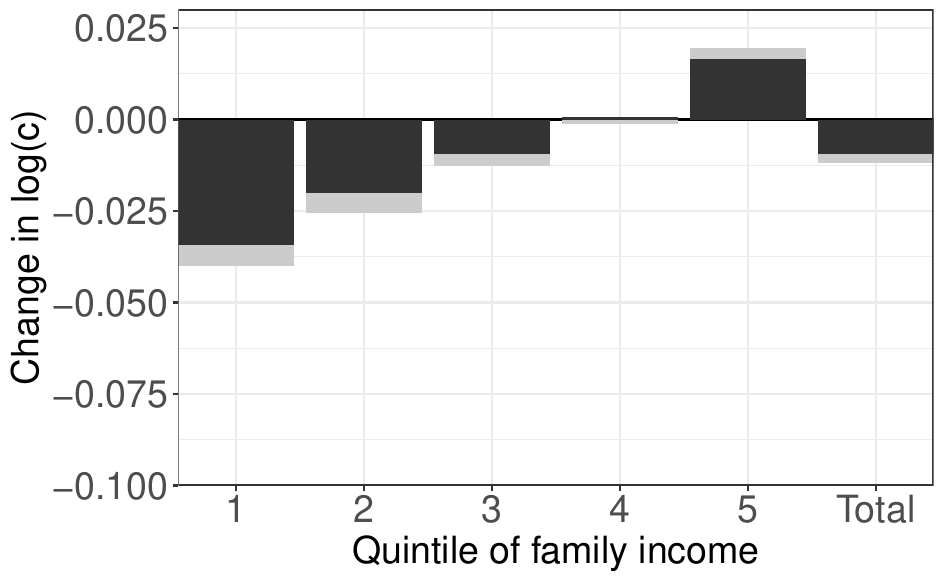}\\
			\multicolumn{3}{c}{DGP with $K=3$} \\ 
			(d) Transitory tax & (e) Permanent tax & (f) Regressivity\\
			\includegraphics[width=5.35cm]{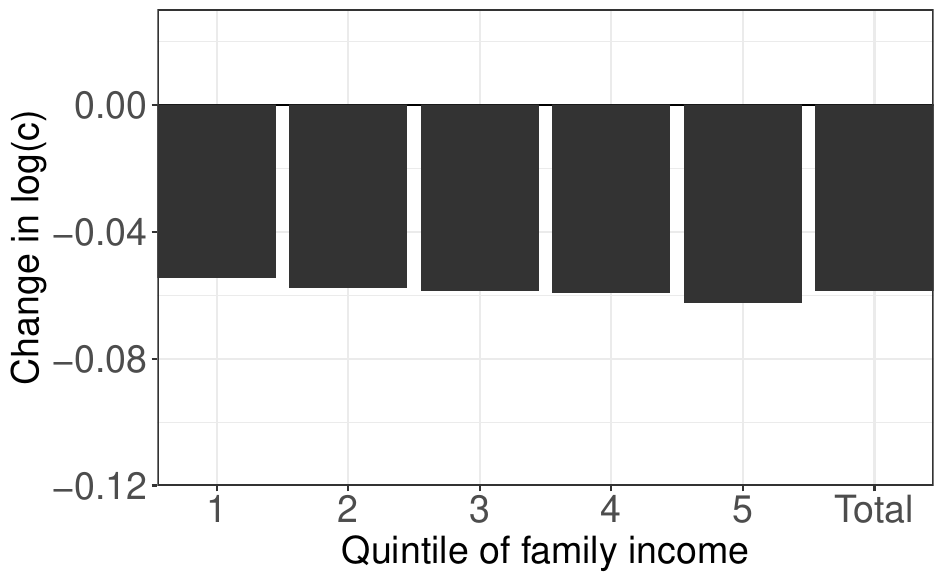}& 
			\includegraphics[width=5.35cm]{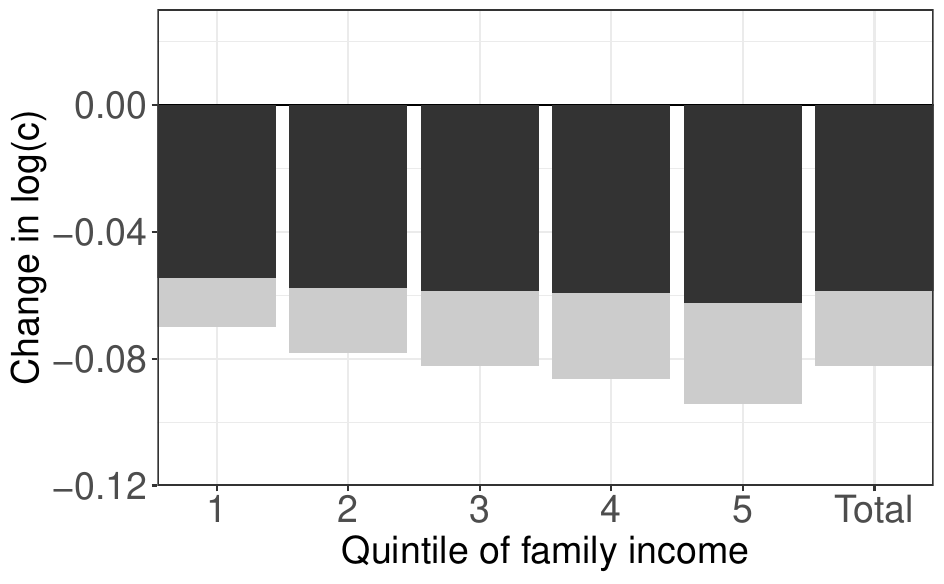}& 
			\includegraphics[width=5.35cm]{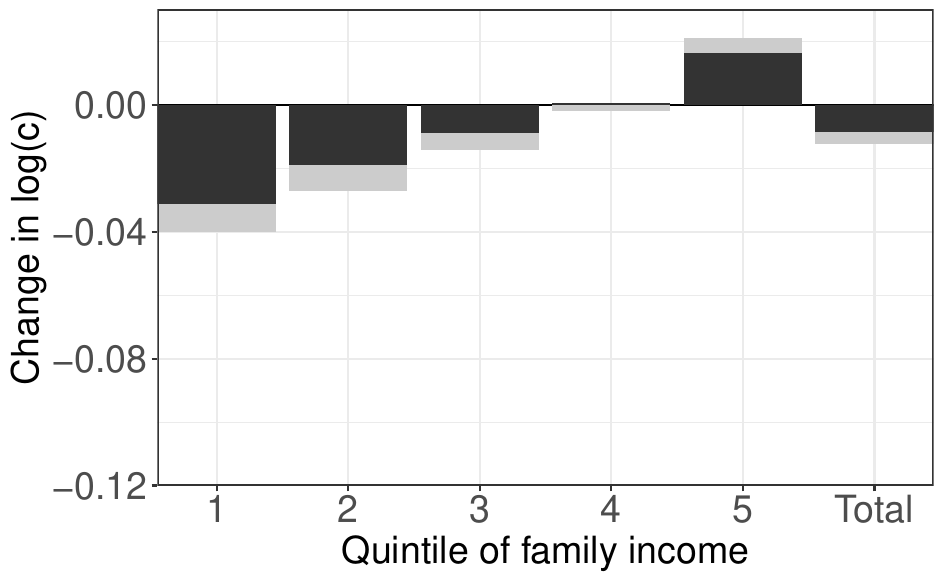}
		\end{tabular}
	\end{center}
	
	{\footnotesize \textit{Notes: See the notes to Figure \ref{fig_decompo_cons_types}. Results averaged across 10,000 simulations based on OLS in first differences, see the estimates reported in Table \ref{tab_counter_mcarlo}. The underlying data generating process is a finite mixture model with $K=2$ types in the top panel and $K=3$ types in the bottom panel.}}
	
\end{figure}

%%%%%%%%%%%%%%%%%%%%%%%%%%%%%%%%%%%%%%%%%%%%%%%%%%%%%%%%%%%%%%%%%

\clearpage

% Policy rules

%%%%%%%%%%%%%%%%%%%%%%%%%%%%%%%%%%%%%%%%%%%%%%%%%%%%%%%%%%%%%%%%%%%%%%%%%%%%%%%%%%%%%%%%%%%%%%%%%%%%%%%%%%%%%%%%%%%%%%%%%%%%%
% Consumption, assets and income

\end{document}